\documentclass[lettersize,journal]{IEEEtran}

\usepackage{fancyhdr}
\usepackage{eso-pic}    % 用于向页面插入文本
\usepackage{lipsum}     % 可选：用于示例文字

\usepackage{algorithm}  
\usepackage{algorithmic}
\usepackage{amsmath}  
\usepackage{amsfonts}
\usepackage{array}
\usepackage[caption=false,font=footnotesize]{subfig}
\usepackage{textcomp}
\usepackage{stfloats}
\usepackage{color}
\usepackage{xcolor}
\usepackage{url}
\usepackage{verbatim}
\usepackage{graphicx}
\usepackage{subfig}
\usepackage{balance}
\usepackage{adjustbox}
\usepackage{bm}
\usepackage{amssymb}
\usepackage[mathscr]{eucal}
\usepackage{flushend}
\usepackage{cite}
\usepackage{soul}
\usepackage{booktabs}
\usepackage{threeparttable}
\usepackage{verbatim}
\usepackage{tablefootnote}

\usepackage{cleveref}   %可以调用 \Cref{fig:xxx}或者 \ref{fig:xxx} 
\newcommand\mynote[1]{\textcolor{black}{#1}}

\newcommand\mynoteblue[1]{\textcolor{black}{#1}}

\begin{document}

\AddToShipoutPicture*{
  \AtPageUpperLeft{
    \put(20,-20){\parbox{\textwidth}{\footnotesize
    \centering
    This work was accepted for publication in IEEE Transactions on Mobile Computing and was published on April 10, 2025. This is the accepted version.\\
    DOI: 10.1109/TMC.2025.3559927 \quad Final published version available at: \url{https://ieeexplore.ieee.org/document/10962318}
    }}
  }
}

\title{\textit{Baton}: Compensate for Missing Wi-Fi Features \\ for Practical Device-free Tracking}
\author{Yiming~Zhao,~\IEEEmembership{}
        Xuanqi~Meng,~\IEEEmembership{}
        Xinyu~Tong,~\IEEEmembership{}
	Xiulong~Liu,~\IEEEmembership{}
        Xin~Xie,~\IEEEmembership{}
	Wenyu~Qu~\IEEEmembership{}

\thanks{Yiming~Zhao, Xuanqi~Meng, Xinyu~Tong, Xiulong~Liu, Xin~Xie and Wenyu~Qu are with the College of Intelligence and Computing, Tianjin University, Tianjin, China, 300350.}
\thanks{E-mail: \{zhaoyiming, xuanqimeng, xytong, xiulong\_liu, xinxie, wenyu.qu\}@tju.edu.cn.}
\thanks{Corresponding author: Xinyu Tong.}}
  
\maketitle

\thispagestyle{fancy}
\fancyhf{}  % 清空原始 header/footer
\renewcommand{\headrulewidth}{0pt}
\renewcommand{\footrulewidth}{0pt}

\fancyfoot[C]{\footnotesize
© 2025 IEEE. Personal use of this material is permitted.  Permission from IEEE must be obtained for all other uses, in any current or future media, including reprinting/republishing this material for advertising or promotional purposes, creating new collective works, for resale or redistribution to servers or lists, or reuse of any copyrighted component of this work in other works.}

\begin{abstract}
Wi-Fi contact-free sensing systems have attracted widespread attention due to their ubiquity and convenience.
The integrated sensing and communication (ISAC) technology utilizes off-the-shelf Wi-Fi communication signals for sensing, which further promotes the deployment of intelligent sensing applications.
However, current Wi-Fi sensing systems often require prolonged and unnecessary communication between transceivers, and brief communication interruptions will lead to significant performance degradation.
This paper proposes \textit{Baton}, the first system capable of accurately tracking targets even under severe Wi-Fi feature deficiencies.
To be specific, we explore the relevance of the Wi-Fi feature matrix from both horizontal and vertical dimensions.
The horizontal dimension reveals feature correlation across different Wi-Fi links, while the vertical dimension reveals feature correlation among different time slots.
Based on the above principle, we propose the \underline{S}imultaneous \underline{T}racking \underline{A}nd \underline{P}redicting (STAP) algorithm, which enables the seamless transfer of Wi-Fi features over time and across different links, akin to passing a baton.
We implement the system on commercial devices, and the experimental results show that our system outperforms existing solutions with a median tracking error of $0.46m$, even when the communication duty cycle is as low as $20.00\%$. \mynote{Compared with the state-of-the-art, our system reduces the tracking error by $79.19\%$ in scenarios with severe Wi-Fi feature deficiencies.}
\end{abstract}

\begin{IEEEkeywords}
Wi-Fi, Wireless Sensing, Channel State Information
\end{IEEEkeywords}

% Abstract
% \begin{abstract}
% Wi-Fi contact-free sensing systems have attracted widespread attention due to their ubiquity and convenience.
% %
% The Integrated Sensing and Communication (ISAC) technology utilizes off-the-shelf Wi-Fi communication signals for sensing, which further promotes the deployment of intelligent sensing applications.
% %
% However, current Wi-Fi sensing systems often require prolonged and unnecessary communication between transceivers, and brief communication interruptions will lead to significant performance degradation.
% %
% This paper proposes \textit{Baton}, the first system capable of accurately tracking targets even under severe Wi-Fi feature deficiencies.
% %
% To be specific, we explore the relevance of the Wi-Fi feature matrix from both horizontal and vertical dimensions.
% %
% The horizontal dimension reveals feature correlation across different Wi-Fi links, while the vertical dimension reveals feature correlation among different time slots.
% %
% Based on the above principle, we propose the \underline{S}imultaneous \underline{T}racking \underline{A}nd \underline{P}redicting (STAP) algorithm, which enables the seamless transfer of Wi-Fi features over time and across different links, akin to passing a baton.
% %
% We implement the system on commercial devices, and the experimental results show that our system outperforms existing solutions with a median tracking error of $0.46m$, even when the communication duty cycle is as low as $20.00\%$.
% \end{abstract}

% Part 1: Introduction
\section{Introduction}
\subsection{Background and Motivation}
\begin{figure}[t]
\centering
\subfloat[\mynote{When the CDC is lower than $100\%$, there may be no packet transmitted in a certain link at a given moment (\textit{e.g.} Link A)}]{\includegraphics[width = 0.95\linewidth]{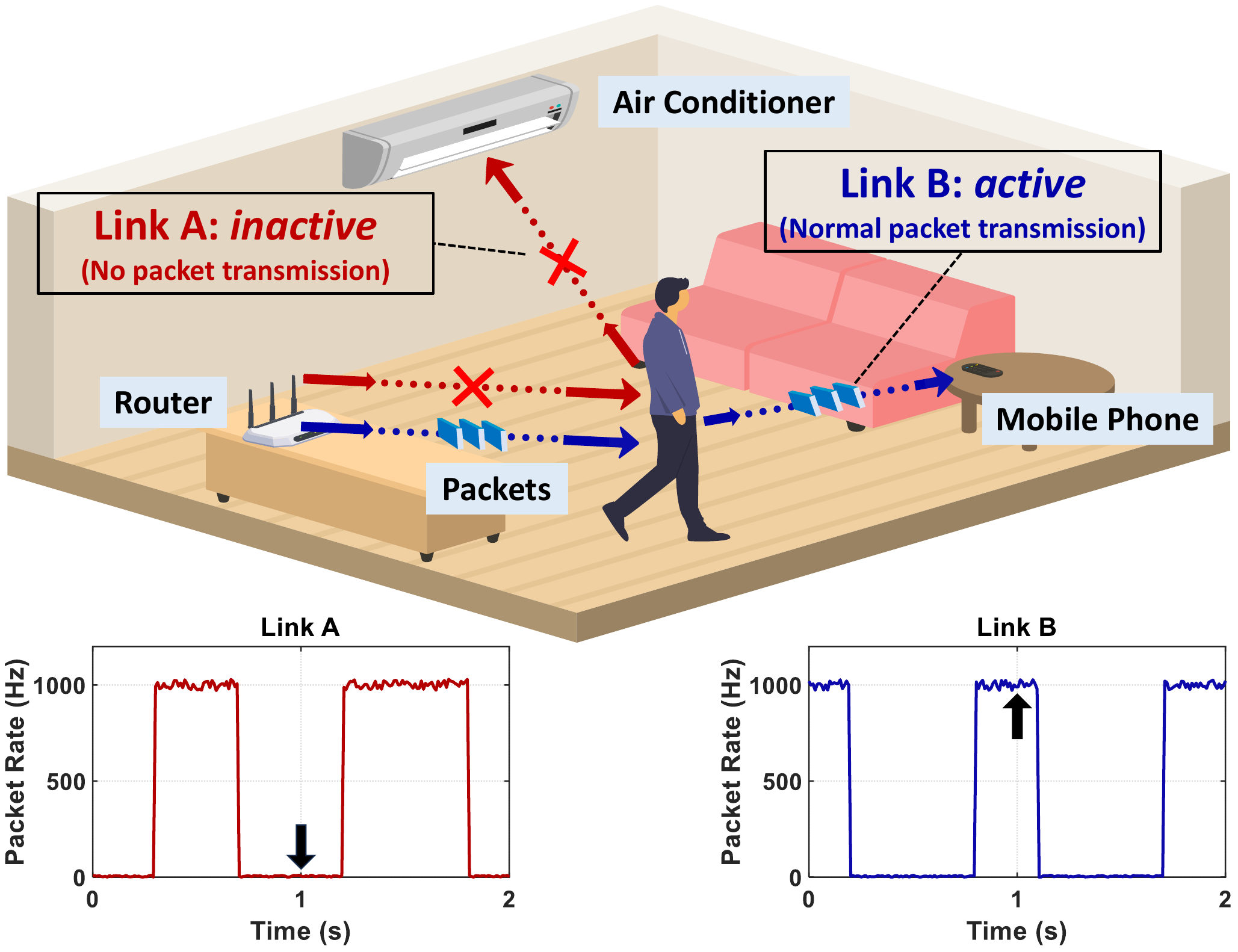}\label{fig:intro_SmartHome}}\\
\subfloat[Fresnel zone model-based tracking results with different CDCs]{\includegraphics[width = 1\linewidth]{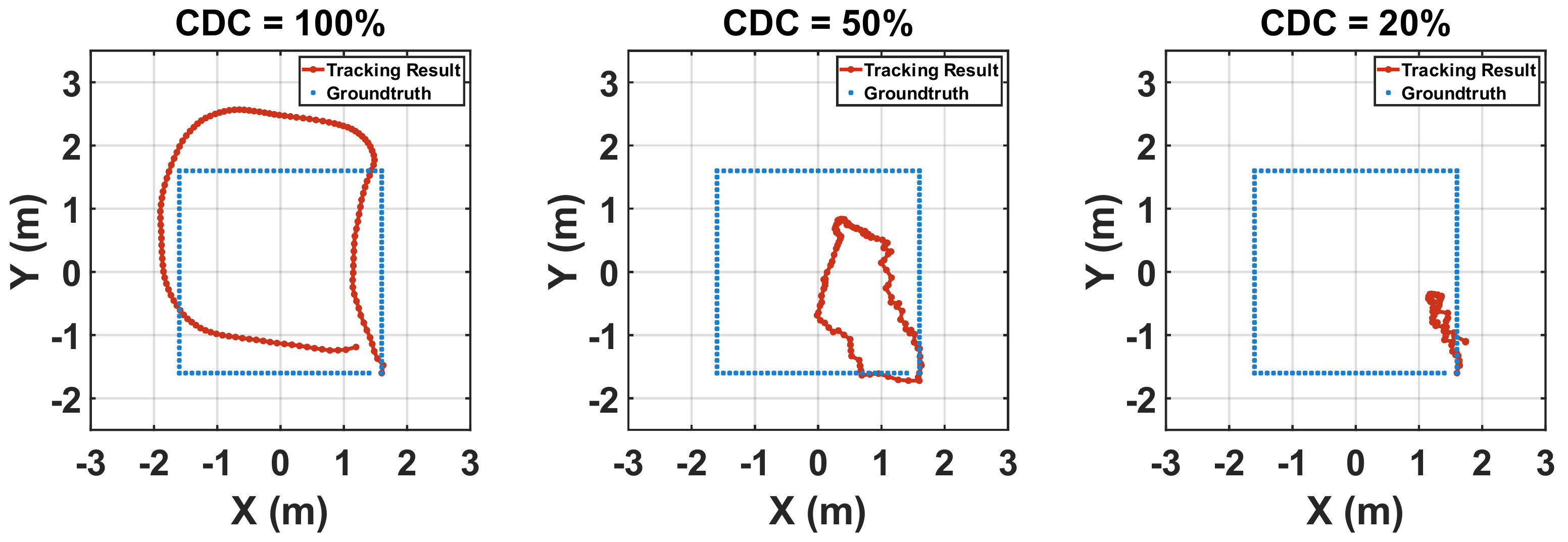}\label{fig:cdc_motivation}}
\caption{Application and motivation.}
\vspace{-3mm}
\end{figure}

\IEEEPARstart{S}{ensing} and communication are two crucial foundations of the Internet of things (IoT), and the integration of these two functions has become a critical issue in the design of IoT \cite{liu2023towards,meng2023securfi,wang2022placement}.
Non-contact sensing, as a typical example of integrated sensing and communication (ISAC)~\cite{cui2021integrating,liu2022integrated}, utilizes communication signals reflected by the human body to sense target behavior without physical contact, and has attracted widespread attention~\cite{wu2022wifi,zeng2019farsense,zheng2019zero,tian2024device}.
Such methods can track users by utilizing packets for communication between transceivers, without requiring them to send additional packets specifically for sensing.
The example is illustrated in Fig.~\ref{fig:intro_SmartHome}, where we can utilize the communication between the transmitter and the receiver to track the user who does not carry the Wi-Fi devices.

However, current systems are not effectively integrating sensing and communication functions due to the following reason: 
\textit{wireless sensing requires continuous and stable communication, whereas IoT traffic is typically characterized by small, fast bursts }\cite{sivanathan2018classifying,tahaei2020rise}.
Specifically, IoT devices have very short traffic flow durations.
While most traffic flows of non-IoT devices have a duration between $1s$ and $1,000s$, IoT devices in smart homes are characterized by very short traffic flows close to $0s$~\cite{mainuddin2021network}.
To provide stable sensing performance, traditional sensing systems \cite{wu2016widir,tian2023wstrack,zhang2023understanding} usually maintain frequent communication packets between each pair of transceivers.
Unfortunately, it is not always feasible to maintain such frequent communication between devices and routers in real applications.
Inevitably, those frequent communications dedicated to sensing (\textit{e.g.}, link A in Fig.~\ref{fig:intro_SmartHome}) will occupy the normal communication resources of the router with other devices (\textit{e.g.}, link B in Fig.~\ref{fig:intro_SmartHome}), so communication and sensing cannot be perfectly integrated.
\mynote{
In fact, intermittent communication between transceivers is typical in real-world IoT devices, which is the cause of missing Wi-Fi features.
Under such a condition, the absence of Wi-Fi features can persist for a while in any communication link.
During this period, there is no packet transmitted in the given link.
Hence, this scenario is different from the case with a low packet sampling rate.
For example, during a period of time (e.g. around $t=1s$) in Fig.~\ref{fig:intro_SmartHome}, there is normal packet transmission in Link B, but no transmission in Link A.
}

To visually demonstrate the influence of intermittent Wi-Fi communication on sensing, we conduct a comparison of tracking performance across various communication duty cycles in Fig.~\ref{fig:cdc_motivation}, where the communication duty cycle (CDC) refers to the effective communication packets that can be used for sensing.
For example, with a CDC as low as $20\%$, there is no packet available for sensing during $80\%$ of a time slot.
The motivational experiments, utilizing the Fresnel zone model-based tracking method, clearly demonstrate a decrease in tracking performance with reduced CDCs. 
When the CDC is reduced to $20\%$, the tracking error has increased to more than $2.5\times$ of the baseline.
The above experiments demonstrate that using non-sequential communication packets for sensing significantly affects tracking performance.

\subsection{\textit{Baton} in a Nutshell}
The inherent conflict between sensing and communication drives us to develop a practical tracking system called \textit{Baton}.
The primary objective is to investigate the correlation among multiple Wi-Fi links and leverage this correlation to compensate for any missing sensing features.
In doing so, we aim to enable the seamless transfer of Wi-Fi features over time, akin to passing a baton.
As the number of Wi-Fi devices in smart homes continues to increase, there is a growing practical significance in exploring the association among multiple Wi-Fi links to compensate for missing sensing features.

\subsection{Challenges and Solutions}
It is not trivial to design such a system, and we need to address the following challenges:

\textit{Challenge and solution 1: how to compensate for missing features while tracking users? }
The accuracy of tracking and feature prediction are mutually dependent.
In other words, accurate tracking relies on the known features, while predicting features requires knowledge of the user's trajectory during the previous moment.
To achieve \underline{S}imultaneous \underline{T}racking \underline{A}nd \underline{P}redicting (STAP), we theoretically and experimentally prove that the signal correlation at different times and across different Wi-Fi links.
In the proposed \mynote{system}, we design a novel reliability matrix to balance different prediction methods, so that we can realize accurate tracking.

\textit{Challenge and solution 2: how to determine the user's initial velocity in the absence of Wi-Fi features?}
For a low CDC, it is also difficult to determine the initial velocity to start the STAP algorithm.
To address this problem, we make full use of the limited non-missing feature data that are available.
By exploiting the continuity of signal features, we can obtain a relatively accurate initial position prediction sequence, from which we can determine the initial velocity of the user.
This partial prediction lays the foundation for the execution of the STAP algorithm.

\subsection{Contributions and Advantages over Prior Arts}
This paper for the first time realizes device-free tracking under discontinuous Wi-Fi links.
The contributions of \textit{Baton} are as follows:

\begin{enumerate}
    \item We explore the essential signal correlations among different time slots and Wi-Fi links.
    Based on these correlations and mathematical modeling, we propose mechanisms to compensate for missing Wi-Fi features in practical device-free tracking.
    \item We propose the STAP algorithm, a novel method to realize simultaneous tracking and predicting, which achieves accurate device-free tracking under severe Wi-Fi feature deficiencies.
    \item We implement the prototype with commercial off-the-shelf (COTS) Wi-Fi devices. The experimental results demonstrate that the median tracking error of \textit{Baton} is $0.46m$ when the CDC is $20.00\%$, which reduces the tracking error by $79.19\%$ compared with the state-of-the-art.
\end{enumerate}

The advantage of the \textit{Baton} system over previous work is as follows:
We realize sensing in non-persistent communication scenarios, thus relaxing the impractical requirements of sensing technology for communication.
Consequently, the proposed method has made significant strides toward ISAC for Wi-Fi-based systems.

\mynote{
The remainder of this paper is organized as follows.
\S Section~\ref{sec:related} reviews the related work, covering both device-based and device-free tracking methods as well as feature interpolation and compensation techniques.
In \S Section~\ref{sec:basicidea}, we discuss the empirical observations and theoretical analysis of the foundational ideas behind the \textit{Baton} system. 
\S Section~\ref{sec:overview} provides an overview of the system design, highlighting the three main stages of \textit{Baton}. 
\S Section~\ref{sec:sysdesign} describes the system's architecture in depth, explaining its three primary stages: signal feature processing, predicting, and tracking.
\S Section~\ref{sec:exp} presents the implementation details and experimental evaluation of \textit{Baton} under various conditions. 
Finally, we conclude the paper in \S Section~\ref{sec:conc}.
}
% The remainder of this paper is organized as follows.
% %
% We discuss the related work in \S Section~\ref{sec:related}.
% %
% We present the empirical observations and basic idea in \S Section~\ref{sec:basicidea}, and
% the system overview in \S Section~\ref{sec:overview}.
% %
% We introduce the proposed system in \S Section~\ref{sec:sysdesign}.
% %
% \S Section~\ref{sec:exp} presents the implementation and the experimental results.
% %
% Finally, we conclude the paper in \S Section~\ref{sec:conc}.

\section{Related Work}\label{sec:related}
Indoor tracking and localization have accumulated much interest in recent decades \cite{yang2013rssi,tian2023wireless}.
The existing proposed tracking methods can be generally categorized into two sets, device-based tracking \cite{kotaru2015spotfi,tong2018fineloc,tong2021mapfi,zhu2021bls} and device-free tracking \cite{qian2017widar,xie2019md,ge2023crosstrack,tian2023wstrack}.
In this section, we provide a review of relevant research efforts.
\subsection{Device-based Tracking}
Since common mobile devices such as cell phones, tablets and laptops can conveniently provide received signal strength indicator (RSSI) as WiFi receivers, RSSI has become a widely used feature for tracking \cite{li2013smartphone,wang2013bluetooth,tong2018fineloc,hoang2019recurrent}.
However, the multi-path effect and the desynchronization between devices make high-precision RSSI tracking difficult to achieve.
\mynote{
Some sensing systems such as Travi-Navi \cite{zheng2014travi} use computer vision for indoor tracking and navigation.
Pallas \cite{luo2016pallas} uses Wi-Fi monitors to realize self-bootstrapping for smartphones.
Beyond tracking for users, SoM \cite{zheng2022sound} realizes wrist tracking with a smart watch.
}

As channel state information (CSI) provides more information with multiple antennas and multiple carriers, accurate tracking is now feasible.
ArrayTrack~\cite{xiong2013arraytrack} builds an antenna array based on multiple-input, multiple-output techniques to analyze the angle of arrival (AoA) of signals and achieve high-accuracy mobile device tracking.

%
%SpotFi \cite{kotaru2015spotfi} utilizes a super-resolution algorithm to estimate the AoA and time of flight between the targets and access points.
%
Inspired by RSSI indoor positioning, several studies demonstrate that CSI fingerprinting can eliminate the effects of multi-path \cite{wang2016csi,gao2020crisloc,wang2015phasefi,tong2020csi}.
PhaseFi \cite{wang2015phasefi} extracts the phase information in CSI as a fingerprint database and designs a deep-learning network for offline training.
\mynote{RoArray \cite{gong2018roarray} is a sensing system under a low Signal-to-Noise Ratio (SNR) with AoA.}
Research \cite{tong2020csi} proposes a scheme to automatically update the CSI fingerprint database to improve the efficiency of localization without field collection of fingerprints.
MapFi\cite{tong2021mapfi} and DLoc \cite{ayyalasomayajula2020deep} construct environmental maps with AoA using the modeling method and deep learning method, respectively.
As a step further, BLS-Location \cite{zhu2021bls} and ResLoc \cite{wang2020indoor} both train the network with a small size of data to minimize the time for manual collection and construction of the fingerprint database.
Although the above methods make efforts to track indoors, they all require users to carry electronic devices in order to work, which undoubtedly reduces users' willingness to use them.

\subsection{Device-free Tracking}
Device-free tracking has significant advantages over device-based tracking since it extracts and analyzes features of the signal without requiring the user to carry any equipment.
Widar \cite{qian2017widar} first infers the user's velocity with the path length change rates of multiple Wi-Fi links.
Indotrack \cite{li2017indotrack} proposes a joint Doppler-MUSIC and Doppler-AOA estimation of the velocity and position of the target to determine the trajectory.
Widar2.0 \cite{qian2018widar2} and md-Track \cite{xie2019md} go further by combining the AoA, time of flight (ToF) and Doppler frequency shift (DFS) to achieve single-link tracking.
WSTrack \cite{tian2023wstrack} combines Wi-Fi DFS and acoustic AoA to realize multi-modal tracking.
Extending this idea, WSTrack+ \cite{tian2024device} achieves tracking and recognizing gait patterns with smart speakers.
\mynote{
Real-time functionality is important for indoor tracking and other sensing applications.
WiDFS \cite{wang2022single} is a real-time passive tracking system for a single target.
Beyond indoor tracking, other sensing systems such as HandFi \cite{ji2023construct} can construct hand skeletons with commercial Wi-Fi devices in real-time.}
CrossTrack \cite{ge2023crosstrack} proposes a method to eliminate the limitation of tracking when users cross Wi-Fi links.
Beyond that, NNE-Tracking~\cite{tong2024nne} demonstrates a wireless sensing framework that can effectively combat environmental noise by generating large-scale data to train the framework while a physical model serves as a supervisor for the training.
Wireless localization at line-of-sight (LoS) has improved dramatically, but the models described above, which are based on near-ideal scenarios, often struggle to cope with the presence of extensive occlusion in the environment.
\mynote{
Several existing works focus on solving the problem of multi-target tracking. For example, MultiAuth \cite{kong2021multiauth} employs AoA to distinguish multipath components in CSI for multi-user authentication. WiPolar \cite{venkatnarayan2020leveraging} utilizes circular polarized antennas for multi-target tracking but relies on dense equipment deployment.}
To break the limitation of obstructions, tracking in non-line-of-sight (NLoS) has also become available.
NLoc \cite{zhang2022toward} models the blocked reflections and virtual direct signals to enable NLoS localization.
HyperTracking~\cite{xu2024hypertracking} proposes an NLoS tracking method combining spatial model features and neural networks to remove the interference of obstructions on localization with the elimination of common paths as the core.
\mynote{
\subsection{Feature Interpolation and Compensation}
The interpolation and compensation of Wi-Fi features play a critical role in enhancing localization accuracy.
Yang \cite{yang2020indoor} proposes a dual-frequency interpolation-based indoor localization system (DFD-CSI) that leverages cubic spline interpolation to upgrade fingerprint databases and improve robustness against device heterogeneity.
While their approach reduces the calibration time and enhances positioning accuracy using spatial interpolation, our work focuses on compensating for missing Wi-Fi features under low communication duty cycles.
Wang et al. \cite{wang2020fingerprinting} propose a fingerprint-based indoor localization system that uses interpolated preprocessed CSI phases to improve positioning accuracy.
Zhao et al. \cite{zhao2016applying} apply universal Kriging interpolation to Wi-Fi fingerprinting systems, generating interpolated RSS values to reduce database construction efforts and improve localization accuracy. 
However, the approaches above primarily focus on spatial interpolation for device-based localization in static conditions.
}

\mynote{
In addition to interpolation, several studies focus on compensating for specific errors in Wi-Fi-based systems.
Chen et al. \cite{chen2019residual} address the issue of residual carrier frequency offset (CFO) in Wi-Fi systems, which distorts the phase of CSI. 
They propose a multi-scale sparse recovery and MUSIC-based algorithm to estimate and compensate for CFO. 
While their work focuses on hardware-induced phase errors, our study addresses incomplete Wi-Fi features caused by low communication duty cycles.
Sun et al. \cite{sun2022simultaneous} propose a Wi-Fi ranging compensation method based on a semiparametric error model and enhanced genetic algorithm (EGA) to address measurement errors in NLoS environments.
Cao et al. \cite{cao2024compensation} propose an indoor positioning algorithm that integrates LoS error compensation and trusted NLoS distance recognition to enhance location estimation accuracy. 
However, their work focuses on static range error correction for device-based localization and lacks consideration for feature interpolation under intermittent signals.
In contrast, our work addresses the challenge of compensating for missing Wi-Fi features dynamically under low CDCs, enabling robust localization for device-free tracking.
}

With the development of IoT technology, there are usually multiple smart devices in daily environment.
Multiple transceiver links allow for the availability of abundant user motion information, but such rich links do not exist continuously.
The above-mentioned device-free tracking systems fail to track users with missing link information, and all of them require continuous link information.
To achieve larger coverage and continuous sensing, \textit{Baton is a device-free tracking system that infers the user's velocity and location from incomplete Wi-Fi link information.}
In contrast to other systems, Baton has no requirement for a fixed transceiver relationship, but rather compensates for the missing Wi-Fi features dynamically in response to the user's motion, which is an attractive alternative to the current work.

\section{Empirical Observations and Basic Ideas}\label{sec:basicidea}

In real smart home scenarios, there is usually no long-term continuous and stable communication between any pair of transceivers especially for wireless sensing.
Accordingly, how to deal with missing signal features becomes a key problem.
Our intuitive idea is to use multiple transceiver links to compensate each other for missing signal features.
In terms of application, it is feasible because there are indeed multiple Wi-Fi-enabled smart devices in homes.
In terms of technology, we have the following observations that make the feature complementarity concept among multiple links more convincing.

\begin{figure*}[t]
\centering
\subfloat[]{\includegraphics[height=0.18\linewidth]{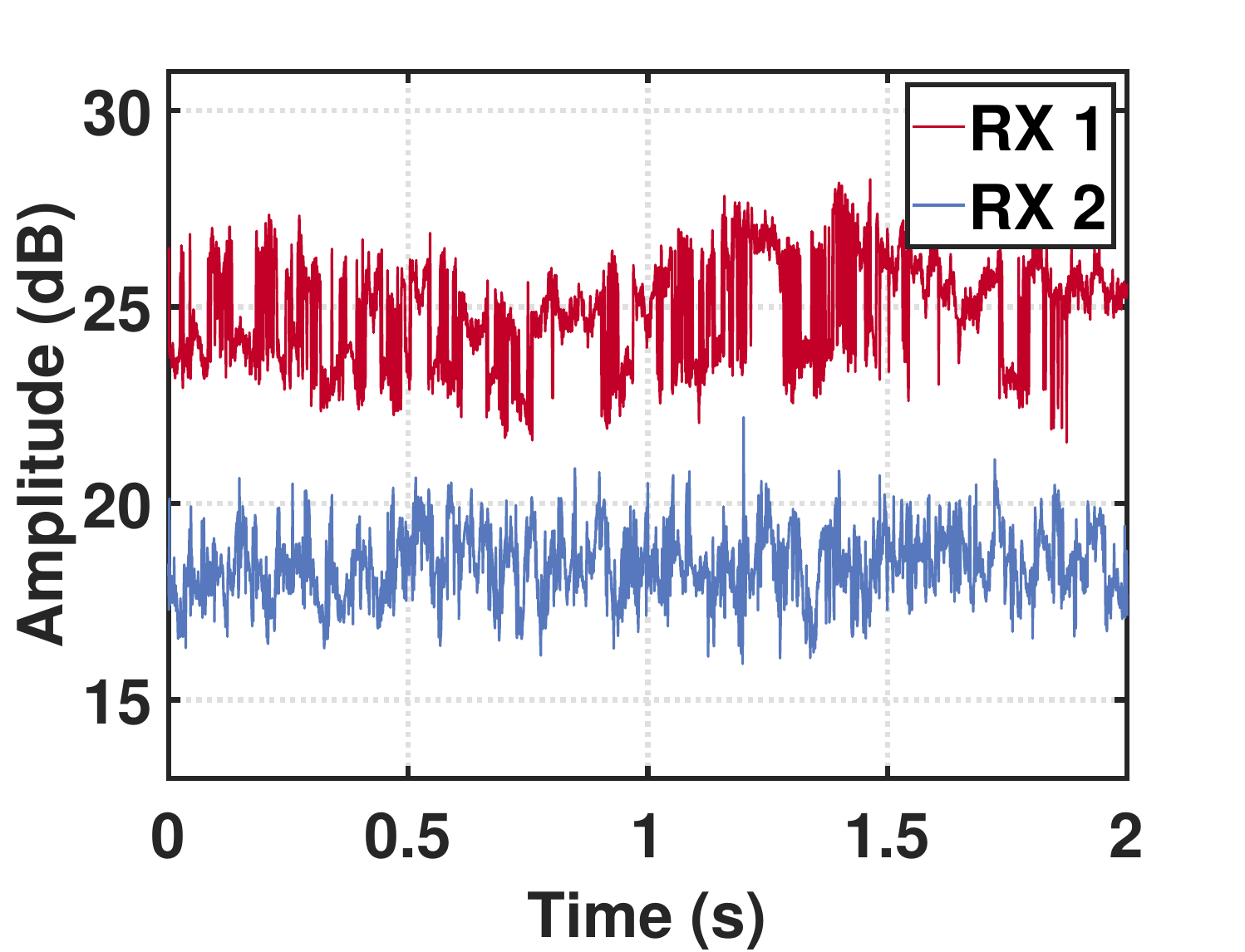}\label{csi_amp}}
\hfill
\subfloat[]{\includegraphics[height=0.18\linewidth]{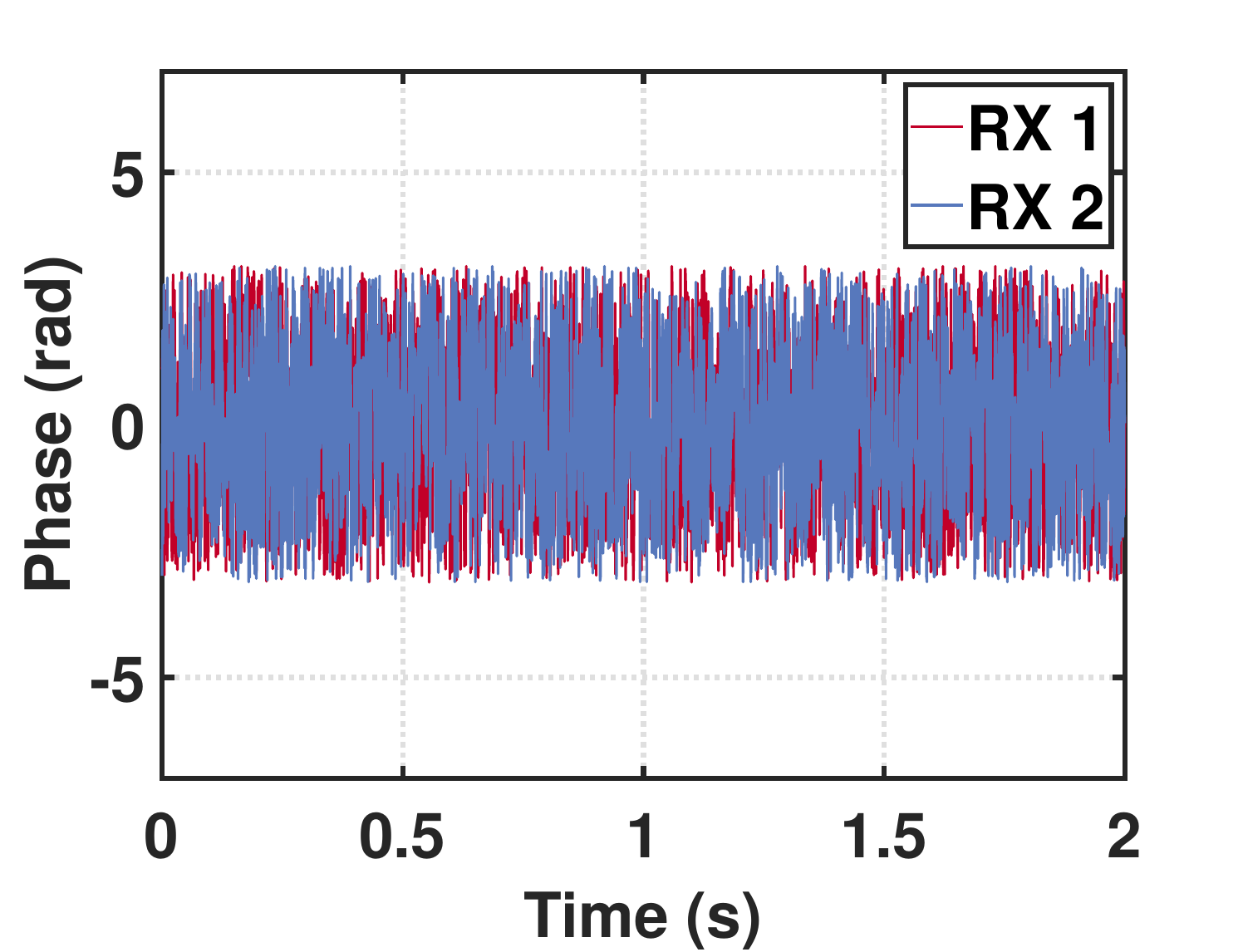}\label{csi_ph}}
\hfill
\subfloat[]{\includegraphics[height=0.18\linewidth]{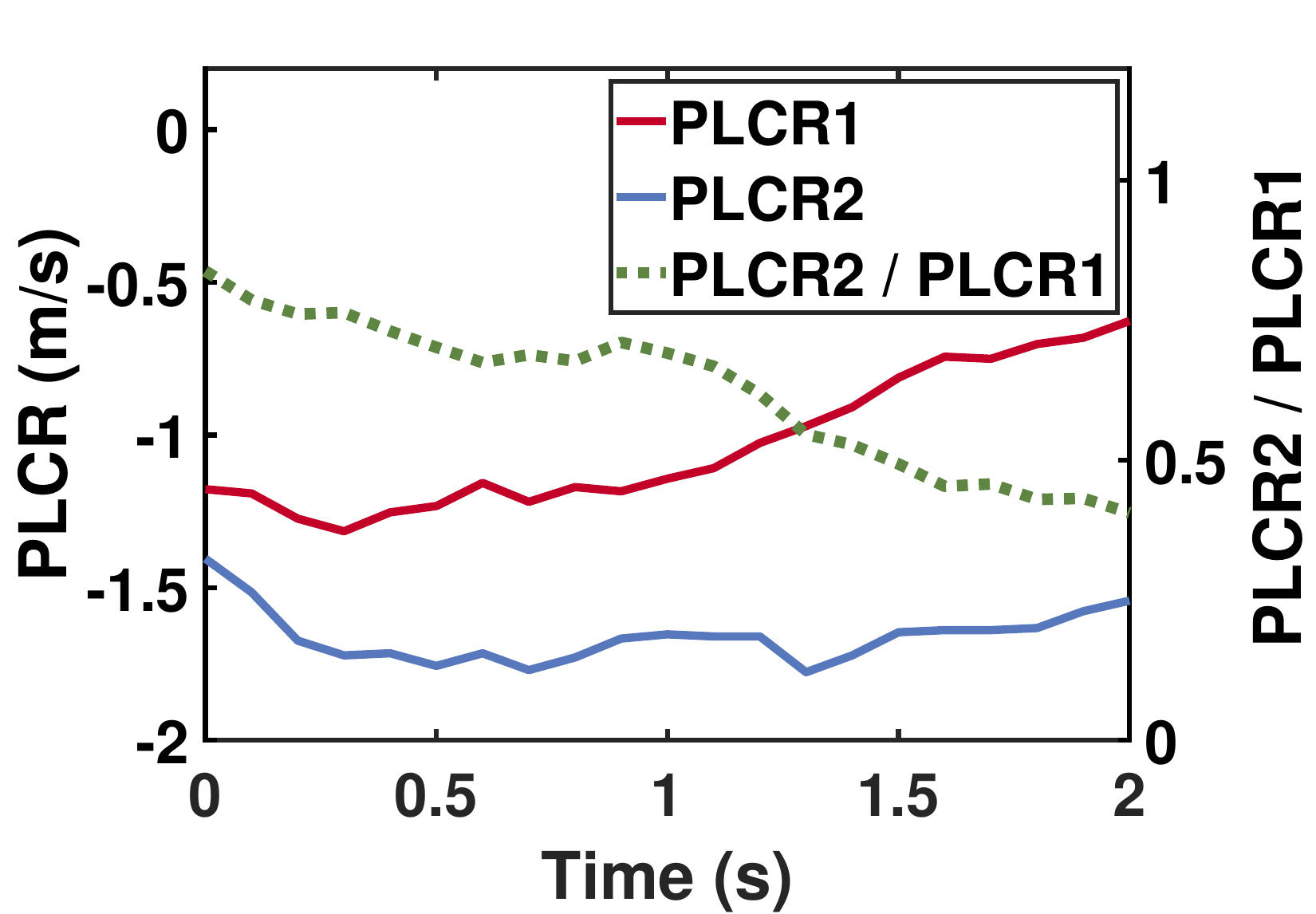}\label{observation_proportion}}
\hfill
\subfloat[]{\includegraphics[height=0.18\textwidth]{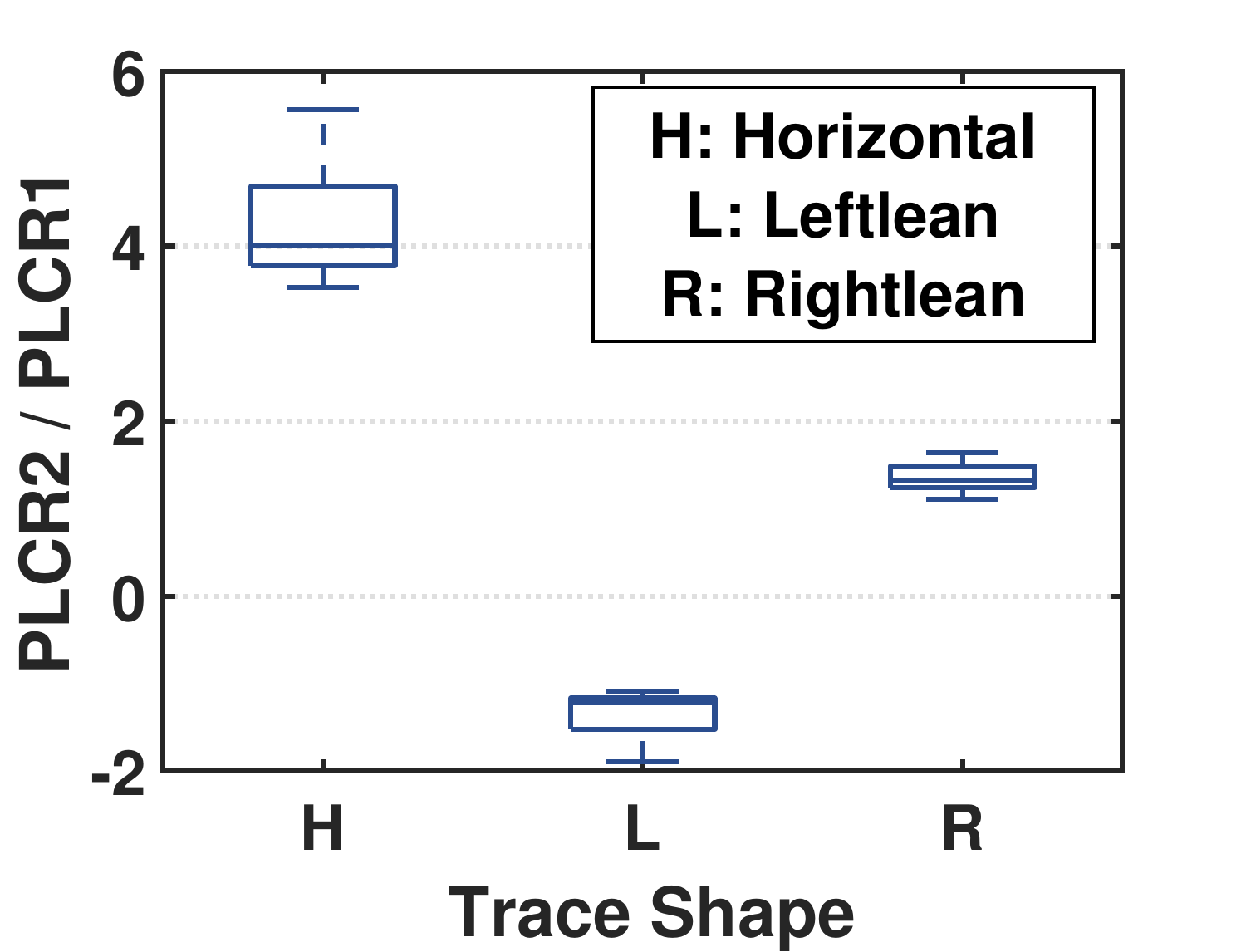}\label{observation_boxplot}}
\caption{(a) and (b) show that the amplitude and phase of CSI do not exhibit obvious characteristics. However, the Wi-Fi features extracted from CSI are more informative -- (c) shows the PLCRs of two links and their ratio maintain stable \mynote{during a relatively small time span, such as 0.5 seconds. (d) shows for different trace shapes, the distribution of the ratio between two PLCRs is quite concentrated within a short period of 0.5 seconds.}}
\label{preliminary}
\vspace{-3mm}
\end{figure*}

\subsection{Observation 1: Correlation among Different Time Slots}\label{sec:difftime}
Our initial observation is that human motion features can be reused in a time-sequential manner within a single link.
This means that the features of human movement in a specific link show a smooth transition and avoid sudden changes over a short period of time.
This continuity in motion features is due to the inertia of human motion, which dictates that the position and speed of individuals do not undergo abrupt changes within brief time intervals.
%
% As the change in motion features is primarily attributed to human motion state, the features present a continuous pattern as well.
%
\mynote{
To validate this, we first extract the real signals during a time span, \textit{e.g.}, $2$ seconds.
}
As shown in Fig.~\ref{csi_amp} and Fig.~\ref{csi_ph}, both the amplitude and phase of signals change drastically.
This is because the correspondence between the original signal and the user's velocity is non-linear.
We then extract the path length change rate (PLCR) from the original signals, which depicts the path change rate reflected off the human body \cite{qian2017widar}.
\mynote{Specifically, wireless signals are transmitted from the transmitter, reflected by the human body, and then received by the receiver. As the user moves, the total path length changes. The rate of change in the reflection path length is referred to as the PLCR, which is the key information in device-free Wi-Fi sensing systems \cite{li2024wifi}.}
\mynote{
Although the raw CSI signals may exhibit abrupt changes, the features derived from these signals change gradually.
As shown in Fig.~\ref{observation_proportion}, the PLCRs in different links (\textit{i.e.} the red and blue lines in the figure) exhibit relatively stable changes.
When the observation window is further divided into smaller intervals of 0.5 seconds, the variations in PLCR within each 0.5-second interval become extremely small, reflecting a high degree of temporal stability in PLCRs of communication links.
Hence, we conclude that the PLCR in a given link remains almost constant during a very brief time span.
As the PLCR changes gradually, we leverage this characteristic by applying a weighted method that assigns higher importance to recent observations, which is described in detail in the following sections.
}
While there are other CSI-based features besides PLCR that might be used for sensing, it is hard to utilize those features in practical tracking settings.
Specifically, AoA and ToF could also be extracted from CSI data.
However, accurate AoA estimation requires manual pre-calibration because of initial phase offset (IPO) \cite{wang2016gait} and strict antenna designs that cannot be easily satisfied by COTS Wi-Fi devices.
In addition, a bandwidth wide enough is needed for estimating ToF accurately, but this conflicts with current Wi-Fi protocols.
Thus, due to practicality in common Wi-Fi environments, our system exclusively employs PLCR for tracking, sidestepping the limitations of AoA and ToF.

% Human movement in a short time span exhibits two characteristics.
% %
% Firstly, the velocity and the direction of speed do not change abruptly, leading to no significant change in $\bm{v}$.
% %
% Secondly, the position of the user do not change drastically, so $a_x$ and $a_y$ do not change sharply either.
% %
% The box plots in Fig. ~\ref{preliminary1sub3} show the variation of $a_x$ and $a_y$ for 50 positions in close proximity to each other (no more than 0.1m in this case).
% %
% It is clear that when the position of the person changes slightly, the parameters $a_x$ and $a_y$ do not deviate significantly.
% %
% Therefore, the change in the PLCR is also very moderate during a short period.
% %
% This observation helps us build a relationship among PLCRs in a given link in a cross-time manner.

\subsection{Observation 2: Correlation across Different Wi-Fi Links}\label{sec:difflink}
Our second observation helps us establish a relationship between the signal features in a cross-device way.
In particular, human movement leads to simultaneous changes in the PLCRs of different links, albeit with varying magnitudes and rates.
Under this situation, we note that the ratio between the PLCRs of individual links is generally consistent over a short time interval.
It is because the person's position remains almost the same during a short period of time.
Accordingly, the direction of the normal vector of the Fresnel zone ellipse does not change significantly.
As a result, the ratio among PLCRs of different links does not change abruptly.

\begin{figure}[!t]
\centering
\subfloat[]{\includegraphics[height=0.2\textwidth]{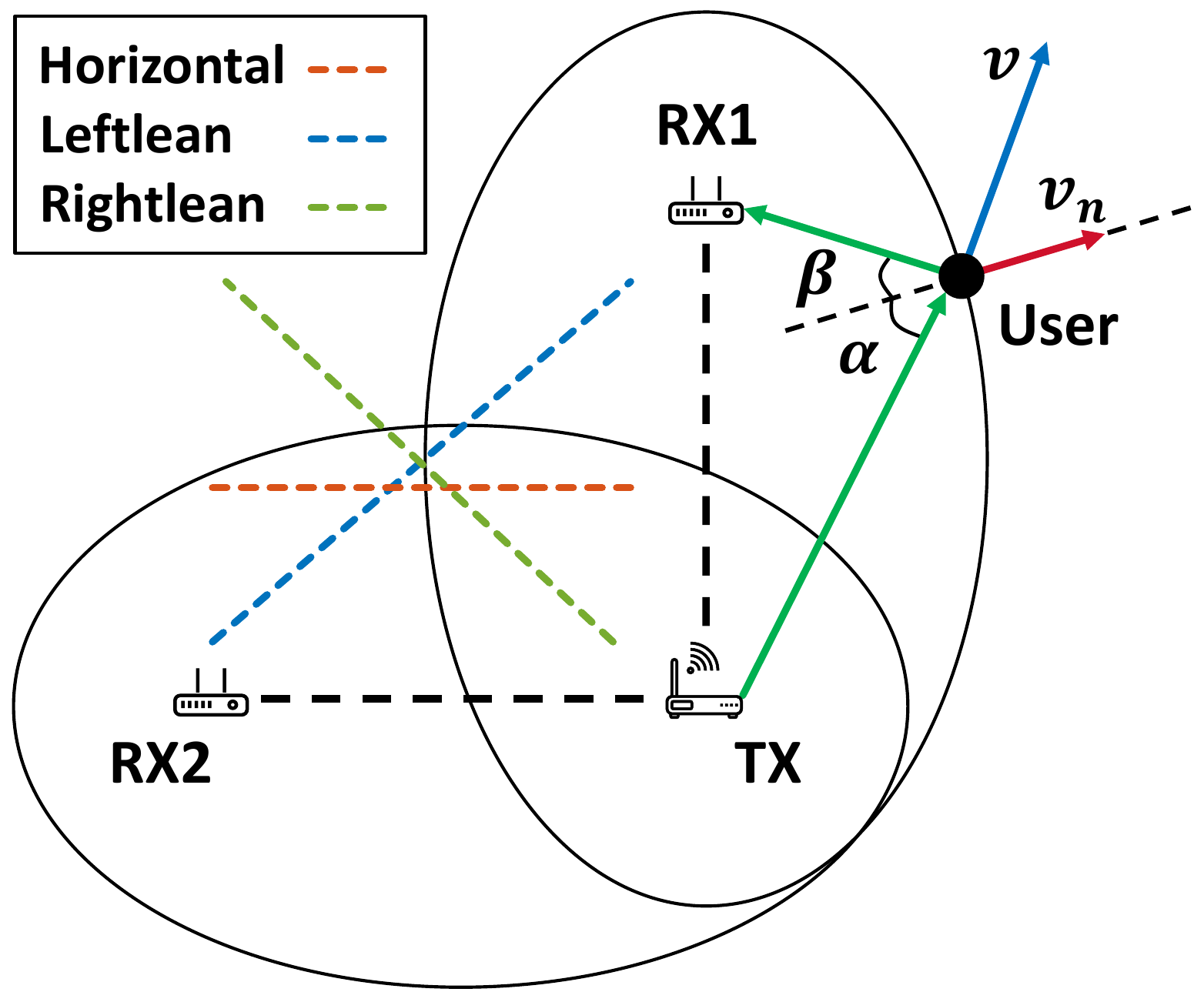}\label{Deployment}}
\subfloat[]{\includegraphics[height=0.2\textwidth]{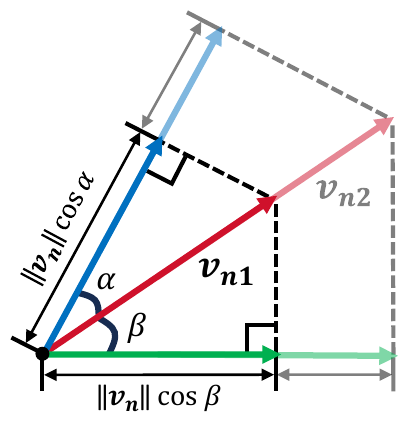}\label{SimilarTriangle}}
\caption{(a) shows that user movements change the length of reflection paths and the traces in the validation experiment.  \mynote{(b) shows the underlying cause of the observation in \S Section~\ref{sec:difflink} -- the magnitude change of $\bm{v_n}$ (from $\bm{v_{n1}}$ to $\bm{v_{n2}}$) leads to changes of two PLCRs in the same proportion.}}
\vspace{-3mm}
\end{figure}

\mynote{
\textbf{Empirical experiments.} 
We calculate the ratio between two PLCRs in a brief time span.
Fig.~\ref{observation_proportion} demonstrates that the ratio between the two PLCRs remains stable during a small interval, as indicated by the green line.
In this validation experiment, the positions of TX, RX1, and RX2 are $(2.4m, -2.4m)$, $(2.4m, 2.4m)$, and $(0m, -2.4m)$, respectively, where TX refers to the transmitter while RXs refer to receivers.
The user moves from $(-2.4m, 0m)$ to $(2m, 0m)$.
To ensure the robustness and generalizability of our second observation, we further analyze the variation of feature ratios for different experiment settings and for different path shapes.
To set up different experimental environments, we put the transceivers at various locations. RX2 is first moved $(-2.4m,2.4m)$. Then, in another experiment, TX, RX1, and RX2 are deployed sequentially in a straight line, with a spacing of $2.4m$ between each pair.
Similar results to Fig.~\ref{observation_proportion} are observed across different experimental configurations, reflecting the consistency and generalizability of the PLCR ratio stability.
To further ensure the robustness of the observation, the experimenter walks along three different paths in various directions and two receivers are deployed, as Fig.~\ref{Deployment} shows.
The user moves from $(-2.4m, 0m)$ to $(2m, 0m)$ for the horizontal shape, and from $(-2.4m, 2m)$ to $(2m, 2.4m)$ for the leftlean shape.
For the rightlean shape, the start position is $(2m,-2m)$ and the end position is $(-2.2m,2.2m)$.
We extract the PLCRs for different paths during a 0.5-second time span, and calculate their ratio.
As shown in Fig.~\ref{observation_boxplot}, for all three path shapes, the aforementioned PLCR ratio changes slightly in a short time span.
In each of the box plots, this ratio for ten consecutive time slots within $0.5s$ clearly demonstrates a high concentration.
The central boxes in the plot are notably narrow, indicating that the majority of the PLCR ratios are tightly clustered together.
This validates that the observation could be applied to different trace shapes.
Accordingly, the proportion of PLCRs in different links remains relatively stable over a brief time span.
}

% \txy{07.16}

\textbf{Theoretical analysis.}
The human velocity can be decomposed into normal and tangential components on the ellipse.
As the PLCR depends only on the normal component $\bm{v_n}$ in Fig.~\ref{Deployment}, the PLCR $r$ \mynote{for a specific link} can be expressed as
\begin{equation}\label{eq:plcr_alpha_beta}
r=\left\|\bm{{v_n}}\right\|(\cos{\alpha}  + \cos{\beta}),
\end{equation}
where $\alpha$ and $\beta$ are the two angles between $\bm{v_n}$ and the reflection path.
Over a short time interval, the person's position can be considered approximately constant.
Therefore, $\alpha$ and $\beta$ can be regarded unchanged \cite{qian2017widar,qian2018widar2}, resulting in
\begin{equation}\label{eq:plcr_vn}
r \propto \left\|\bm{{v_n}}\right\|.
\end{equation}
When the person's speed changes from $\bm{v_{1}}$ to $\bm{v_{2}}$, its normal component changes from $\bm{v_{n1}}$ to $\bm{v_{n2}}$, but its direction can be regarded as unchanged within a small time span.
According to Eq.~\ref{eq:plcr_vn}, the PLCR can be visually expressed as the sum of the lengths of the two right-angled edges in \mynote{Fig.~\ref{SimilarTriangle}}.
The lengths of the two edges vary proportionally due to the change from $\bm{v_{n1}}$ to $\bm{v_{n2}}$, as the two sets of right triangles are similar. 
\mynote{
Considering a given link, we denote its PLCR at the previous moment as $r_1$, and the PLCR after a very short time interval on the same link as $r_2$.}

\begin{figure}[t]
\centering
\includegraphics[width=0.95\linewidth]{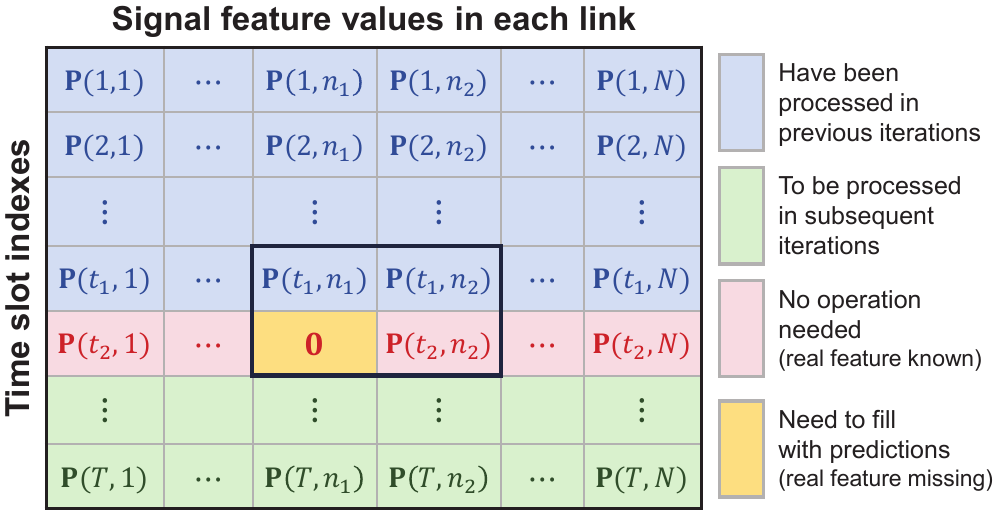}
\caption{Structure of the signal feature table maintained by the \mynote{system}.}
\label{tablemaintained}
\vspace{-3mm}
\end{figure}

\mynote{
According to Eq.~\ref{eq:plcr_alpha_beta} and Eq.~\ref{eq:plcr_vn}, the proportion between the previous PLCR ${r_1}$ and the subsequent PLCR ${r_2}$ in a given link is
\begin{equation}\label{plcr_proportion}
\frac{{r_2}}{{r_1}}
=\frac{\left\|\bm{{v_{n2}}}\right\|  (\cos{\alpha}  + \cos{\beta})}{\left\|\bm{{v_{n1}}}\right\| (\cos{\alpha}  + \cos{\beta})}
=\frac{\left\|\bm{{v_{n2}}}\right\|}{\left\|\bm{{v_{n1}}}\right\|}
=\frac{\left\|\bm{{v_{2}}}\right\|}{\left\|\bm{{v_{1}}}\right\|}
\approx \mathrm{const}.
\end{equation}
For clarity, in the equation above, $r_1$ and $r_2$ are the PLCRs in the same link for two adjacent moments, rather than the PLCRs for two different links.
We can further consider an example with three links for clarification, where there are Link A, B and C.
The PLCR in Link A is ${r_1^A}$ in the first moment, and is ${r_2^A}$ in the second moment.
The PLCR in Link B is ${r_1^B}$ in the first moment, and is ${r_2^B}$ in the second moment.
The PLCRs for Link C in two adjacent moments are similarly defined as ${r_1^C}$ and ${r_2^C}$.
Based on Eq.~\ref{plcr_proportion}, we have
\begin{equation}\label{two_link_proportion}
\frac{{r_2^A}}{{r_1^A}} = \frac{{r_2^B}}{{r_1^B}} = \frac{{r_2^C}}{{r_1^C}}.
\end{equation}
By reformulating Eq.~\ref{two_link_proportion}, we have
\begin{equation}\label{reformulate}
\frac{{r_2^A}}{{r_2^B}} = \frac{{r_1^A}}{{r_1^B}} = k_{AB}; \frac{{r_2^A}}{{r_2^C}} = \frac{{r_1^A}}{{r_1^C}} = k_{AC}; \frac{{r_2^B}}{{r_2^C}} = \frac{{r_1^B}}{{r_1^C}} = k_{BC}.
\end{equation}
In Eq.~\ref{reformulate}, the coefficients $k_{AB}$, $k_{BC}$ and $k_{BC}$ are not necessarily the same, as they depend on the specific pair of links being considered.
However, for any two given links, the ratio of PLCRs at two adjacent moments remains consistent.
Hence, the feature above can be utilized when one of the PLCRs is missing, and the missing value can be compensated with the help of known values.
Nevertheless, utilizing this feature in practice needs some care, because PLCR data is signed rather than always positive.
When a specific PLCR, such as $r_2$ in Eq.~\ref{plcr_proportion}, changes from positive to negative at two adjacent moments, it signifies a transition in the reflection path length from increasing to decreasing.
Meanwhile, $r_1$ may stay positive during this time span.
As a result, the proportion in Eq.~\ref{plcr_proportion} will change its sign and even its absolute value is not informative.
Thus, it is advisable to utilize this feature when both adjacent PLCRs in a communication link keep their signs unchanged to ensure accuracy.
}

\begin{figure*}[!t]
\centering
\includegraphics[width=7in]{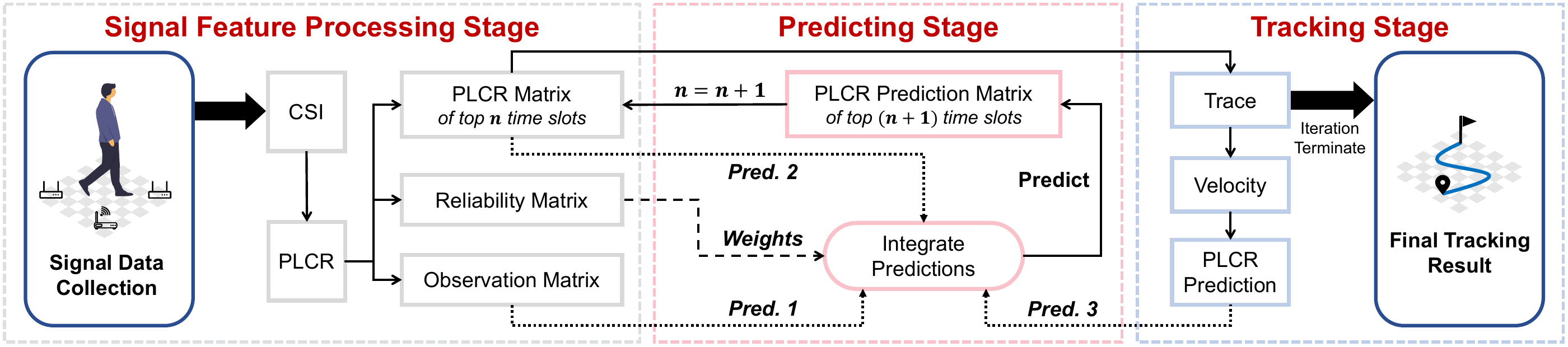}
\caption{System overview.}
\label{SystemOverview}
\vspace{-3mm}
\end{figure*}

\subsection{Insights: Simultaneous Tracking and Predicting}
These two observations reveal the connections of location-related features between links across two dimensions: time and Wi-Fi network connections.
Specifically, we can model the Wi-Fi features using the 2D matrix shown in Fig.~\ref{tablemaintained}, where the row denotes the features at different time slots, and the column denotes the features with respect to different Wi-Fi links.
The first observation helps compensate for missing Wi-Fi features across neighboring time slots within the same link, so we can fill in the missing features vertically.
The second observation unveils the proportional relationship among features of different links at a given time slot, which can help us complement missing features horizontally.
These observations have laid the foundation for our \mynote{system} by providing a way to make predictions directly from observed signal features.

To compensate for the missing features when all Wi-Fi links are temporarily unavailable, \mynote{the system} further leverages human movement speed, obtained through tracking, along with the corresponding location information.
By employing mathematical deductions based on the velocity and direction of human movement, we can also deduce the PLCR features of different links from mathematical models, rather than predict directly from observed feature values.
\mynote{
For clarity, in this paper, we refer to the overall system framework as \textit{Baton}, while STAP denotes the specific algorithm used within the system for feature compensation and tracking.
}

To implement the STAP algorithm, we adopt an incremental tracking approach, as opposed to considering all PLCR values for all time points as the input.
This means that at a specific time slot, we have already gathered all necessary information of earlier time slots in previous iterations.
We can obtain the final PLCR predictions for missing values by combining different types of predictions.
To accurately fill in the missing features, we combine the predicted values with the recent observations in the same link by assigning appropriate weights to them.
In this way, we mitigate errors that might arise from relying solely on one specific type of prediction.

% Part 3: System Overview
\section{System Overview}\label{sec:overview}
The \textit{Baton} system, depicted in Fig.~\ref{SystemOverview}, comprises three stages: signal feature processing, predicting, and tracking.

\textbf{\textit{Signal Feature Processing Stage.}}
This stage reads raw Wi-Fi signals and analyzes the CSI readings. 
Wi-Fi features, namely the PLCRs, are then extracted from CSI data.
Subsequently, observation-based and reliability matrices are computed, which aid in predicting PLCR values in the next stage. 
Finally, the system goes into a loop in which it alternates between tracking and predicting operations.

\textbf{\textit{Predicting Stage.}}
The goal is to achieve the PLCR prediction for facilitating the tracking stage.
Three types of predictions are combined in this stage.
The observation-based prediction (\textit{Pred.1}) is directly calculated from raw PLCRs based on the initial observation.
The proportionate prediction (\textit{Pred.2}) is determined based on the correlation across different Wi-Fi links.
The model-based prediction (\textit{Pred.3}) is obtained by mathematical modeling.
The reliability matrix obtained in the signal feature processing stage is used for determining the weight of \textit{Pred.1} when calculating the final PLCR prediction.
Considering the degree of signal feature missing, the above three predictions are weighted and combined to obtain the ultimate PLCR prediction.

\textbf{\textit{Tracking Stage.}}
This stage determines the user's velocity and predicts signal features for the next moment.
A neural network provides the partial trace for the top several time slots. 
Then, the velocity is obtained by calculating the differences between adjacent trace points.
We can derive the model-based PLCR prediction for the most recent moment through the Fresnel zone model.
This prediction is reused at the next time slot due to the continuity of human movement.
The \textit{Baton} system uses an incremental tracking approach, ensuring all necessary information of previous time slots is obtained in each iteration.
The system then proceeds to the predicting stage for the next iteration to process data of the following time slot, until it gets the final trajectory.
%
% Part 4: System Design
\section{System Design}\label{sec:sysdesign}
\subsection{Signal Feature Processing Stage}
\begin{figure}[!t]
\centering
\includegraphics[width=0.95\linewidth]{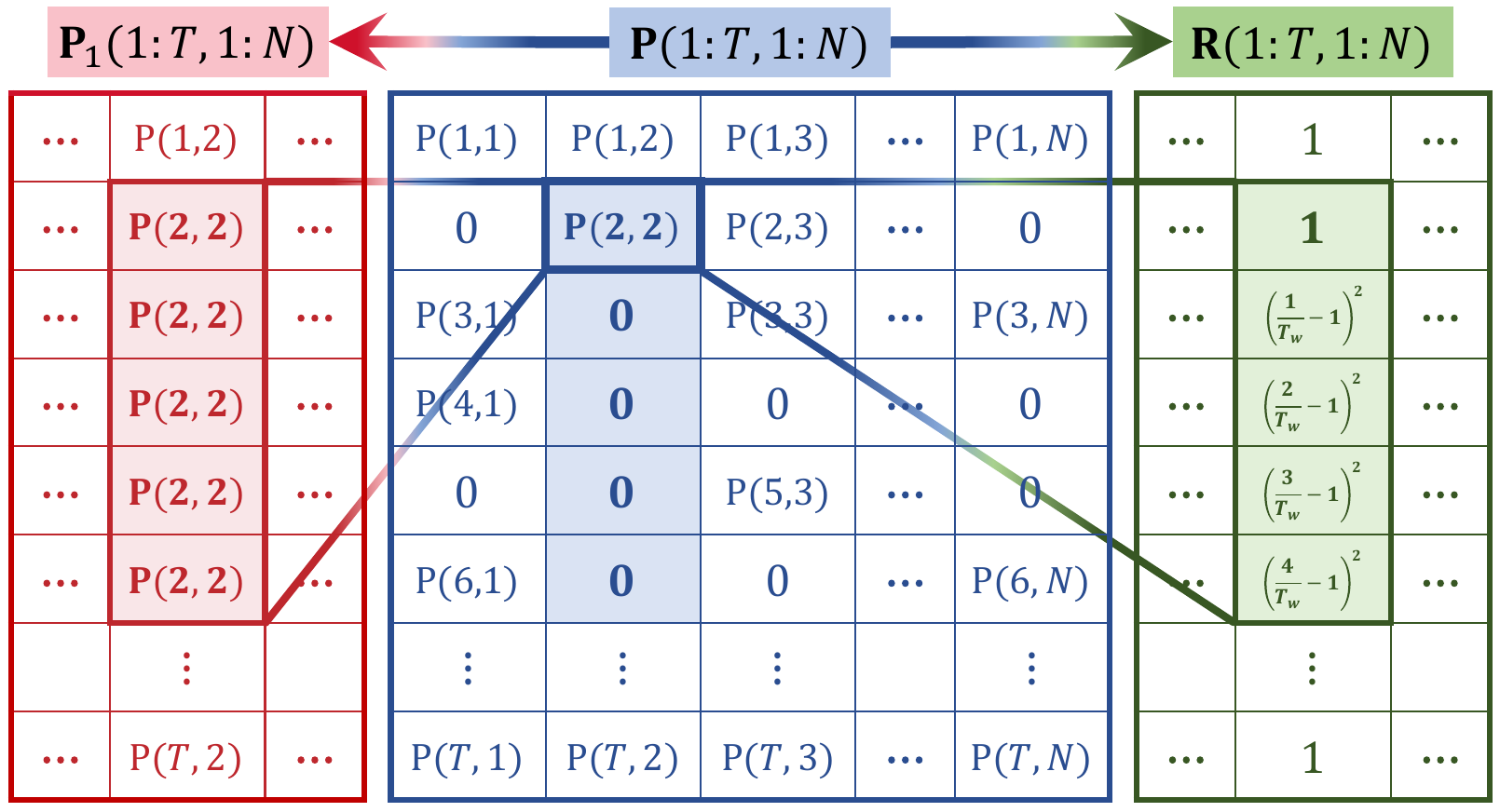}
\caption{The observation matrix ($\mathbf{P}_1$) and reliability matrix ($\mathbf{R}$) are calculated from the raw PLCR matrix ($\mathbf{P}$).}
\label{threemats}
\vspace{-3mm}
\end{figure}

The purpose is to generate the basic feature matrix for the predicting and tracking stages.
To this end, we design three matrices shown in Fig.~\ref{threemats}, including the PLCR matrix, the observation matrix, and the reliability matrix:
\textbf{(i)} The \textit{PLCR matrix} stores the PLCR data that can be measured for all Wi-Fi links.
However, a significant amount of data in this matrix is missing because there is not always stable and continuous communication;
\textbf{(ii)} The \textit{observation matrix} simply reuses the PLCR features in previous time slots, so that we can obtain a basic matrix without missing features;
\textbf{(iii)} Over time, the features reused in the observation matrix will gradually become unreliable, so we use an additional \textit{reliability matrix} as an indicator to characterize this problem.
Now, we will delve into the detailed process of obtaining these matrices.

% By utilizing available values in the PLCR Matrix, we fill in the PLCR data at the missing positions and obtain the \textit{Observation Matrix}.
%
% Finally, we acquire the \textit{Reliability Matrix} by analyzing the locations of the available values in the PLCR Matrix.
% %
% This matrix is used to assign weights to three independent predicted values and combine them to obtain the final PLCR prediction.

\subsubsection{The PLCR matrix}
In this part, we first acquire the CSI and then extract the PLCR from the raw CSI readings by performing the short-time Fourier transform (STFT), which has been described in detail in Widar \cite{qian2017widar}.
Denoting the wavelength as $\lambda$, the attenuation as $A(f,t)$, and the phase shift as $e^{-j 2 \pi \frac{L(t)}{\lambda}}$, then the CSI at time $t$ with a frequency of $f$ can be mathematically represented as
\begin{equation}
H(f, t)
=H_s(f, t)+H_d(f, t)
=H_s(f, t)+A(f, t) e^{-j 2 \pi \frac{L(t)}{\lambda}},
\end{equation}
where $H_s(f, t)$ represents the static phasor component, and $L(t)$ is the path length of the dynamic phasor component $H_d(f, t)$.
The PLCR is mathematically defined as $r \triangleq \frac{\mathrm{d} L(t)}{\mathrm{d} t}$, where $L(t)$ is the dynamic path length at time $t$.
As a result of the Doppler effect, the PLCR is a constant multiple of the DFS and we have
\begin{equation}
f_D=-\frac{1}{\lambda} \frac{\mathrm{d}L(t)}{\mathrm{d}t}=-\frac{1}{\lambda} r,
\end{equation}
where $f_D$ and $r$ denote the DFS and PLCR respectively.
However, due to problems such as clock asynchrony in COTS devices, CSI is disturbed by unknown phase shifts $\kappa (f,t)$.
In addition, when multiple dynamic paths exist in the scenario, the final signal is a superposition of multi-path signals.
Denoting $P_d$ is the total path number, the CSI can be rewritten as
\begin{equation}
H(f, t) = e^{-j \kappa(f, t)} (H_s(f)+\sum_{u \in P_d} A_u(f, t) e^{-j 2 \pi \frac{L_u(t)}{\lambda}}) .
\end{equation}
After removing the initial phase offset's effect with conjugate multiplication \cite{li2020wiborder}, we can perform high-pass filtering, low-pass filtering and STFT to get the PLCR data.

The PLCR matrix can be represented by a two-dimensional $T \times N$ matrix $\mathbf{P}$, where $T$ is the total number of time slots and $N$ is the number of links. 
The row index $t$ represents the time slot and the column index $n$ represents the signal features across different Wi-Fi links, \textit{i.e.}, received by different receivers.
When there is no Wi-Fi communication at time $t$ and link $n$, we set the corresponding value denoted by $\mathbf{P}(t,n)$ as $0$.
This matrix is then updated continuously in each iteration of the predicting stage.

% PLCR data are not available when signal features are missing.
% %
% Under this circumstance, we set zero in corresponding positions in the PLCR matrix to indicate this missing condition.
% %
% Despite the substantial data missing, we can use limited information to make basic speculations for the values in these missing positions and get the Observation Matrix.
%

\subsubsection{The observation matrix}
The observation matrix $\mathbf{P}_1$ is the same size as the PLCR matrix and can be exported directly from the PLCR matrix.
In particular, we compensate for missing Wi-Fi features based on the 
most recent real PLCR feature that can be observed in the same Wi-Fi link.
As a result, the elements in this matrix can be regarded as basic, though not delicate, predictions for the missing features.
In fact, according to our first observation in \S Section~\ref{sec:difftime}, the signal features for short periods of time are indeed believable.
For long-term missing features, we will use the reliability matrix and other prediction methods to deal with them.

% In fact, these predictions are based on the first observation we presented in this paper.
% %
% Because the signal features in the same link do not change drastically in a brief time span, it is reasonable to use the most recent observed value as the predicted value for the current position when we encounter a missing one in the raw PLCR Matrix.
% %

To calculate the observation matrix, we examine each element in the raw PLCR matrix.
When hitting a missing value in it, we scan upward from this position until we encounter a non-missing PLCR value, \textit{i.e.}, the most recent observed value.
Then we fill this value at the missing point as the prediction in the observation matrix.
The elements in this matrix are referred to as observation-based values, \textit{i.e.}, \textit{Pred. 1} in Fig.~\ref{SystemOverview}.
%
% In addition to this prediction, we also introduce predictions obtained by other approaches.
% %
% In order to combine multiple predictions to mitigate errors, we introduce the Reliability Matrix for assigning weights to different predictions and for obtaining the final PLCR prediction.
% %

\subsubsection{The reliability matrix}
This reliability matrix is denoted as $\mathbf{R}$, which is the same size as the above two matrices, while the elements in it are not PLCRs but weights between $0$ and $1$.
Although we have obtained a basic observation-based prediction, it is not reasonable to rely solely on this to achieve tracking.
To improve the accuracy of the final prediction, we introduce two other predictions obtained via other techniques, which are represented as \textit{Pred. 2} and \textit{Pred. 3} in Fig.~\ref{SystemOverview}.
One is the predicted value obtained by performing a proportional prediction based on the observation in \S Section~\ref{sec:difflink};
the other is obtained based on mathematical models.

To effectively and accurately combine the three predictions, we assign weights to them and then calculate the final predicted PLCR.
As the PLCRs in a given link maintain stability over a short time interval, we are more inclined to trust observation-based predictions (\textit{Pred. 1}) than other predictions (\textit{Pred. 2} and \textit{Pred. 3}) as long as signals in this link are not missing for too long.
%
% Therefore, we design a weighting method in which one weight is assigned to \textit{Pred. 1}, while the other is assigned to either \textit{Pred. 2} or \textit{Pred. 3}.
% %
To measure the extent to which we can rely on the observation-based prediction, each element in the reliability matrix is a weight assigned to \textit{Pred. 1}, representing the extent of our trust in it at the corresponding time slot.

When the actual PLCR can be measured at a specific position, the weight for \textit{Pred. 1} is 1, as this observation is accurate.
However, if signal feature loss occurs continuously, the reliability of past PLCR observations diminishes.
Nonetheless, recent observations of the same link remain meaningful.
Hence, the weight assigned to the observation-based prediction gradually decreases over time until reaching 0 after $T_w$ time slots.
Empirically, we adopt a quadratic function to reduce this weight when a link experiences continuous feature loss:
\begin{equation}\label{calweight}
w = \left\{
\begin{aligned}
& (\frac{1}{T_w}t-1)^2 & &, 0 < t < T_w, \\
& 0 & &, t \geq T_w,
\end{aligned}\right.
\end{equation}
where $w$ denotes the weight discussed above and this weight decreases to zero at the time slot $t=T_w$.
The complementary weight, obtained by subtracting the elements from 1 in the reliability matrix, is assigned to either \textit{Pred. 2} or \textit{Pred. 3} based on the discussed circumstances in the predicting stage.

\subsection{Predicting Stage} \label{sec:predstage}
In this section, we introduce how to obtain the three predictions and combine them to get final PLCR predictions.

\begin{figure*}[!t]
\centering
\includegraphics[width=1\linewidth]{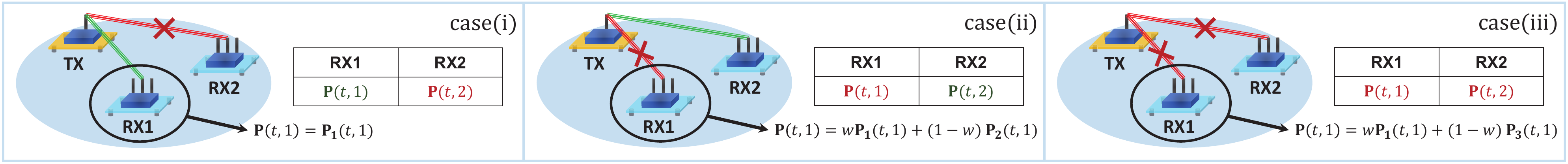}
\caption{This schematic diagram illustrates that how to calculate the final prediction depends on the feature missing condition.}
\label{3cases}
\vspace{-3mm}
\end{figure*}

\subsubsection{Observation-based predictions (Pred. 1)}
For each position where the real PLCR observation is missing in the raw PLCR matrix $\mathbf{P}$, we scan upwards until hitting a non-missing value. 
The most recent observed PLCR value of the same link is taken as the observation-based prediction for this location.
Denoting the $t$-th element of the $n$-th link in the observation matrix as $\mathbf{P}_1(t,n)$, the corresponding observation-based prediction can be calculated as
\begin{equation}\label{cale1}
\mathbf{P}_1(t,n)=
\left\{
\begin{aligned}
& \mathbf{P}(t,n) & & , \text{$\mathbf{P}_1(t,n) \neq 0$}, \\
& \mathbf{P}(t',n) & & , \text{$\mathbf{P}_1(t,n) = 0$},
\end{aligned}\right.
\end{equation}
where $t'$ is the minimal index such that $\mathbf{P}(t',n) \neq 0$ and $1 \leq t' < t$.

\subsubsection{Proportionate predictions (Pred. 2)}

According to the observation in \S Section~\ref{sec:difflink}, we design the proportionate prediction matrix denoted by $\mathbf{P}_2$.
The proportion of PLCRs in two given links is approximately constant for two adjacent moments $t=t_1$ (the preceding moment) and $t=t_2$ (the subsequent moment) shown in Fig.~\ref{tablemaintained}.
Suppose the system is processing the data corresponding to $t=t_2$ in the current iteration, then we confirm that the PLCR of each link at its preceding moment $t=t_1$ has been attained in the previous iteration.
Therefore, as long as there exists at least one non-missing PLCR at $t=t_2$ in link $n_2$, we are able to fill in the missing data in link $n_1$ by exploiting the proportional relationship.
Denoting the missing PLCR at $t=t_2$ as $\mathbf{P}_2(t_2,n_1)$, we can compute \textit{Pred. 2} with three non-missing values as
\begin{equation}\label{cale2}
\mathbf{P}_2(t_2,n_1)=\frac{{\mathbf{P}(t_2,n_2)} \cdot {\mathbf{P}(t_1,n_1)}}{\mathbf{P}(t_1,n_2)}.
\end{equation}
In practice, we observe improvement in tracking accuracy when replacing $\mathbf{P}_2(t_1,n_1)$ and $\mathbf{P}_2(t_1,n_2)$ with estimated model-based PLCRs around this time slot, which are obtained with the partial tracking result before $t=t_2$.
Such estimated values can help mitigate errors caused by the inaccuracy of individual PLCRs in the signal feature table, thus improving the overall accuracy of the system.
Since \textit{Pred. 2} cannot be obtained when all PLCR data at $t=t_2$ are missing, we have to resort to model-based predictions under such circumstance as described below.

\subsubsection{Model-based predictions (Pred. 3)}
The model-based prediction matrix denoted by $\mathbf{P}_3$ is derived from mathematical modeling. 
In previous works, mathematical models have been used to obtain traces from PLCR data \cite{qian2017widar}. 
However, we can also utilize the mathematical model, in turn, to deduce PLCRs from the historical trace.
Considering a specific link, suppose the position of the transmitter and the receiver are $\bm{l}_t=\left(x_t, y_t\right)$ and $\bm{l}_r=\left(x_r, y_r\right)$ respectively, the current human position as $\bm{l}_h=\left(x_h, y_h\right)$, and the human velocity as $\bm{v}=\left(v_{x}, v_{y}\right)^\mathbf{T}$, then the model-based PLCR prediction \textit{Pred. 3} can be calculated as
\begin{equation}\label{cale3}
\mathbf{P}_3(t,n)=\mathbf{A} \times \bm{v} = a_x v_{x}+a_y v_{y} ,
\end{equation}
where
\begin{equation}
\left\{
\begin{aligned}
& \mathbf{A} = (a_x,a_y), \\
& a_x=\frac{x_h-x_t}{\left\|\bm{l}_h-\bm{l}_t\right\|}+\frac{x_h-x_r}{\left\|\bm{l}_h-\bm{l}_r\right\|}, \\
& a_y=\frac{y_h-y_t}{\left\|\bm{l}_h-\bm{l}_t\right\|}+\frac{y_h-y_r}{\left\|\bm{l}_h-\bm{l}_r\right\|}.   
\end{aligned}\right.
\end{equation}
Specifically, if we successfully obtain the trajectory prediction corresponding to the first $t$ time slots in a given iteration, we are then able to derive the velocity prediction for the first $t$ moments by taking approximate derivatives of the position sequence.
With this velocity prediction, we can inversely deduce the PLCRs for the first $t$ time slots.
Due to the continuous character of the PLCR demonstrated in \S Section~\ref{sec:difftime}, the PLCR at the $(t+1)$-th moment can be approximately fitted with the model-based prediction of the $t$-th time slot.
Therefore, even if all PLCR data at a specific time slot are missing, we are still able to make out a prediction based on previous position sequences.

\subsubsection{Integrate predictions}
The purpose is to effectively integrate the three predictions above to obtain an accurate final prediction.
First, the observation-based prediction can help the system avoid getting biased predictions because it is based on real measured values.
Second, the proportionate prediction presents a trade-off between relying on the PLCR-motion relationship and depending on observation values.
Third, the model-based prediction is derived from trajectories and relies entirely on the PLCR-motion relationship without integrating any observations.
Accordingly, although the latter two predictions effectively implement compensation for missing signal features, errors can accumulate when partial trajectory predictions are not accurate enough.
Hence, we need to rectify the bias with the aid of observation-based predictions to prevent errors from accumulating.

\mynote{Based on the above analysis}, we combine the three predictions by the following weighting method.
Suppose that in the current iteration, the first $t$ rows have been processed and the $(t+1)$-th row is now under processing.
Then, for each value in the $(t+1)$-th row,
\textbf{(\romannumeral1)} if this value is non-missing, we can simply use \textit{Pred. 1}, the observed value;
\textbf{(\romannumeral2)} if this value is missing, and there is at least one non-missing PLCR value that can be measured in the $(t+1)$-th row, then \textit{Pred. 1} and \textit{Pred. 2} are combined;
\textbf{(\romannumeral3)} if this value is missing, and all values in the  $(t+1)$-th row are missing, then \textit{Pred. 1} and \textit{Pred. 3} are combined.

After getting three predictions, each element in the final PLCR prediction matrix $\mathbf{P}$ can be updated as follows:
\begin{equation}
\mathbf{P}(t,n)= \left\{
\begin{aligned}
& \mathbf{P}_1(t,n) & &, \mathrm{{(case\ \romannumeral1})}, \\
& w \cdot \mathbf{P}_1(t,n) + (1-w) \cdot \mathbf{P}_2(t,n) & & , \mathrm{{(case\ \romannumeral2})}, \\
& w \cdot \mathbf{P}_1(t,n) + (1-w) \cdot \mathbf{P}_3(t,n) & & , \mathrm{{(case\ \romannumeral3})},
\end{aligned}\right.
\end{equation}
where $w$ denotes the weight of the observation-based prediction. Then we fill in the missing features with the ultimate PLCR prediction obtained by weighting.
In this way, we not only achieve simultaneous tracking and predicting, but also correct previously built-up errors as the STAP algorithm operates.
Fig.~\ref{3cases} shows the computation of the final prediction in an example case with two links.
After integrating three predictions, we have the complete PLCR matrix for the first $(t+1)$ time slots.
Then, the system sets $t$ as $(t+1)$ and continues processing the next line of the PLCR matrix.
The system loops until it finishes processing the last row of the PLCR matrix, and we obtain the final trajectory.

\begin{figure}[!t]
\centering
\includegraphics[width=0.99\linewidth]{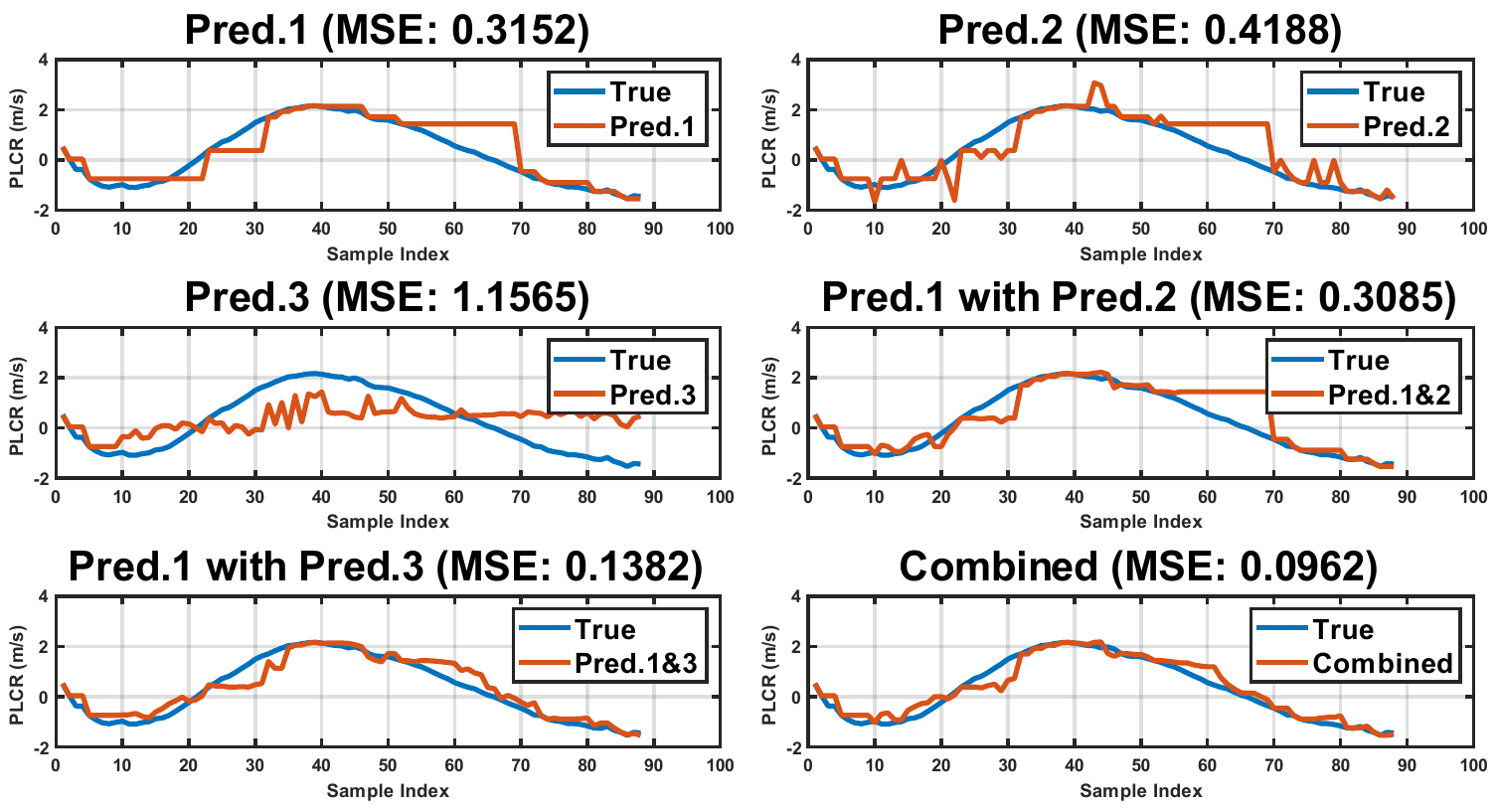}
\caption{\mynote{By combining three different PLCR predictions with a weighted strategy, we achieve a significantly more accurate PLCR prediction.}}
\label{combine-3-preds}
\vspace{-3mm}
\end{figure}

\mynoteblue{
In this process, we combine three predictions obtained from different methods, instead of using only one of them, to gain a final PLCR prediction. The necessity of this weighted combination strategy can be demonstrated through both theoretical analysis and experiments. The scenarios with missing signal features can be categorized into three cases, as shown in Fig.~\ref{3cases}. In case (\romannumeral1), the actual PLCR value in the link is non-missing, meaning this value is inherently accurate. Therefore, we retain the observed value without introducing additional predictions. While \textit{Pred.1} is highly reliable for short-term predictions, its accuracy gradually deteriorates as signal absence persists. In case (\romannumeral2) and (\romannumeral3), because the PLCR value is missing, the system cannot determine which prediction (\textit{Pred.1}, \textit{Pred.2} or \textit{Pred.3}) will be the most accurate under the current conditions. The reliability of each prediction method does not change abruptly but evolves progressively over time. Therefore, a weighted combination based on the duration of the missing features is necessary to adaptively balance their contributions and ensure reliable predictions. We also conduct experiments comparing the accuracy of PLCR predictions when using \textit{Pred.1}, \textit{Pred.2}, and \textit{Pred.3} individually, as well as the proposed weighted combination strategy.
As shown in Fig.~\ref{combine-3-preds}, the PLCR predictions from four approaches are compared against the ground-truth values in a communication link.
In the first three sub-figures in Fig.~\ref{combine-3-preds}, each prediction method is applied individually.
Notably, since \textit{Pred.2} depends on the proportional relationship between active links, it cannot be computed when all links experience feature missing. In such cases, the system defaults to \textit{Pred.1} as a fallback method.
The experimental results demonstrate that \textit{Pred.1} achieves a mean squared error (MSE) of 0.3152, while \textit{Pred.2} and \textit{Pred.3} yield higher errors of 0.4188 and 1.1565, respectively.
In contrast, the combined method achieves a significantly lower MSE of 0.0962, representing a $69.5\%$ improvement over \textit{Pred.1}, a $77.0\%$ improvement over \textit{Pred.2}, and a $91.7\%$ improvement over \textit{Pred.3}.
Combining two methods can improve accuracy but still cannot achieve optimal results.
When \textit{Pred.1} and \textit{Pred.2} are combined in a weighted approach, the result is slightly better than using \textit{Pred.1} alone, since \textit{Pred.2} cannot handle conditions when all links' features are missing.
Combining \textit{Pred.1} and \textit{Pred.3} achieves a quite good result, but the MSE is still higher than our proposed approach.
The results show that no single method consistently achieves the best accuracy. By contrast, the weighted combination strategy using three predictions dynamically adapts to varying conditions and significantly improves the overall accuracy.}

\begin{figure}[!t]
\centering
\includegraphics[width=0.95\linewidth]{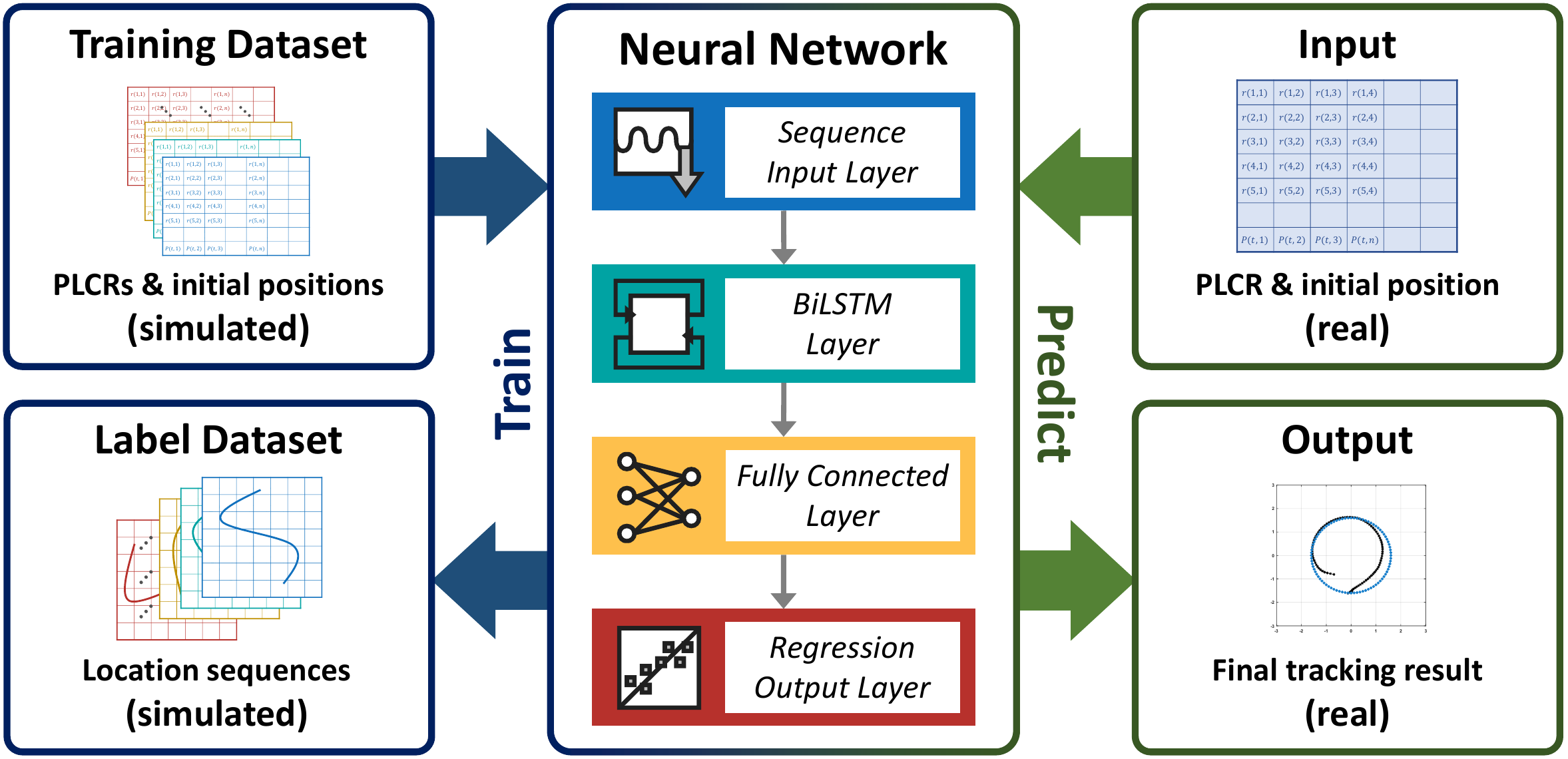}
\caption{The tracking stage uses a neural network. The training process uses simulated data and the model is used for trace prediction.}
\label{nn}
\vspace{-3mm}
\end{figure}

%\txy{07.21}
\subsection{Tracking Stage}
We use a neural network to map between the PLCR (input) and the user's trajectory (output), as shown in Fig.~\ref{nn}.
Since it is not feasible to employ a large amount of real trace data as the training set, we train the network by simulating datasets.
In the simulation process, we adopt a step-by-step simulating method to mimic the real human walking scenario.
This can ensure that the simulated traces in the training set are remarkably comparable to real trajectories.

For each simulated individual sequence in the trace dataset generation process, we first randomly generate an initial position within the defined area and a random direction.
Walking speed and step length are set within predefined ranges. 
Then the process iteratively calculates new positions based on the initial parameters.
For each new position, the process adjusts step length and speed slightly to introduce variability, and changes the direction randomly within a predefined range to simulate turning movements.
Once a complete trajectory is generated, we check if the trajectory stays within the designated area.
In this way, this data generation process results in a dataset that closely mimics the behavior of real-world trajectories.

The STAP algorithm is based on the idea that while tracking, we can inversely predict the PLCR through partial trajectories.
By performing simultaneous tracking and predicting with continuous iterations, we are able to compensate for the missing PLCR features progressively.
To implement this idea, we have to obtain a ``relatively accurate" trajectory prediction at the beginning.
Otherwise, without the initial trajectory prediction, it is infeasible to inversely compensate for the missing features.
Subsequently, we gradually adjust and optimize the partial prediction by executing the STAP algorithm.

To obtain the initial partial trace prediction, we can only rely on the observation-based prediction at first, for we can get this PLCR prediction without knowing the location of the person.
Hence, we take out the top $N_f$ rows of the observation matrix and use them as the input to the neural network, which then gives us a preliminary trajectory prediction.
\mynote{Accordingly, the system relies solely on observation-based predictions for the first $N_f$ rows, without incorporating the other two types of predictions (proportionate and model-based).}
Empirically, $N_f$ equals the number of time slots within a one-second time slice.
Note that this trajectory is not complete, since we have only gained the location sequence corresponding to the first $N_f$ time slots.
Then, the system begins to iteratively execute the STAP algorithm, as shown in Algorithm~\ref{alg:alg1}.

\begin{figure}[!t]
\centering
\includegraphics[width=1\linewidth]{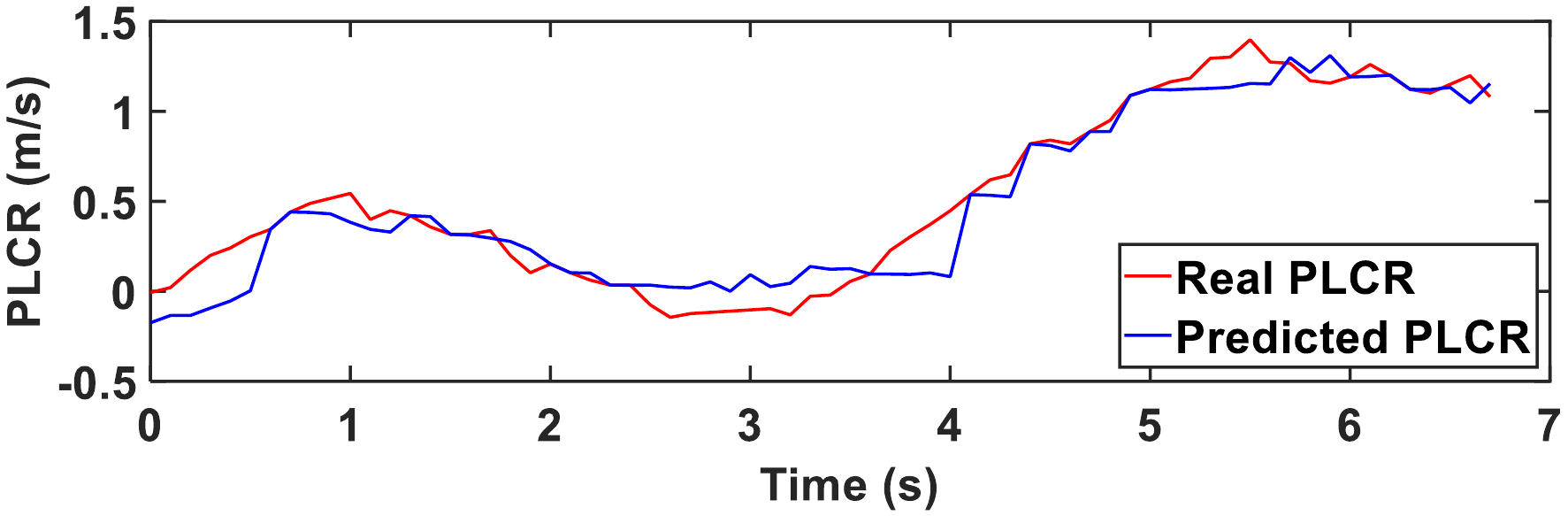}
\caption{\mynote{The predicted PLCRs are close to the real values. This helps the system obtain an accurate tracking result.}}
\label{plcr_pred_acc}
\end{figure}

Since the STAP algorithm takes an incremental approach, we can fill in one row of missing data in the raw PLCR matrix after each iteration.
By using the prediction integrating method in \S Section~\ref{sec:predstage}, we can acquire the final PLCR prediction in each iteration for the next moment.
In a single iteration, after obtaining the trajectory by tracking, we can get the PLCR prediction for the next moment.
With this prediction, we can compensate for the missing signal features and can then re-predict more complete trajectories until the iteration ends.

\begin{figure}[!t]
\centering
\includegraphics[width=1\linewidth]{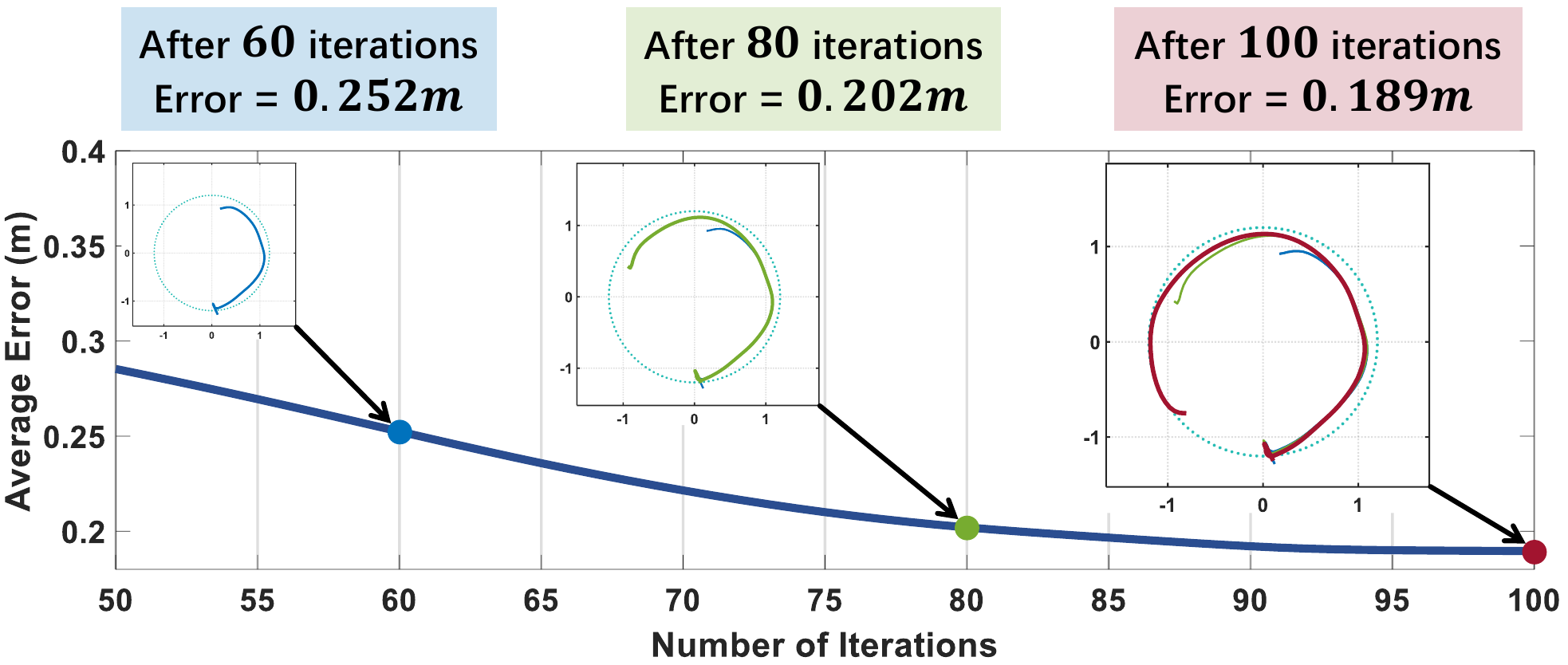}
\caption{As the iterations proceed, the STAP algorithm can rectify the errors it made in previous iterations. The average error decreases with the execution of the algorithm.}
\label{selfcorrect}
\end{figure}

\begin{algorithm}
\caption{Simultaneous Tracking And Predicting}\label{alg:alg1}
\begin{algorithmic}
    \STATE \textbf{Input:} $\mathbf{P}$ (raw PLCR matrix)
    \STATE \textbf{Output:} $\mathbf{T}$ (final tracking result)
    \STATE
    \STATE  Compute $\mathbf{P}_1$ based on Eq.~\ref{cale1}
    \STATE $\mathbf{T} \gets$ {\textsc{PREDICT}}$(\mathbf{P}[1:N_f,:])$
    \STATE $\mathbf{P}[1:N_f,:] \gets \mathbf{P}_1[1:N_f,:]$
    \STATE  Compute $\mathbf{R}$ based on Eq.~\ref{calweight}
    \FOR{$i \gets (N_f+1)$ \textbf{to} $T$}
        \STATE $\mathbf{T} \gets$ {\textsc{PREDICT}}$(\mathbf{P}[1:i,:])$
        \STATE $\textbf{V} \gets$ \textsc{DIFF}$(\mathbf{T})$
        \FOR{$j \gets 1$ \textbf{to} $N$}
        \STATE Compute $\mathbf{P}_2[i,j]$, $\mathbf{P}_3[i,j]$ based on Eq.~\ref{cale2} and Eq.~\ref{cale3}
        \STATE $\mathbf{P}[i,j] \gets$ \textsc{WEIGHT} $(\mathbf{P}_1,\mathbf{P}_2,\mathbf{P}_3,\mathbf{R})$
        \ENDFOR
    \ENDFOR
    \RETURN $\mathbf{T}$
\end{algorithmic}
\label{alg1}
\end{algorithm}

In fact, the tracking accuracy of the system depends on how similar the predicted PLCR values resemble true values.
More accurate PLCR predictions help the neural network gain better partial trace predictions.
Fig.~\ref{plcr_pred_acc} shows that the predicted PLCRs (when PLCRs are severely missing with a CDC of $20\%$) are quite close to the real measured values (when there is no PLCR deficiency).
This shows the STAP algorithm can precisely compensate for missing PLCR values with its signal feature predictions, helping the system obtain an accurate final tracking result.
In addition, experimental results show that the STAP algorithm is capable of self-correcting.
\mynote{
That is, as the algorithm proceeds after a number of iterations, previous location predictions with relatively high errors can be corrected in subsequent iterations.
The updated tracking result in each iteration is not only based on the information before this time slot.
By contrast, new PLCR predictions are introduced in each iteration, thus improving the overall tracking accuracy as the algorithm proceeds.
As a result, the system can achieve a good tracking performance when the CDC is as low as $20\%$.
As shown in Fig.~\ref{selfcorrect}, after a certain number of iterations, the average tracking error decreases steadily when we gradually obtain complete trace results.}
The system tracks and predicts simultaneously until the end of iterations, and we eventually obtain the final tracking result.

% Part 5: Evaluation

\begin{figure*}
\centering
\subfloat[Transceivers]{\includegraphics[height=0.16\textwidth]{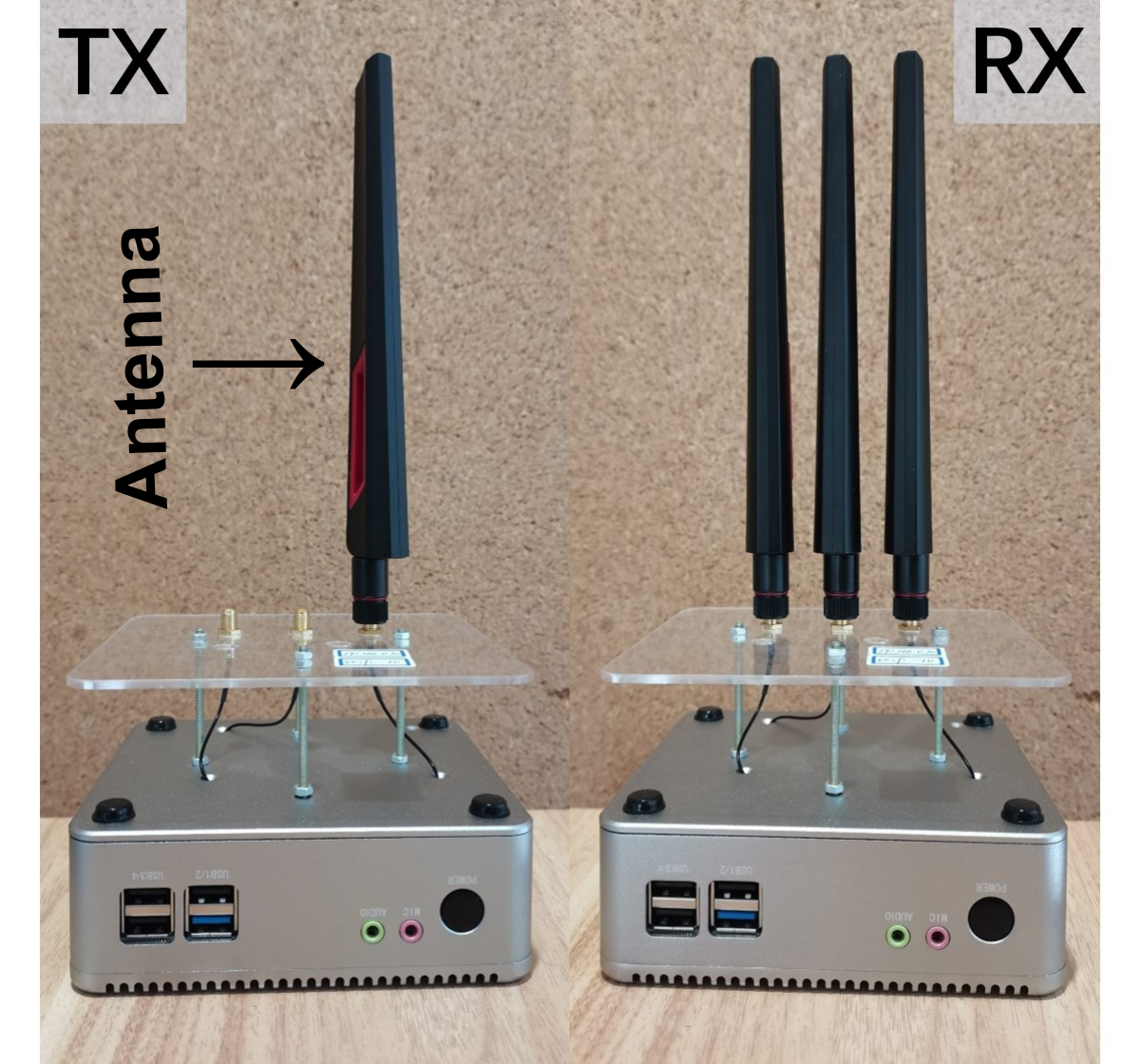}\label{exp_devices}}
\hspace{0.4cm}
\subfloat[Scenario 1: Open Space]{\includegraphics[height=0.16\textwidth]{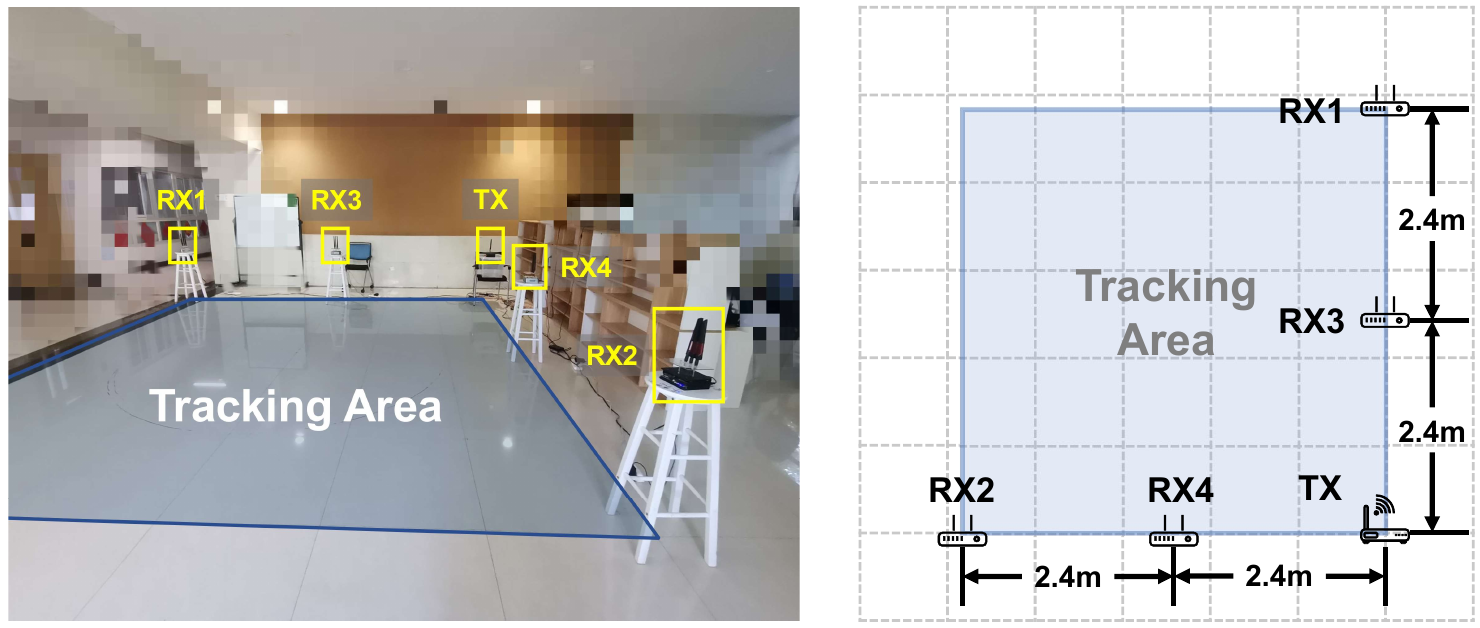}\label{open_space}}
\hspace{0.4cm}
\subfloat[Scenario 2: Cafeteria]{\includegraphics[height=0.16\textwidth]{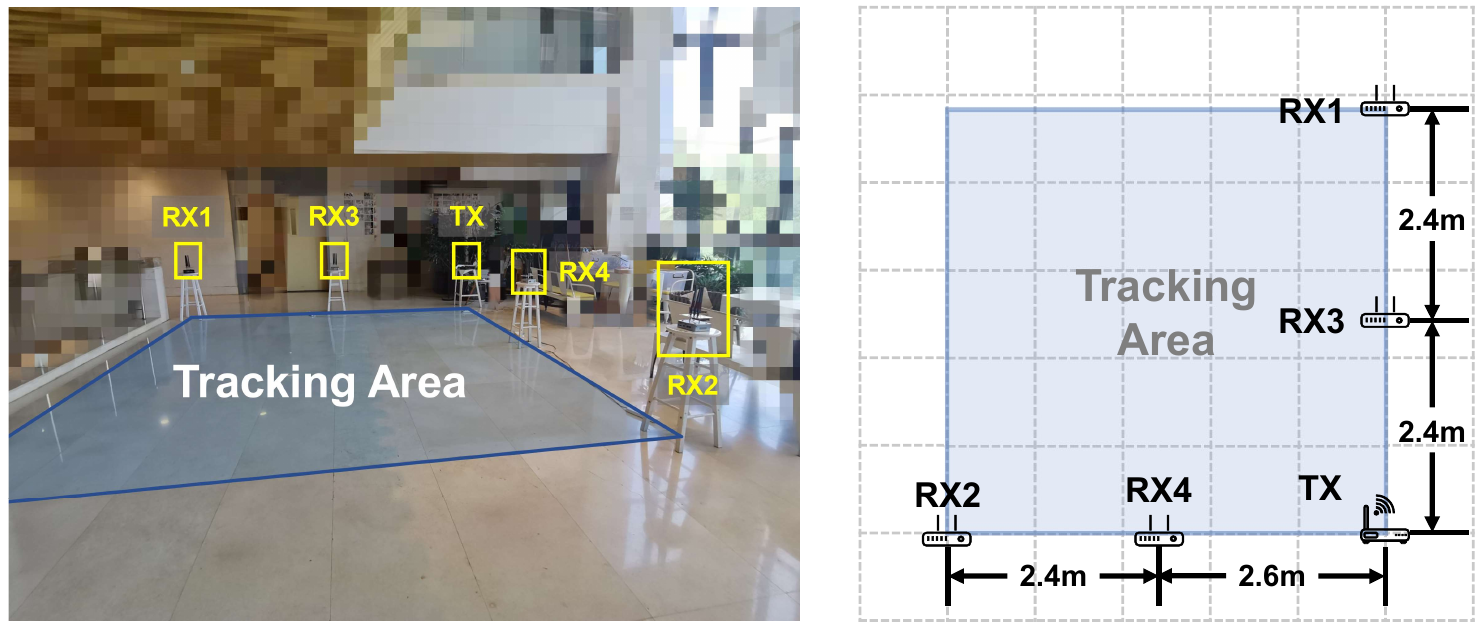}\label{cafe}}
% \caption{Experimental devices and \mynote{LoS} environments.}
\vspace{-1mm}
\\
\subfloat[\mynote{Scenario 3: Scenario with Obstacles}]{\includegraphics[height=0.16\textwidth]{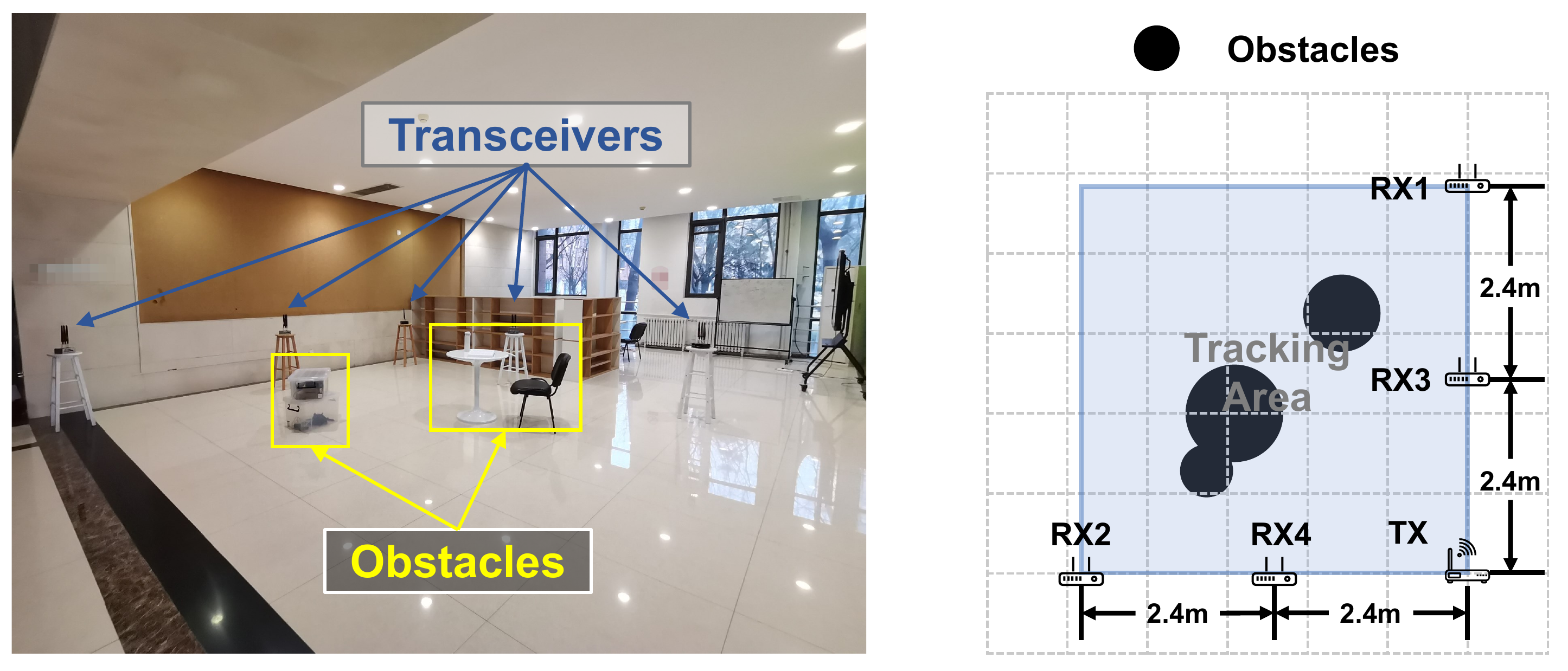}\label{obstacle_scenario}}
\hspace{1.0cm}
\subfloat[\mynote{Scenario 4: NLoS Scenario}]{\includegraphics[height=0.16\textwidth] {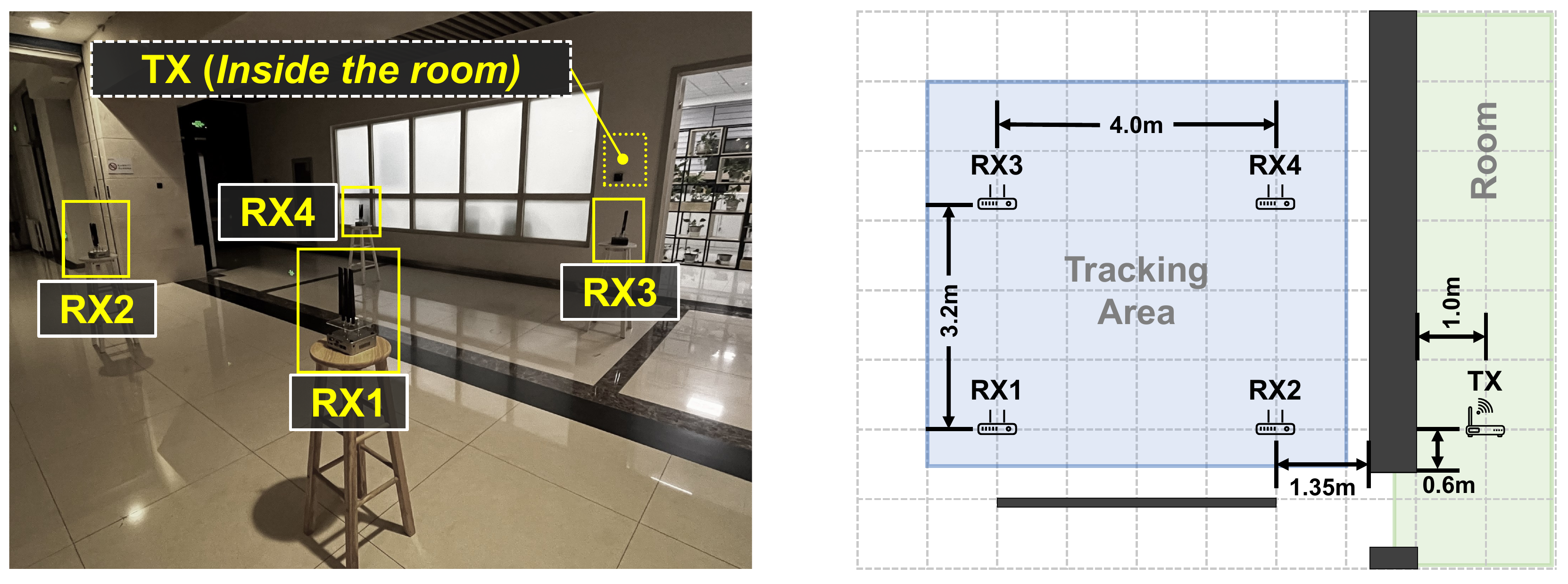}\label{nlos_scenario}}
\caption{\mynote{Experimental devices and environments.}}
\label{exp_deploy}
\vspace{-1mm}
\end{figure*}

\begin{figure*}[ht]
\centering
  \subfloat[Straight]{{\includegraphics[width=0.159\textwidth]{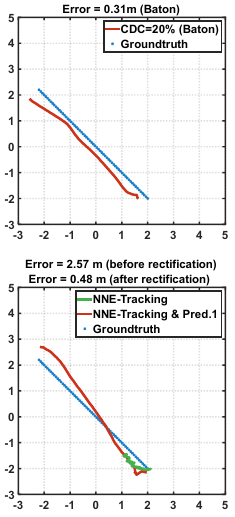}}\label{straight}}
  \hspace{0.15em}
  \subfloat[Turn]{{\includegraphics[width=0.159\textwidth]{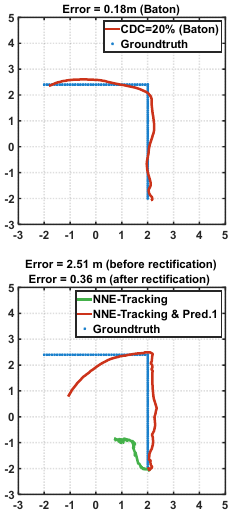}}\label{turn}}
  \hspace{0.15em}
  \subfloat[Circle]{{\includegraphics[width=0.159\textwidth]{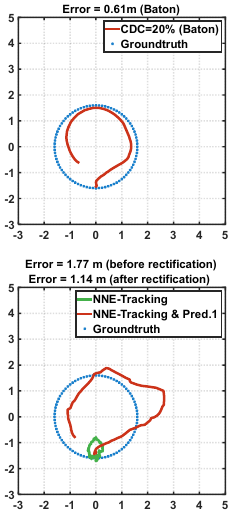}}\label{circle}}
  \hspace{0.15em}
  \subfloat[N-shape]{{\includegraphics[width=0.159\textwidth]{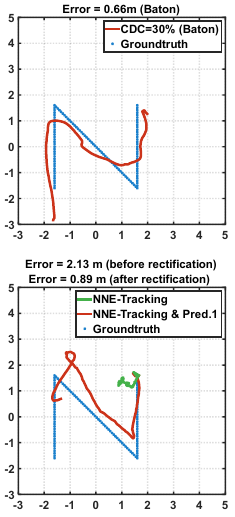}}\label{n-shape}}
  \hspace{0.15em}
  \subfloat[Triple-turn]{{\includegraphics[width=0.159\textwidth]{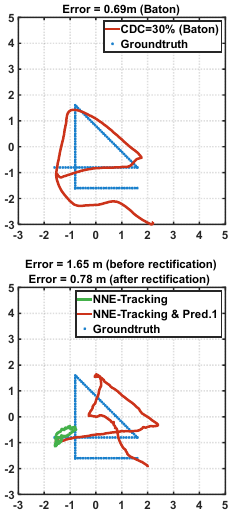}}\label{triple-turn}}
  \hspace{0.15em}
  \subfloat[Eight-shape]{{\includegraphics[width=0.159\textwidth]{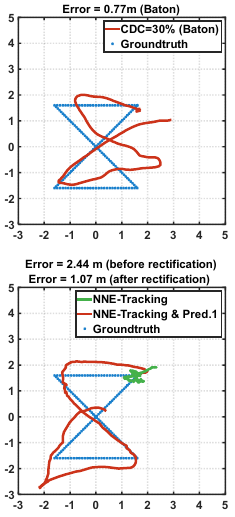}}\label{eight-shape}}

\caption{Tracking results for different trajectories when the CDC is merely $20\%$ for simple trace shapes (a-c) and $30\%$ for complex trace shapes (d-f). We can clearly observe that \textit{Baton} outperforms the state-of-the-art system \textit{NNE-Tracking}, even after rectifying the latter by our observation-based predictions (\textit{Pred. 1}). \mynote{The units for both the x-axis and y-axis are in meters ($m$).}}

\label{trace_example}
\vspace{-2mm}
\end{figure*}

\section{Implementation and Evaluation}\label{sec:exp}
\subsection{Implementation}
\textit{Baton} is deployed using five commercial off-the-shelf devices equipped with Intel 5300 NICs. 
Among these devices, one is designated as the transmitter, featuring a single antenna.
The remaining four serve as receivers, each equipped with three linear array antennas, as shown in Fig.~\ref{exp_devices}.
The antennas are spaced $2.5cm$ apart, and CSI Tool is installed to gather CSI readings. 
The packets are transmitted at a rate of $1000$Hz.
The transmitter is configured in inject mode, while the receivers operate in monitor mode at channel $64$ and  $5.32$GHz. 
\mynote{
The subjects include ten individuals with varying heights and weights, four females and six males.
}

\begin{table}[t]
\caption{Improvement of \textit{Baton} over \textit{NNE-Tracking}. \label{tab:table1}}
\centering
\begin{threeparttable}
% \vspace{0.5em}\centering\wuhao
\begin{tabular}{ccccccc}
\toprule
\textbf{Shape}\tablefootnote{Trace shapes are corresponded to presented samples in Fig.~\ref{trace_example}.} & (a) & (b) & (c) & (d) & (e) & (f)\\
\midrule
\textbf{Improve}\tablefootnote{This row presents the improvement of \textit{Baton} over \textit{NNE-Tracking} after the latter has been rectified with \textit{Pred. 1}. \textit{NNE-Tracking} results without rectification are not compared since they deviate significantly from the ground truths.} & 35.4\% & 50.0\% & 46.5\% & 25.8\% & 11.5\% & 28.0\%\\
\bottomrule
\end{tabular}
\end{threeparttable}
\vspace{-3mm}
\end{table}

\mynote{
The experiments are performed in four distinct settings. For LoS scenarios, we conduct experiments in an open space and a cafeteria measuring $4.8m \times 4.8m$ and $5m \times 4.8m$ respectively, as depicted in Fig.~\ref{open_space} and~\ref{cafe}.
To test the system's robustness, some obstacles are placed in the tracking area as another setting, which is shown in Fig.~\ref{obstacle_scenario}.
Furthermore, we evaluate the system's performance in a completely NLoS scenario where no LoS paths exist, as illustrated in Fig.~\ref{nlos_scenario}.
For the open space and the cafeteria scenarios, the results are presented in \S Section~\ref{sec:overall_performance}.
Detailed results and analysis for scenarios with obstacles and the NLoS scenario are provided in \S Section~\ref{sec:robustness_test}, as part of the robustness tests.
Around the tracking areas, pedestrians other than the subject sometimes pass by, which can cause some disturbance to the signals.}

\subsection{Overall Performance}\label{sec:overall_performance}
We collect trajectories with eight trace types to evaluate the performance of \textit{Baton}.
Among them, there are simple trajectories (\textit{e.g.}, straight, turn and circle) and complex ones (\textit{e.g.}, N-shape, triple-turn, square and eight-shape), \mynote{as depicted in Fig.~\ref{trace_example}}.
For each trace shape tested, we conduct repetitive experiments for 5 times.
\mynote{We compare the performance of our system with both data-driven methods and model-based approaches.}
The state-of-the-art device-free tracking system, \textit{NNE-Tracking} \cite{tong2024nne}, uses neural network-enhanced techniques, namely to use the model-based method to supervise data-based approaches.
We use the implementation of \textit{NNE-Tracking} as a comparison with our \textit{Baton} system.
\mynote{
In addition to this, we compare our system with \textit{WiTraj} \cite{wu2021witraj}, a recent model-based tracking system, which is based on DFS extracted from Wi-Fi CSI.
}
%
%We use the implementation of tracking systems as a comparison with \textit{Baton}.
%

\begin{figure}[ht]
\vspace{-3mm}
\centering
  \subfloat[CDC = $20\%$]{{\includegraphics[width=0.24\textwidth]{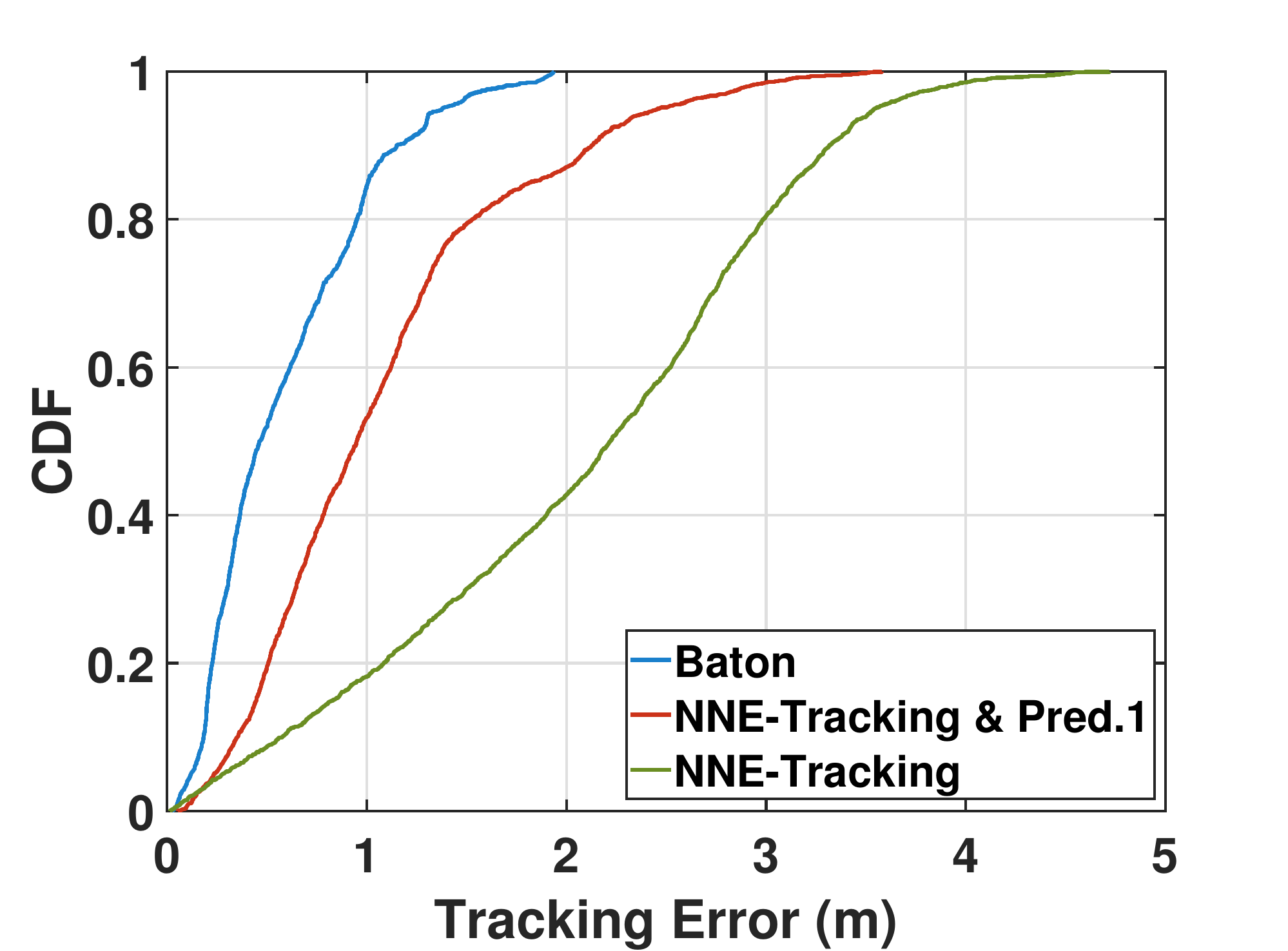}}\label{baton_vs_nne_80}}
  \subfloat[CDC = $40\%$]{{\includegraphics[width=0.24\textwidth]{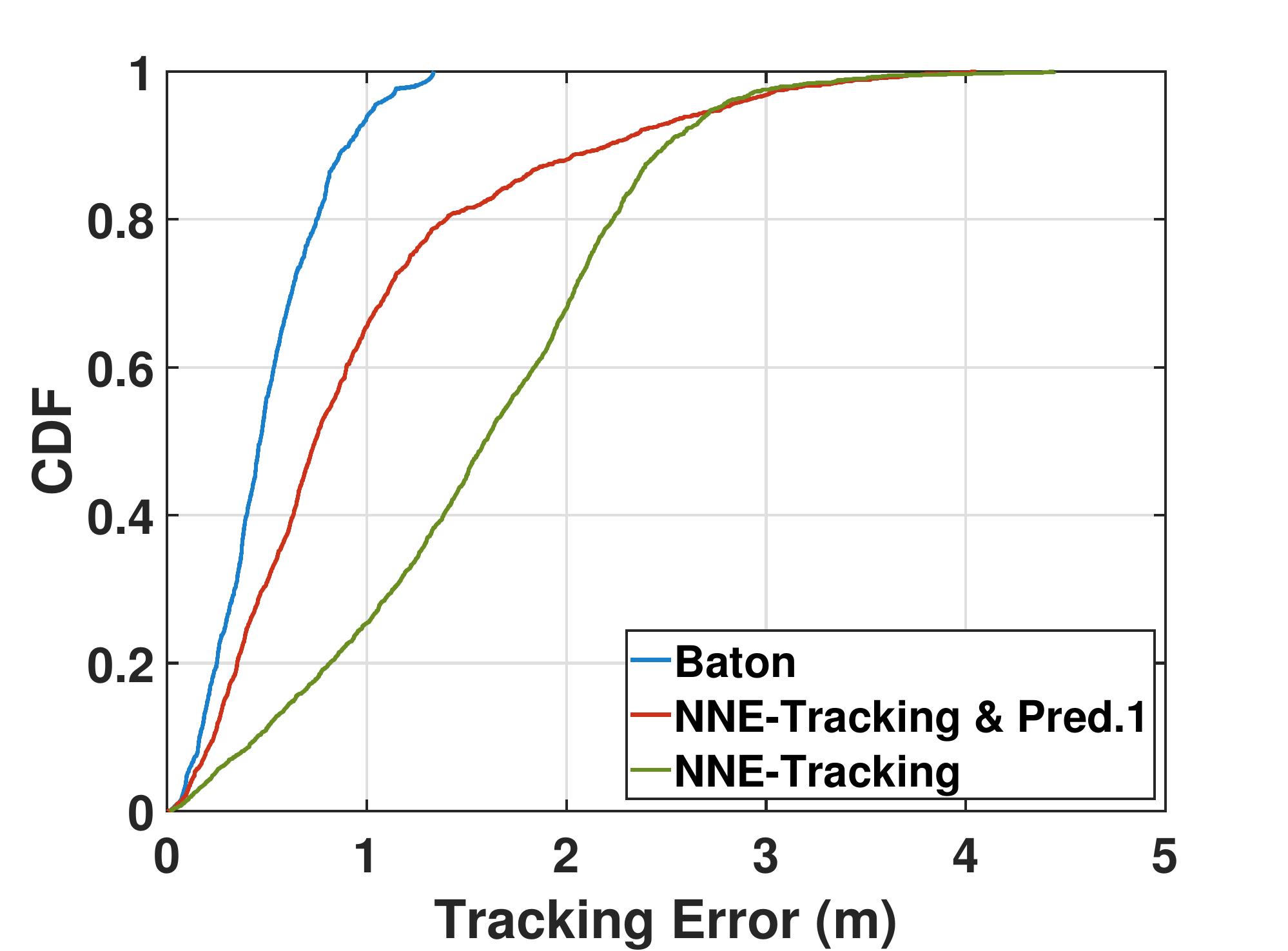}}\label{baton_vs_nne_60}}
\caption{CDF figures of \textit{Baton}'s tracking accuracy \mynote{compared with \textit{NNE-Tracking}}.}
\label{baton_vs_nne}
\end{figure}

\textbf{Tracking performance comparison with current state-of-the-art.}
In experiments, we reduce the CDC to $20\%$ for simple trace shapes and $30\%$ for more complicated ones, indicating that as many as $70\%$ or $80\%$ of signal features are missing.
Fig.~\ref{trace_example} presents sample tracking results under these conditions.
It can be seen that \textit{Baton} successfully recovers the missing signal features and eventually obtains ideal tracking results even when the CDC is as low as $20\%$ to $30\%$.
By contrast, \textit{NNE-Tracking} fails to achieve acceptable results when the CDC is significantly reduced, as its results visually deviate from the ground truths, as illustrated by the green lines.
Even after rectifying the latter by using our observation-based predictions (\textit{Pred. 1}), as depicted by the red lines, \textit{Baton} still presents a superior performance.
For each sample presented in Fig.~\ref{trace_example}, the improvement in tracking accuracy of \textit{Baton} over \textit{NNE-Tracking} (measured by the reduction in median tracking error) is listed in TABLE~\ref{tab:table1}.
\mynote{
Fig.~\ref{baton_vs_nne_80} and Fig.~\ref{baton_vs_nne_60} are the cumulative distribution function (CDF) figures of our system's tracking errors when the CDC is $20\%$ and $40\%$, respectively, which are compared with \textit{NNE-Tracking}.
Results of \textit{NNE-Tracking} after rectifying with our observation-based predictions are also compared with \textit{Baton}.
Obviously, the \textit{Baton} system achieves a higher tracking accuracy.
}

\begin{figure}[ht]
\vspace{-3mm}
\centering
  \subfloat[CDC = $20\%$]{{\includegraphics[width=0.24\textwidth]{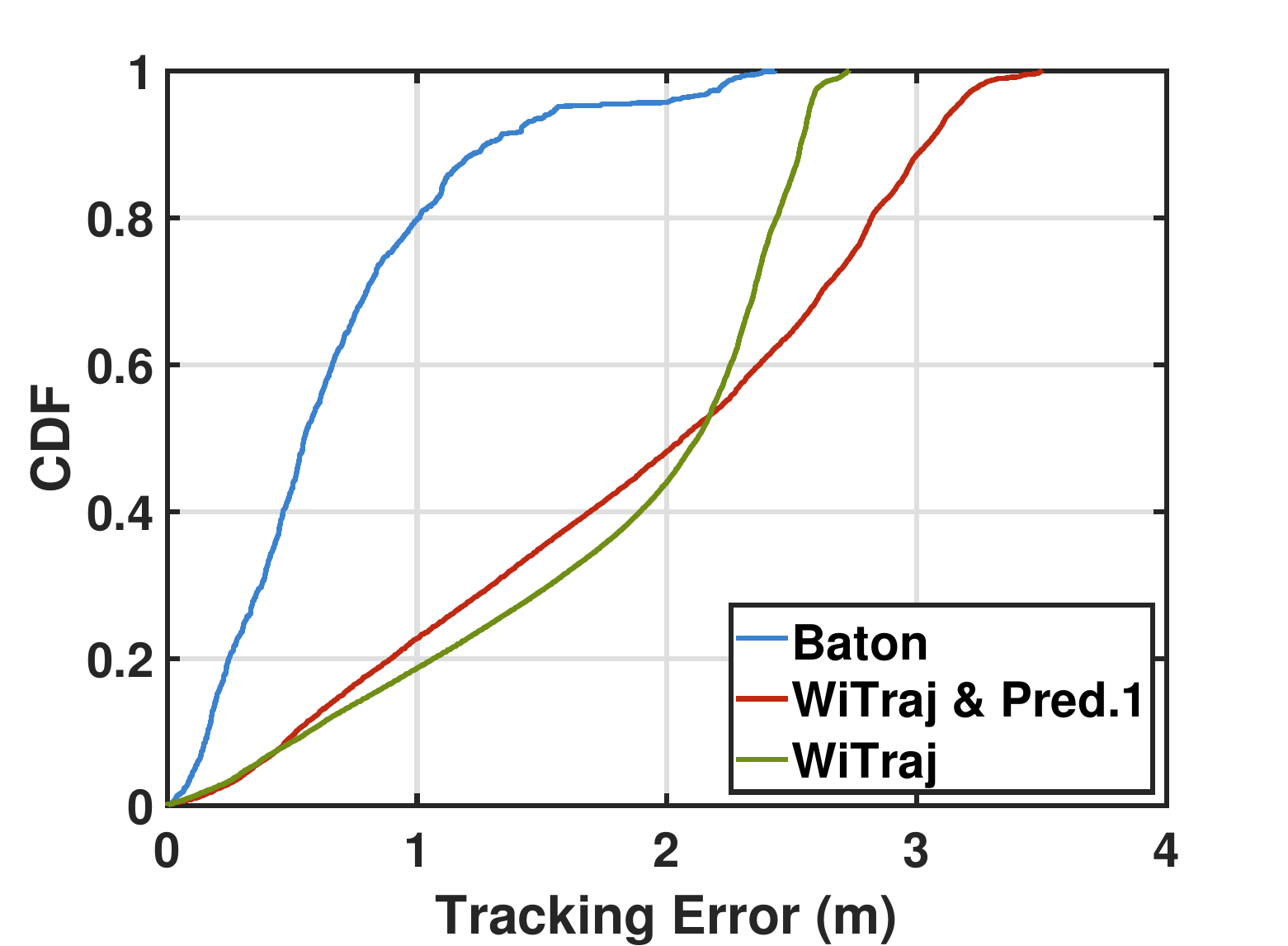}}\label{baton_vs_witraj_80}}
  \subfloat[CDC = $40\%$]{{\includegraphics[width=0.24\textwidth]{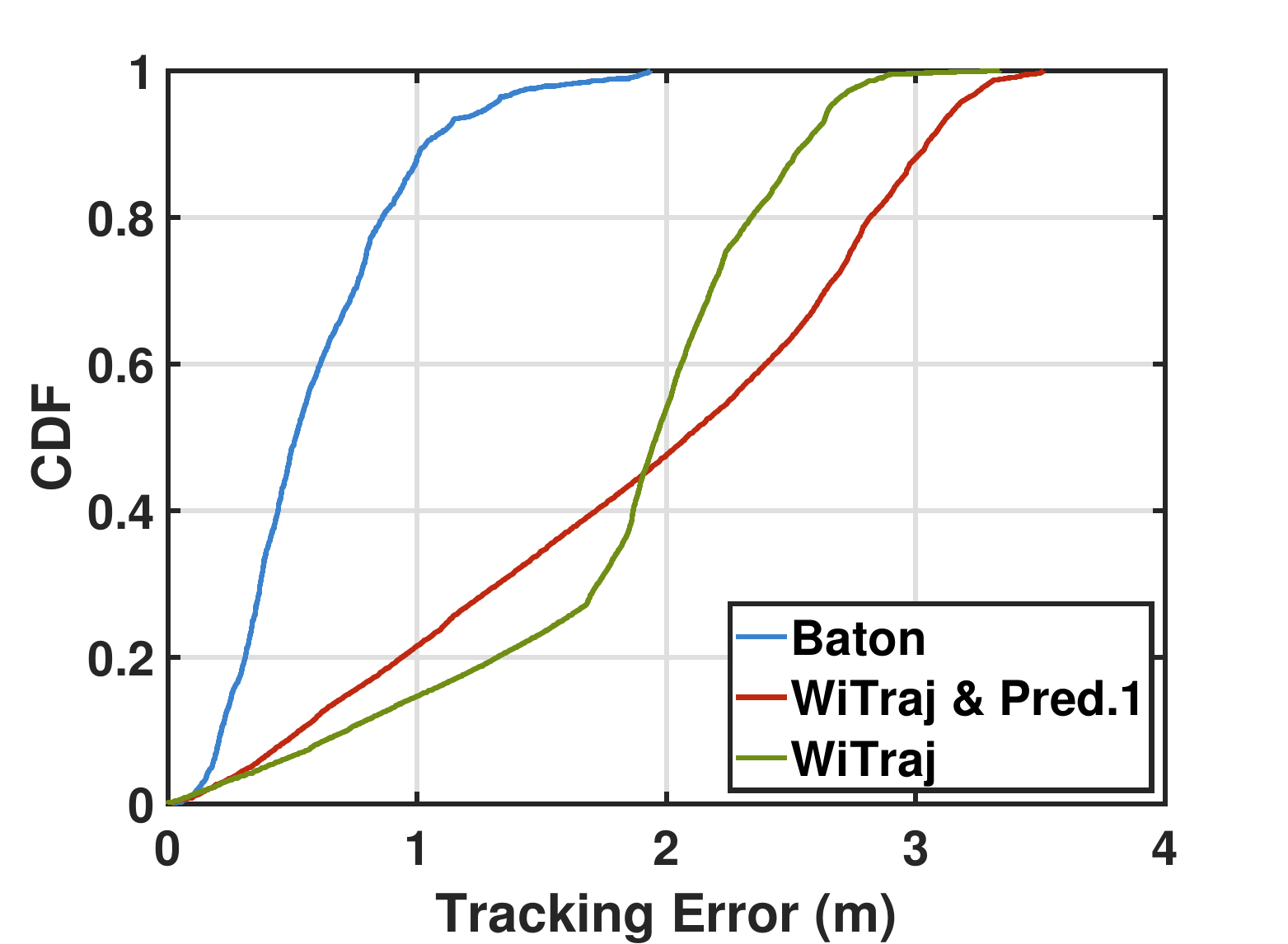}}\label{baton_vs_witraj_60}}
\caption{\mynote{CDF figures of \textit{Baton}'s tracking accuracy compared with \textit{WiTraj}}.}
\label{baton_vs_witraj}
\end{figure}

\mynote{
\textbf{Tracking performance comparison with model-based tracking system.}
To further verify the performance of \textit{Baton}, we compare it with \textit{WiTraj} \cite{wu2021witraj}, a DFS-based device-free Wi-Fi tracking system.
It adopts a model-based approach using DFS extracted from Wi-Fi CSI, which does not require large-scale antenna arrays or high bandwidth.
One transmitter and three receivers are used in this set of experiments.
When testing the performance of \textit{WiTraj}, one of the receivers is put below the transmitter, as recommended by the original paper of \textit{WiTraj}.
As illustrated in Fig.~\ref{baton_vs_witraj_80} and Fig.~\ref{baton_vs_witraj_60}, the median tracking errors of \textit{Baton} are $0.55m$ and $0.51m$, respectively, when the CDC is $20\%$ and $40\%$.
By contrast, \textit{WiTraj} has a median error of $2.12m$ when CDC=$20\%$, and $1.96m$ when CDC=$40\%$.
We also test whether the model-based method's performance could be improved by combining it with our proposed observation-based predictions.
However, this combination does not yield a significant improvement in performance.
The results demonstrate that our system consistently outperforms the model-based approach, achieving significantly lower median tracking errors under varying CDC conditions.
}

\begin{figure*}[ht]
\centering
  \subfloat[\mynote{Overall accuracy}]{{\includegraphics[width=0.198\textwidth]{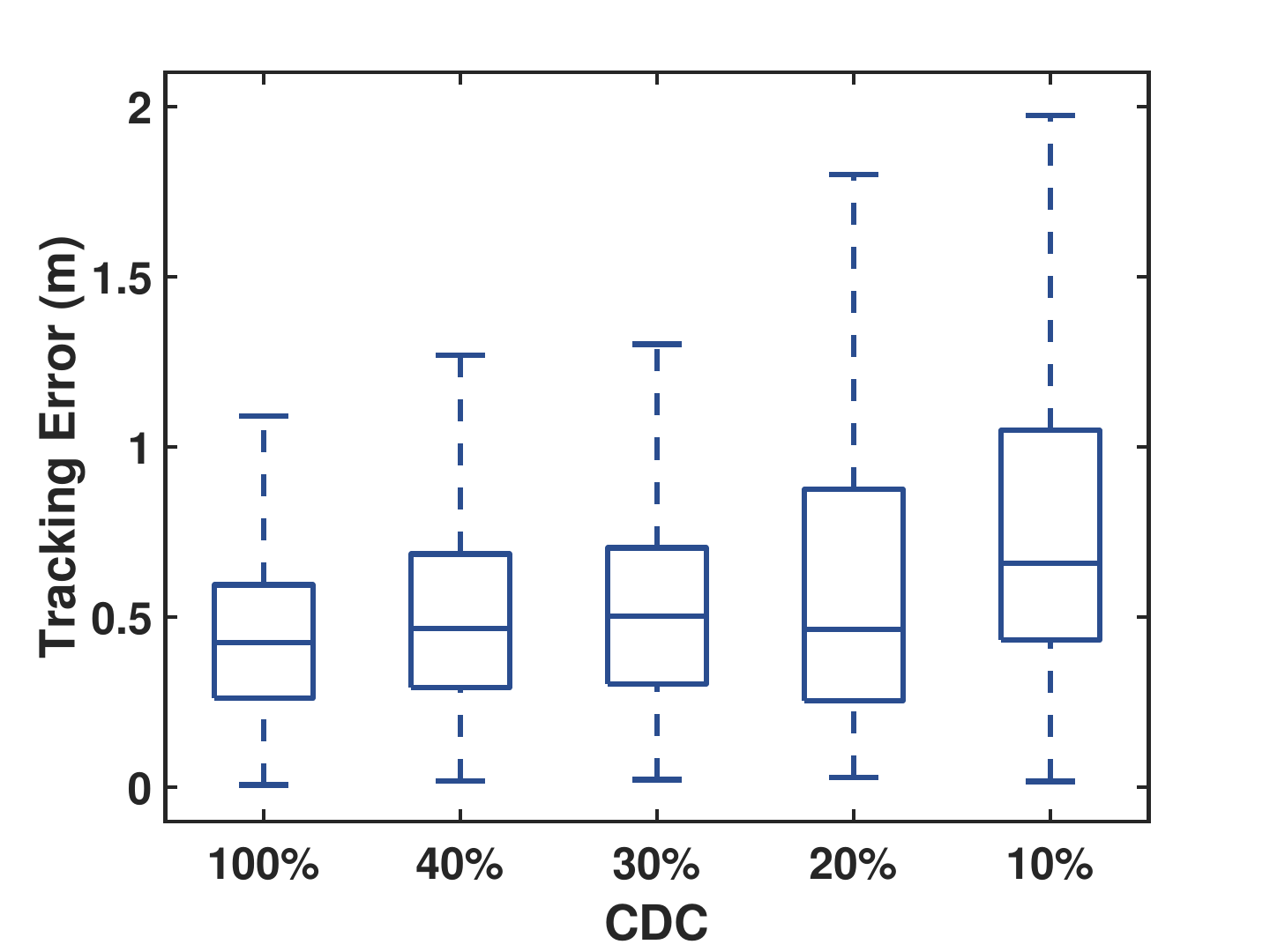}}\label{overall}}
  \subfloat[Experimental scenarios]{{\includegraphics[width=0.198\textwidth]{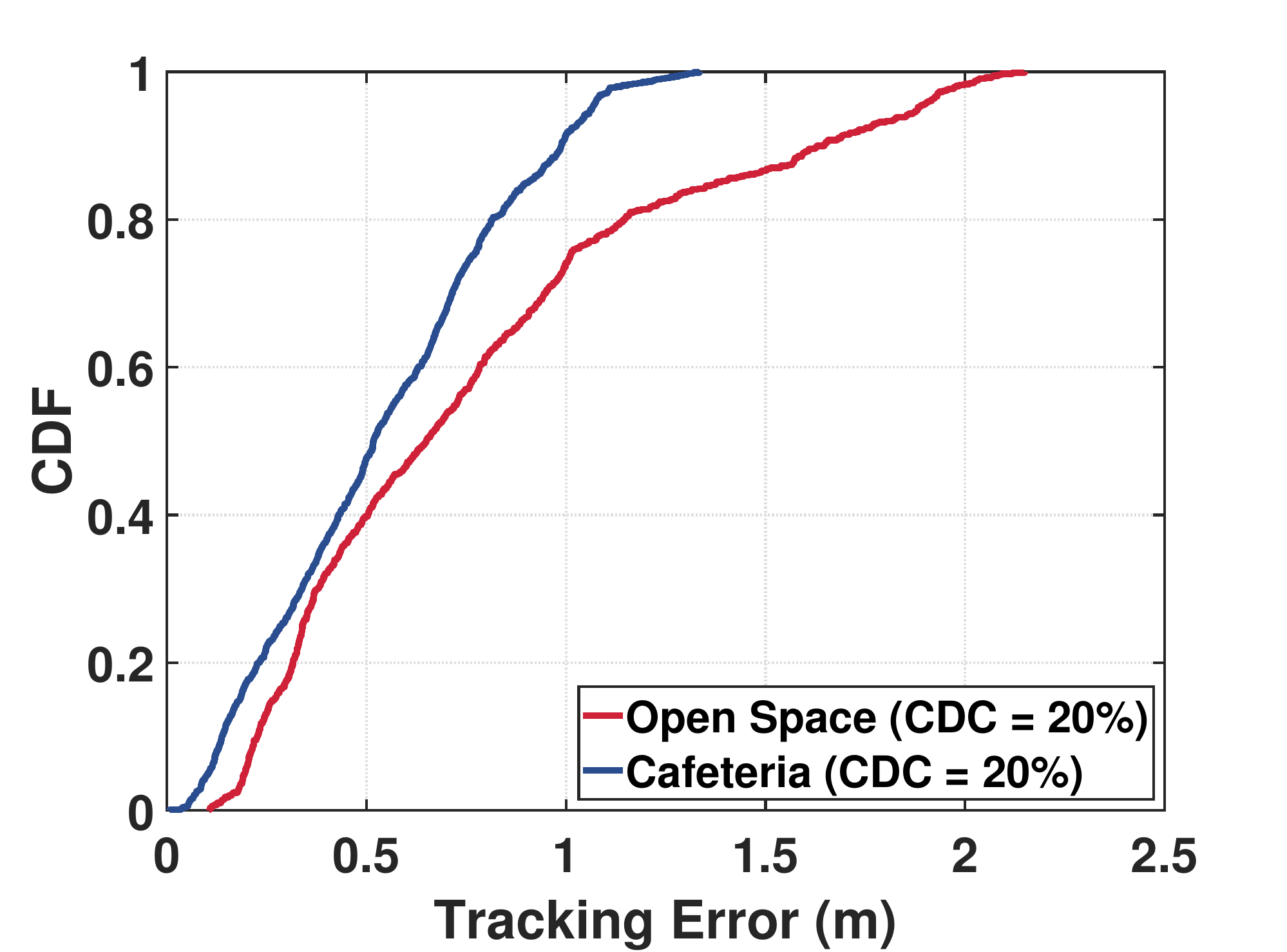}}\label{scenario_comp}}
  \subfloat[\mynote{Impact of users}]{{\includegraphics[width=0.198\textwidth]{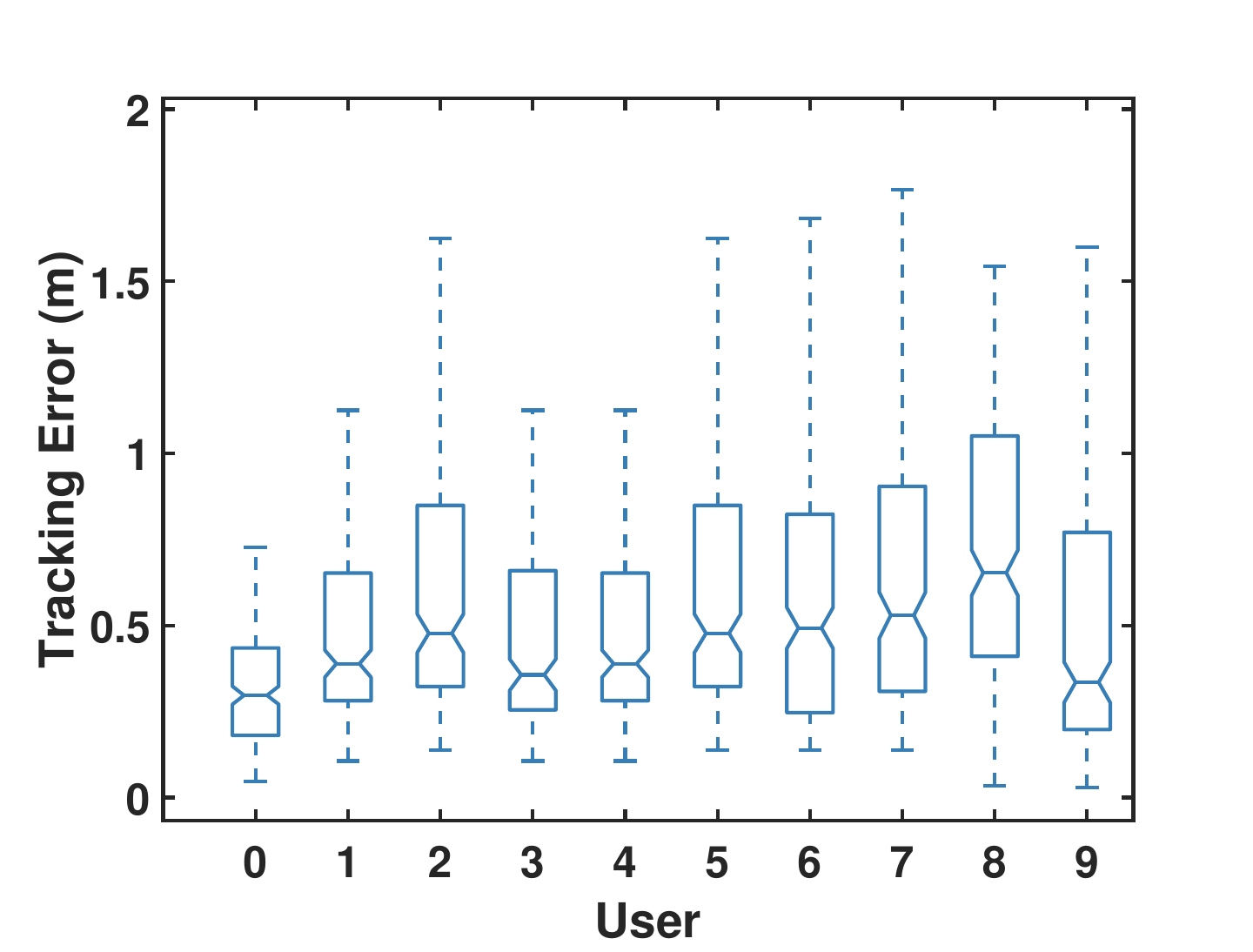}}\label{user_comp}}
  \subfloat[Impact of path shape]{{\includegraphics[width=0.198\textwidth]{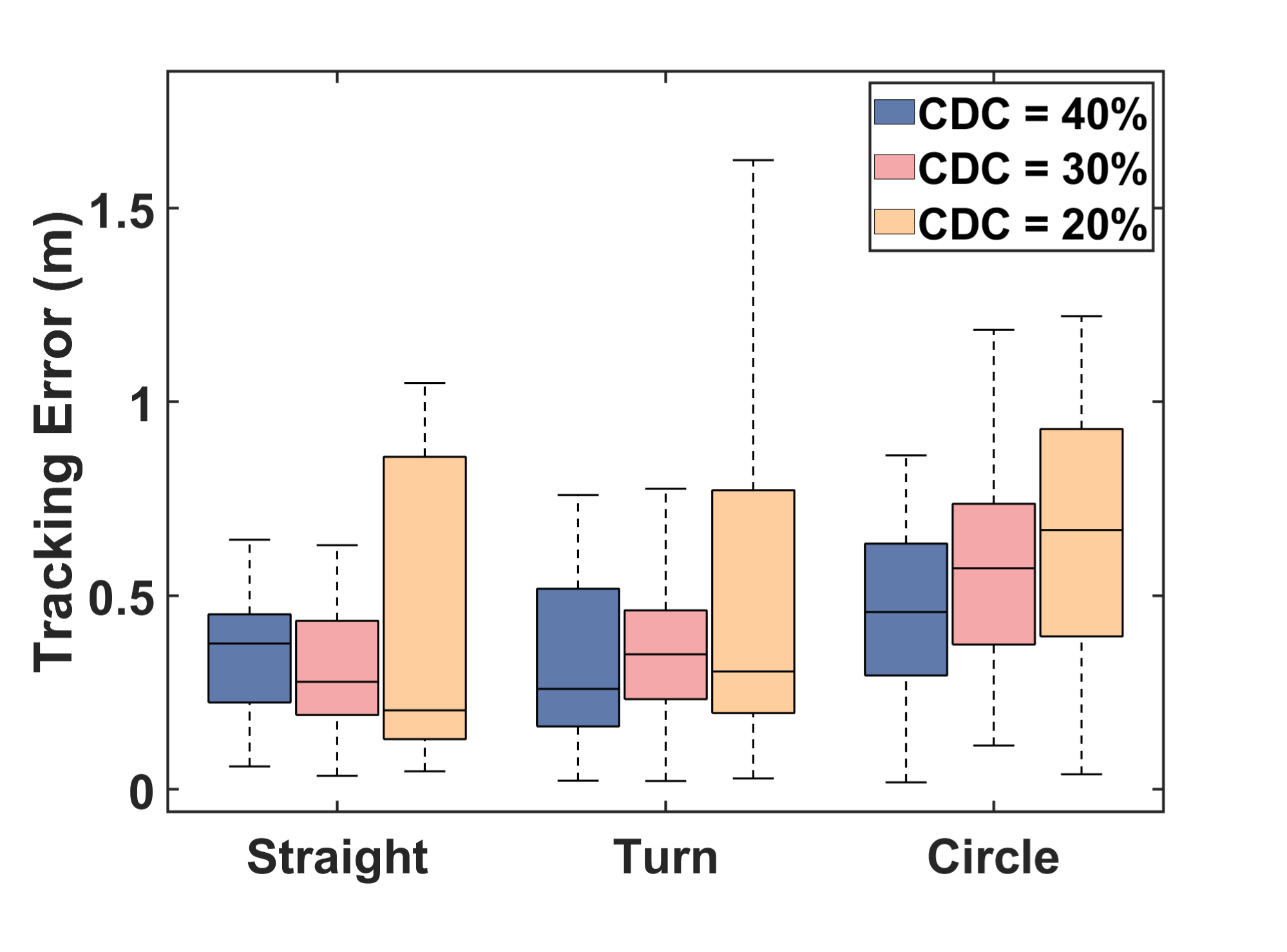}}\label{trace_comp}}
  \subfloat[Impact of path length]{{\includegraphics[width=0.198\textwidth]{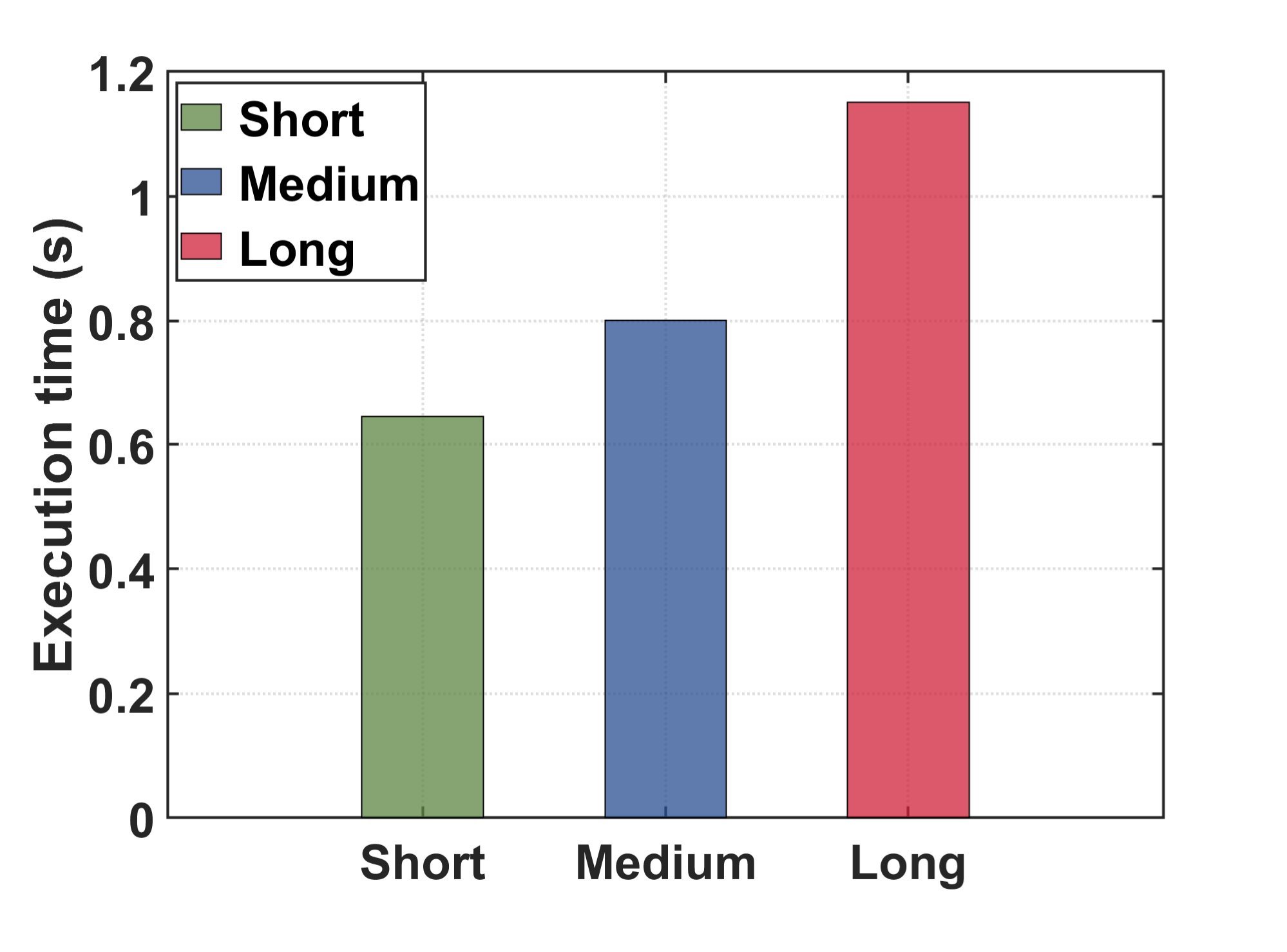}}
  \label{exectime}}
\caption{Overall performance.}
\label{overall_performance}
\vspace{-3mm}
\end{figure*}

\textbf{Overall tracking accuracy.}
The overall tracking performance of \textit{Baton} is illustrated in Fig.~\ref{overall}.
\mynote{The median tracking errors are $0.47m$, $0.50m$, $0.46m$ and $0.65m$ when the CDC is as low as $40\%$, $30\%$, $20\%$ and $10\%$, respectively.
Generally, the overall tracking accuracy of \textit{Baton} decreases as the CDC reduces, based on the box plots.}
Moreover, as shown in Fig.~\ref{scenario_comp}, the accuracy in the cafeteria scenario is higher than in the open space.
This is because the open space is more crowded than the cafeteria, and more passers-by exacerbate the multi-path effect of signals in the cafeteria.
We also evaluate the accuracy of our signal feature predictions, \textit{i.e.} the final predicted PLCRs in the signal feature table after getting the tracking results.
PLCR prediction accuracy is evaluated with the mean squared error (MSE) between estimated PLCRs and real values.
For different trajectory shapes, the average MSE of PLCR prediction accuracy for four different links is $0.199m/s$.
\mynote{
In addition, the accuracy is almost comparable for different users, as illustrated in Fig.~\ref{user_comp}.
This highlights the reliability of our system in maintaining consistent performance regardless of user variability.
}

\textbf{Accuracy and computation efficiency for different traces.}
As depicted in Fig.~\ref{trace_comp}, simpler paths generally correspond to better tracking accuracy.
Also, the accuracy decreases more obviously when the CDC reduces for more complicated paths (\textit{e.g.}, circle) when compared with simpler ones (\textit{e.g.}, straight and turn).
We also explore the execution time of a single trajectory for different trace lengths, as Fig.~\ref{exectime}.
The average execution time for short ($8s$), medium ($10s$) and long ($12s$) traces are $0.65s$, $0.80s$ and $1.15s$, respectively.
\mynote{
In addition, the proposed system can achieve real-time tracking.
This is because the tracking result at a given moment is only based on the information before this moment, rather than relying on future data.
The window size to gather PLCR data and obtain a new location prediction is $0.1s$, which means we can update the user's location ten times per second.
The average time required for the system to process each second of data is 0.086 seconds.
Therefore, the nature of the tracking algorithm and the fast processing speed fully enable real-time applications.
}

\begin{figure*}[ht]
\centering
  \subfloat[Sampling rate]{{\includegraphics[width=0.198\textwidth]{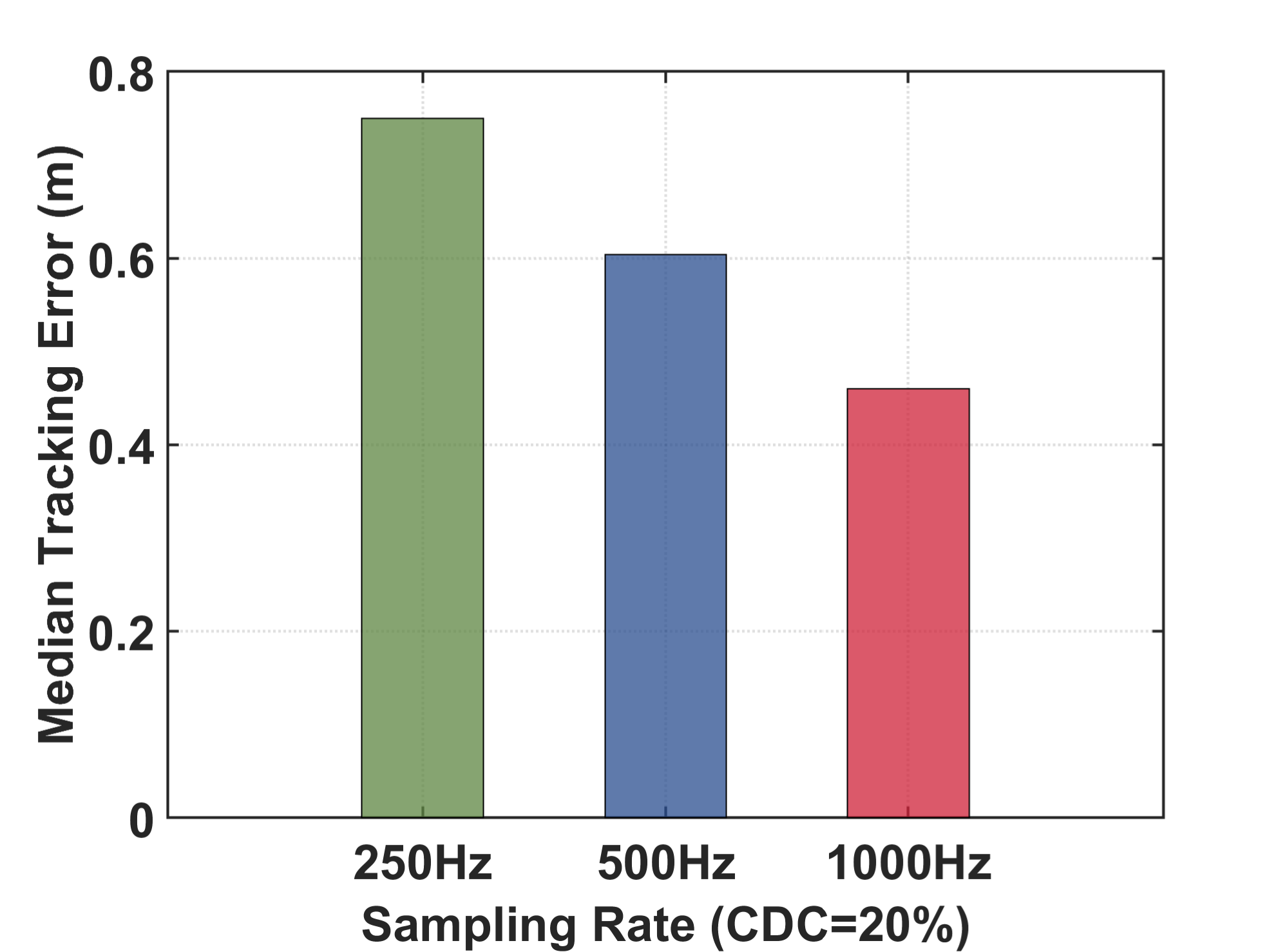}}\label{sr_comp}}
  \subfloat[Training set size]{{\includegraphics[width=0.198\textwidth]{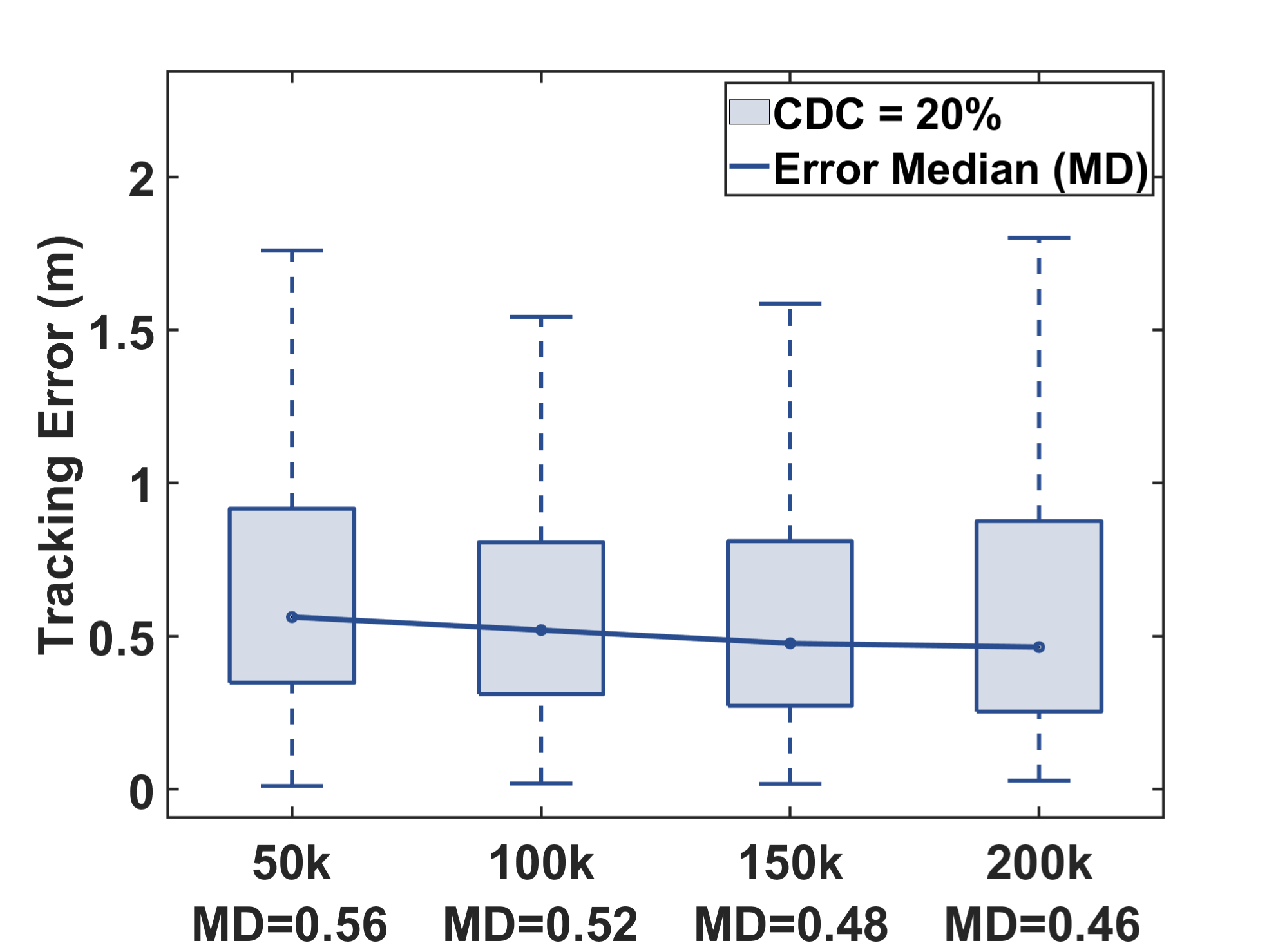}}\label{nntrain_comp}}
  \subfloat[Network parameter]{{\includegraphics[width=0.198\textwidth]{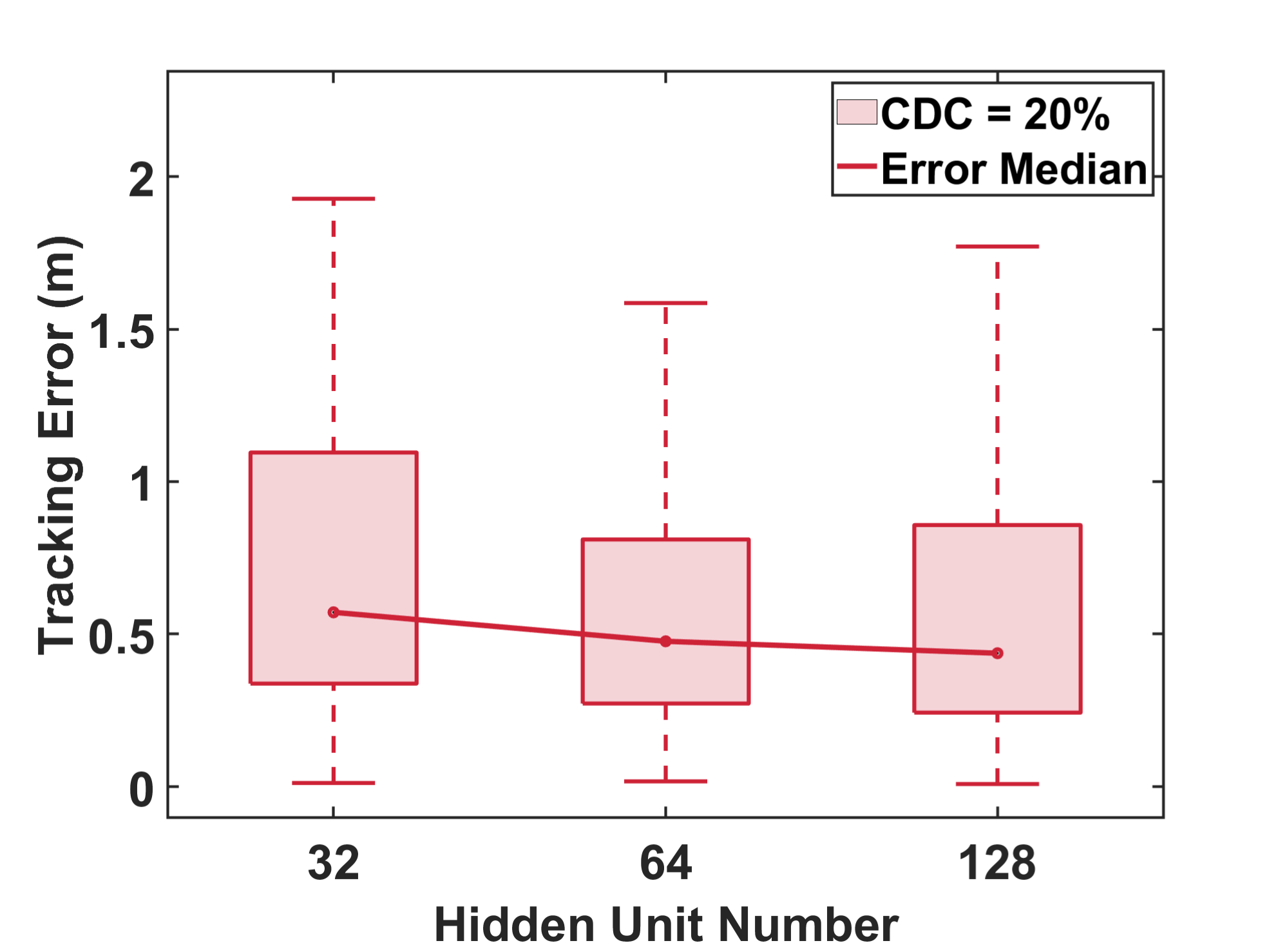}}\label{nnparam_comp}}
  \subfloat[Receiver number]{{\includegraphics[width=0.198\textwidth]{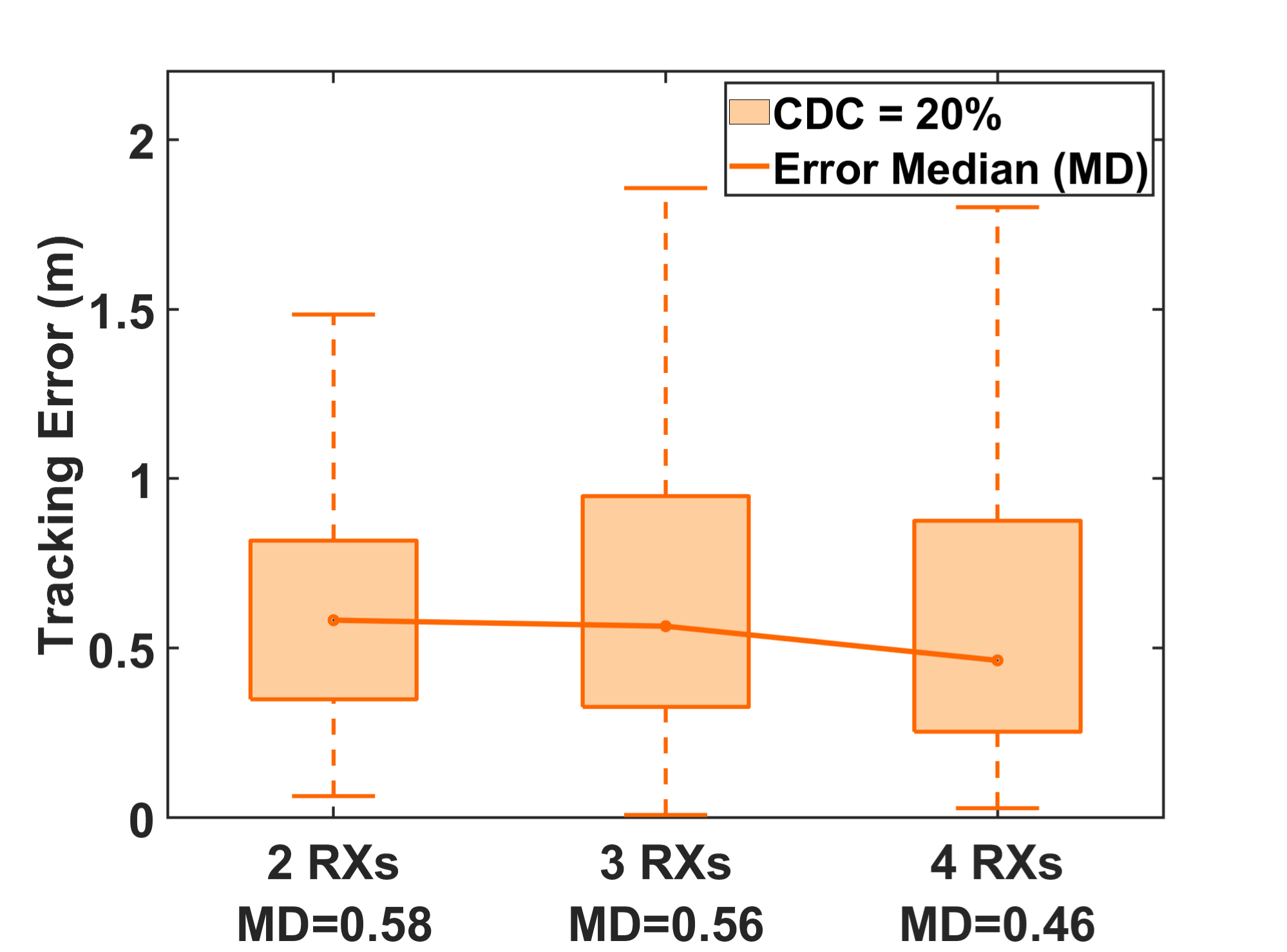}}\label{dev_comp}}
  \subfloat[Walking velocity]{{\includegraphics[width=0.198\textwidth]{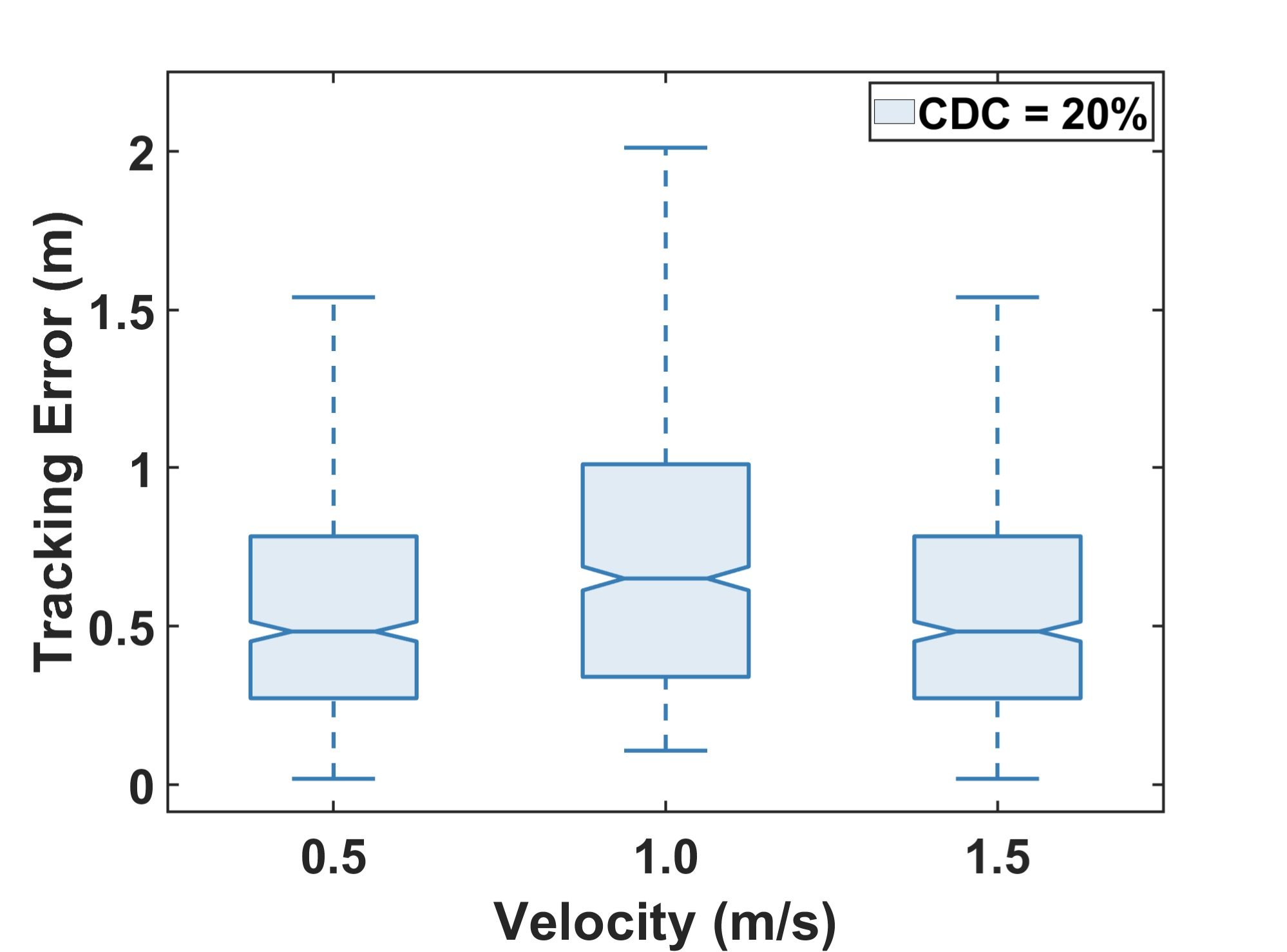}}\label{velocity_comp}}
\caption{Parameter study.}
\label{param_study}
\vspace{-3mm}
\end{figure*}

\subsection{Parameter Study}
\textbf{Impact of sampling rate.}
In a real-world scenario, channel resources can be limited and packet rates may vary.
Lower packet rates can pose negative impacts to tracking systems.
To evaluate the system's performance at various packet rates, we conduct experiments at $1000$Hz, $500$Hz and $250$Hz.
The system's performance degrades as the packet rate drops.
Fig.~\ref{sr_comp} shows that the tracking accuracy is $0.60m$ and $0.75m$ at $500$Hz and $250$Hz, respectively.
The reason is that when the packet rate is lower, the system receives updates about the subject's position less frequently.
This indicates longer intervals between consecutive updates, during which the subject's actual position could change significantly.
Thus, higher packet rates are preferable for more accurate tracking. 

\textbf{Impact of parameters related to the neural network.}
As depicted in Fig.~\ref{nntrain_comp}, a more extensive training dataset leads to a better average performance.
When the number of hidden units increases, the network's average tracking accuracy improves, as shown in Fig.~\ref{nnparam_comp}. 
This is because increased hidden units enhance the network's representation capacity and generalization performance, enabling it to better capture complex patterns and features in the input data.
However, the maximum error increases when using a more extensive training set or more hidden units.
Accordingly, it is crucial to choose moderate parameters to avoid getting biased models.

\textbf{Impact of receiver number.}
For the experiments above, we deploy four devices as receivers. 
When there are fewer receivers, fewer Wi-Fi signal features are available.
In experiments, the receiver number is reduced to three (RX1, RX2 and RX3 are retained in Fig.~\ref{exp_deploy}) and two (RX1 and RX2 are retained).
With fewer receivers, Fig.~\ref{dev_comp} shows that the system still presents an accuracy comparable to the baseline which uses four receivers.
Using only two receivers with the CDC as low as $20\%$, \textit{Baton} achieves an impressive accuracy with a median error of lower than $0.59m$.
This shows less receivers do not cause severe tracking accuracy degradation on the \textit{Baton} system.

\textbf{Impact of velocity.}
Typically, human walking velocities do not exceed $2m/s$ \cite{wang2016gait}.
Participants are asked to walk at different average speeds in this study. Fig.~\ref{velocity_comp} presents the tracking error distribution corresponding to diverse walking velocities.
Despite the varying walking speeds, \textit{Baton} consistently achieves accurate tracking results with an error of less than $0.65m$ when the CDC is as low as $20\%$.
This confirms different walking speeds do not compromise the system’s tracking accuracy.

\mynote{
\subsection{Key Factors Influencing Tracking Accuracy}
\textbf{Impact of the parameter $N_f$ for initial positions prediction.}
To start the system and utilize three types of PLCR predictions, we need to determine the initial velocity and location sequences. At this starting point, no information about the actual initial locations is known. To address this, we make use of the limited non-missing features in the starting stage of the algorithm. For the top $N_f$ rows of the feature table, we only use the observation-based prediction to get a relatively accurate initial prediction. After the $N_f$ rows, we can start using all three predictions and combine them to achieve better results.
Empirically, we choose the number of time slots in one second as $N_f$. In our experiments, ten PLCR values are extracted per second, so that $N_f$ is 10. This is an empirical choice based on our evaluation across different trace samples. For example, Fig.~\ref{nf} illustrates the changes of mean tracking errors of the whole traces with different $N_f$ choices for three trace types. The system achieves the lowest tracking error when $N_f$ is set as 11, 9 and 11, for three different shapes of samples, respectively. In fact, across all samples of different shapes and lengths, we observe that the system can typically achieve the best performance when $N_f$ is around 10. When $N_f$ is too small, further predictions cannot be made based on reasonable initial position predictions, and errors would accumulate and propagate throughout the tracking process. By contrast, when $N_f$ is too large, the system cannot make use of our proportionate and model-based predictions to rectify observation-based predictions. Therefore, a balanced selection of $N_f$ is crucial to optimize the trade-off between initial prediction quality and the system's ability to correct errors through subsequent prediction stages.
}

\mynote{
\begin{figure}[!t]
\centering
\includegraphics[width=1\linewidth]{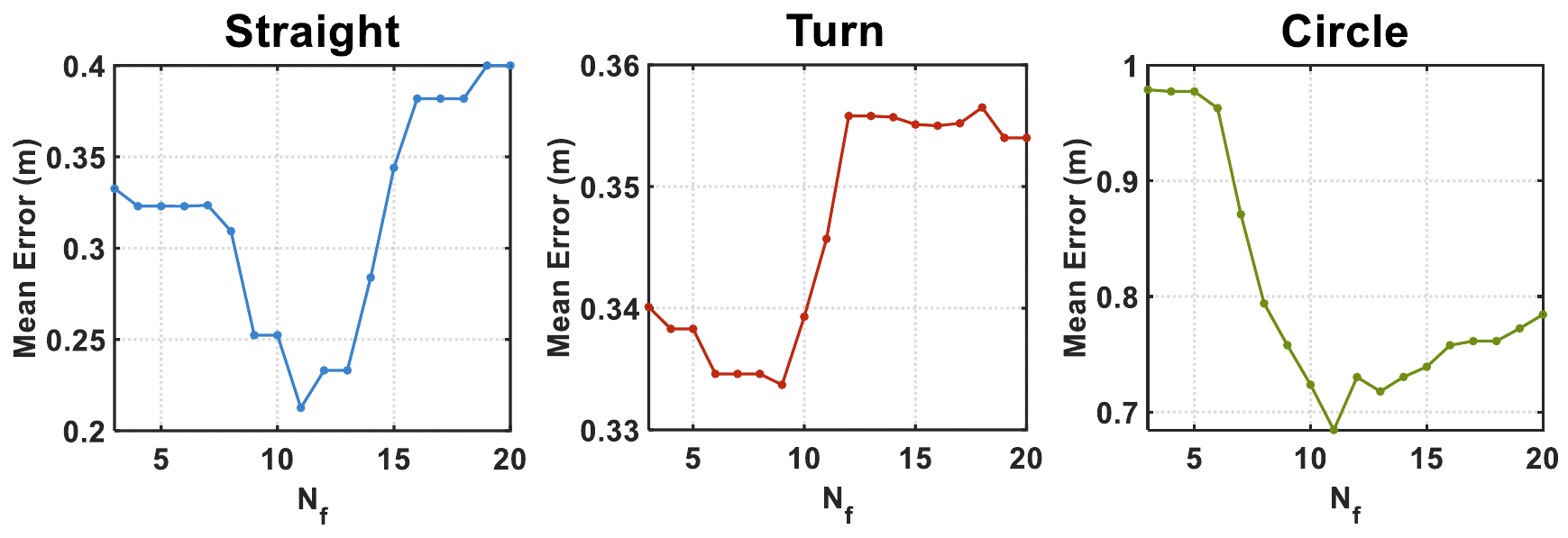}
\caption{\mynote{The system's performance is typically the best when $N_f$ is around 10, which is the number of time slots in a second based on our experiments.}}
\label{nf}
\vspace{-3mm}
\end{figure}
}

\mynote{
\textbf{Addressing error accumulation.} 
As discussed in the previous sections, our algorithm demonstrates a self-correction capability that becomes evident after a certain number of iterations. In Fig.~\ref{accumulate}, we present the average errors within each adjacent 0.5-second time slot for various trace shapes. In our experiments, 10 PLCRs are extracted per second, meaning one second corresponds to 10 iterations in the algorithm. As shown in the figure, the average error per 0.5-second interval peaks after the first 3 seconds of the trace have been processed. That is, the average error gradually increases in the first 30 iterations. However, after about 30 iterations, the error starts to decrease steadily as the system begins to compensate for missing features effectively.
As the self-correction process progresses, previously accumulated errors are rectified through iterative refinements. Early iterations are more prone to errors due to limited observations and incomplete trajectory data. 
As more iterations are processed, the algorithm incrementally integrates more reliable feature predictions with the help of proportionate predictions and model-based predictions.
The self-correction mechanism allows for the adjustment and refinement of earlier errors, ultimately converging to a more accurate trajectory.
}

\mynote{
\textbf{Impact of non-missing PLCR value distributions.} 
In previous experiment settings, the distribution of missing PLCR values in the matrix is random. In other words, the probability of any element in the PLCR matrix being missing is the same.
We further explore how non-missing feature coordinates affect the system's accuracy.
As a baseline, $80\%$ of the values in the PLCR matrix are removed for trajectories in the shape of Turn, indicating a CDC of $20\%$. 
The duration of the trajectories is 8 seconds, and four links are used.
Under this condition, the mean tracking error for all samples is $0.1764m$.
We compare the tracking errors with the baseline when more missing values are concentrated around a given time span and in certain links.
Firstly, to investigate the impact of temporally clustered missing points on tracking accuracy, we remove all non-missing values within a given continuous time period. During this time interval, all links experience continuous feature loss.
The mean errors are $0.3024m$, $0.6702m$ and $0.7125m$, respectively, when these samples are further subjected to continuous losses of $1s$, $2s$ and $3s$ based on the baseline.
Secondly, continuous feature loss in specific links also leads to a decline in accuracy.
When all features in one link are missing, the mean error is $0.5128m$.
When all features in two links are missing, the mean error further rises to $0.7281m$.
Therefore, in practice, if a link is known to be experiencing continuous feature loss, a suitable approach would be to exclude data from this link. In this way, we can only utilize the links where missing features are distributed more evenly. As shown in the experiments in Fig.~\ref{dev_comp}, reducing the number of links does not have a significant effect on localization accuracy.
To summarize, a relatively uniform distribution of non-missing features leads to better tracking accuracy.
}

\mynote{
\begin{figure}[!t]
\centering
\includegraphics[width=0.8\linewidth]{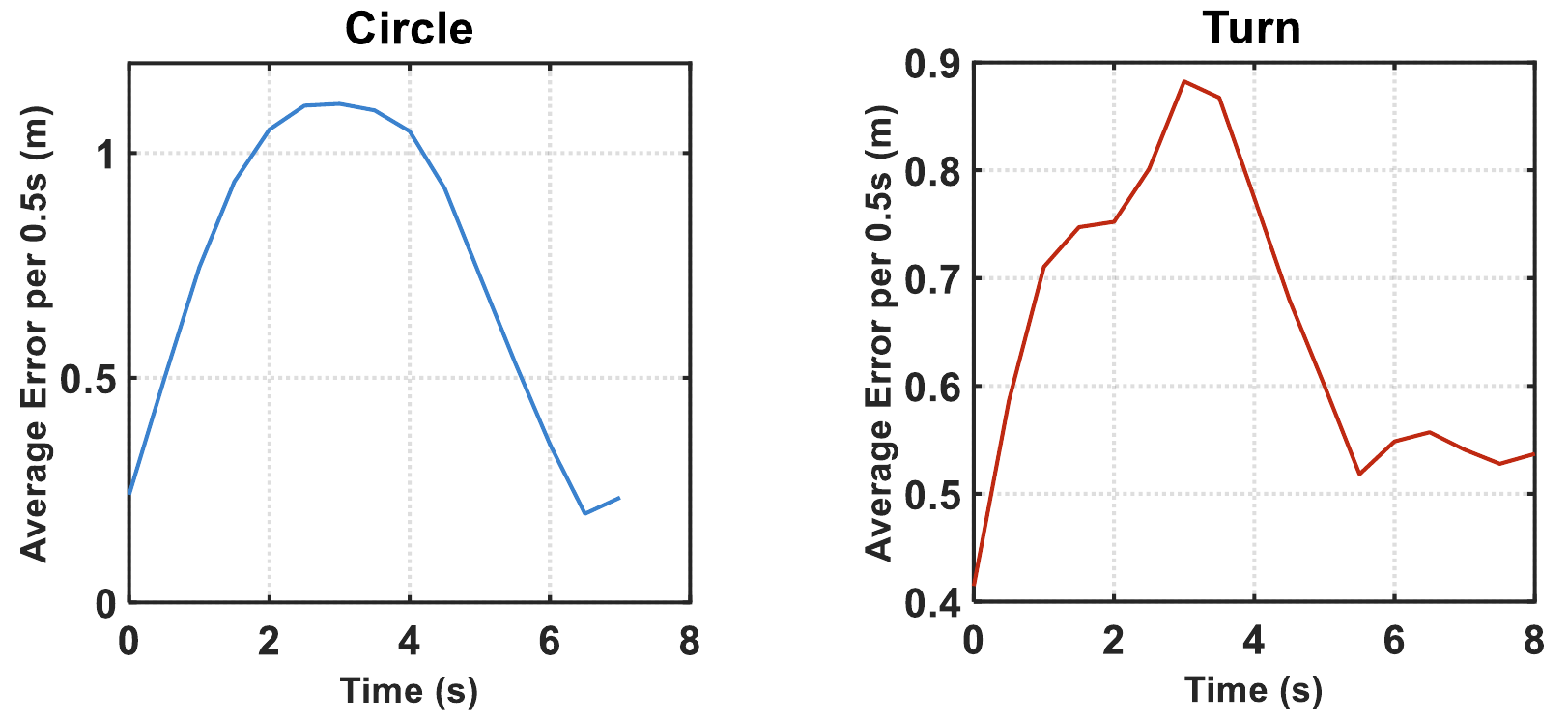}
\caption{\mynote{The average tracking error per 0.5s increases after starting, and then decreases with the system's processing.}}
\label{accumulate}
\vspace{-3mm}
\end{figure}
}

\begin{figure*}[ht]
\centering
  \subfloat[Tracking with obstacles]{{\includegraphics[width=0.33\textwidth]{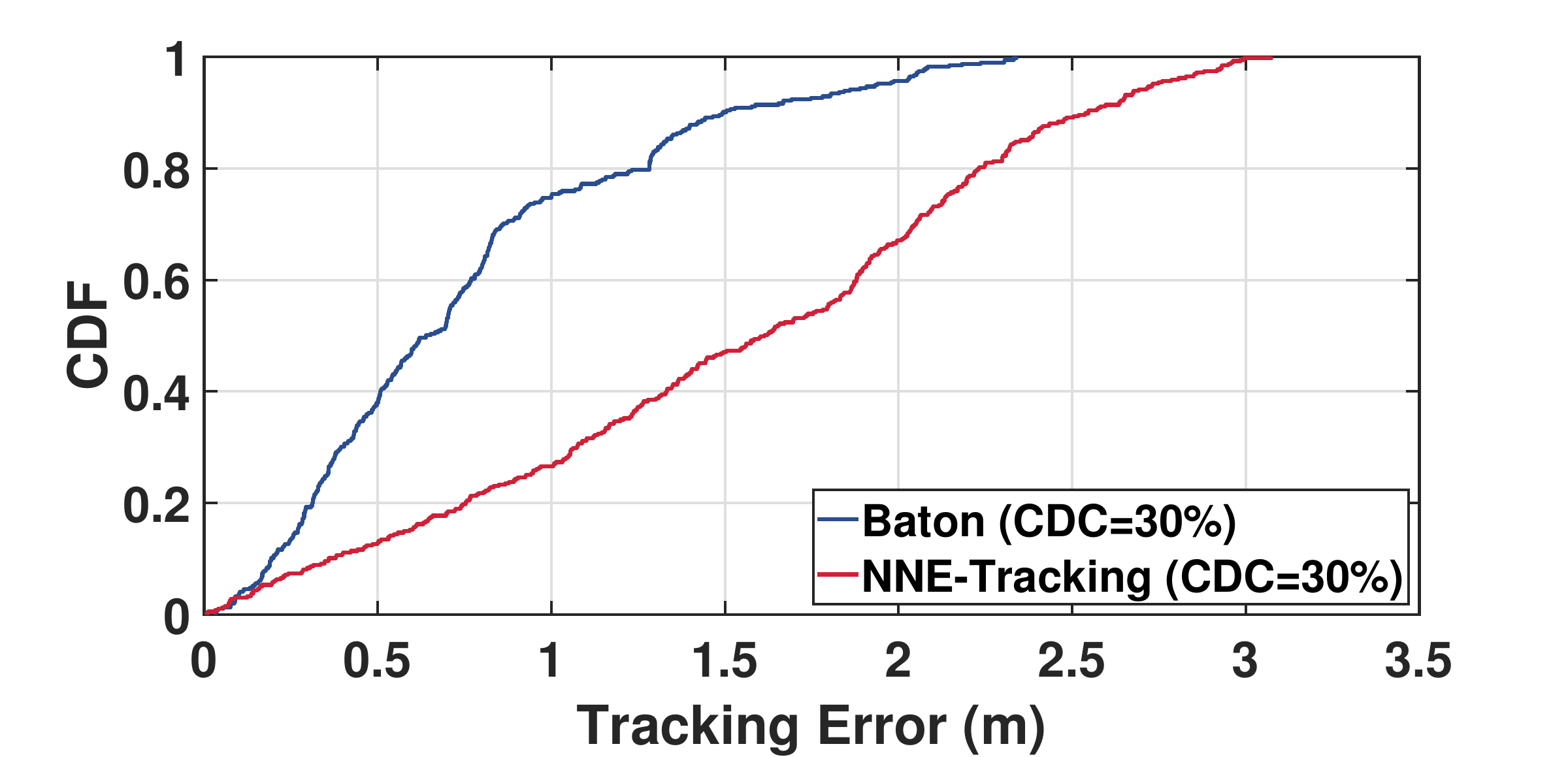}}\label{obstacle_cdf}}
  \subfloat[Non-uniform walking]{{\includegraphics[width=0.33\textwidth]{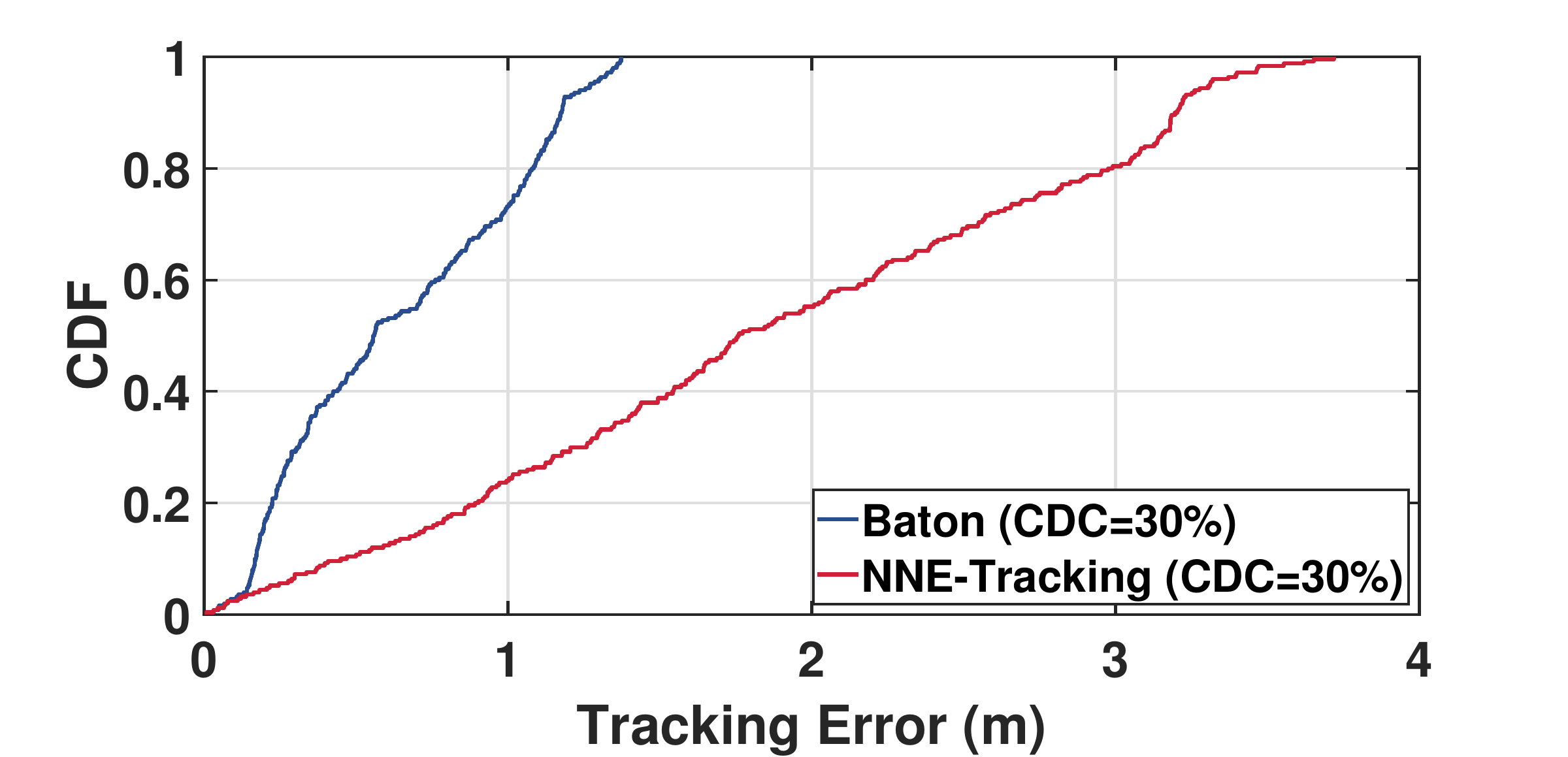}}\label{nonuni_cdf}}
  \subfloat[NLoS tracking]{{\includegraphics[width=0.33\textwidth]{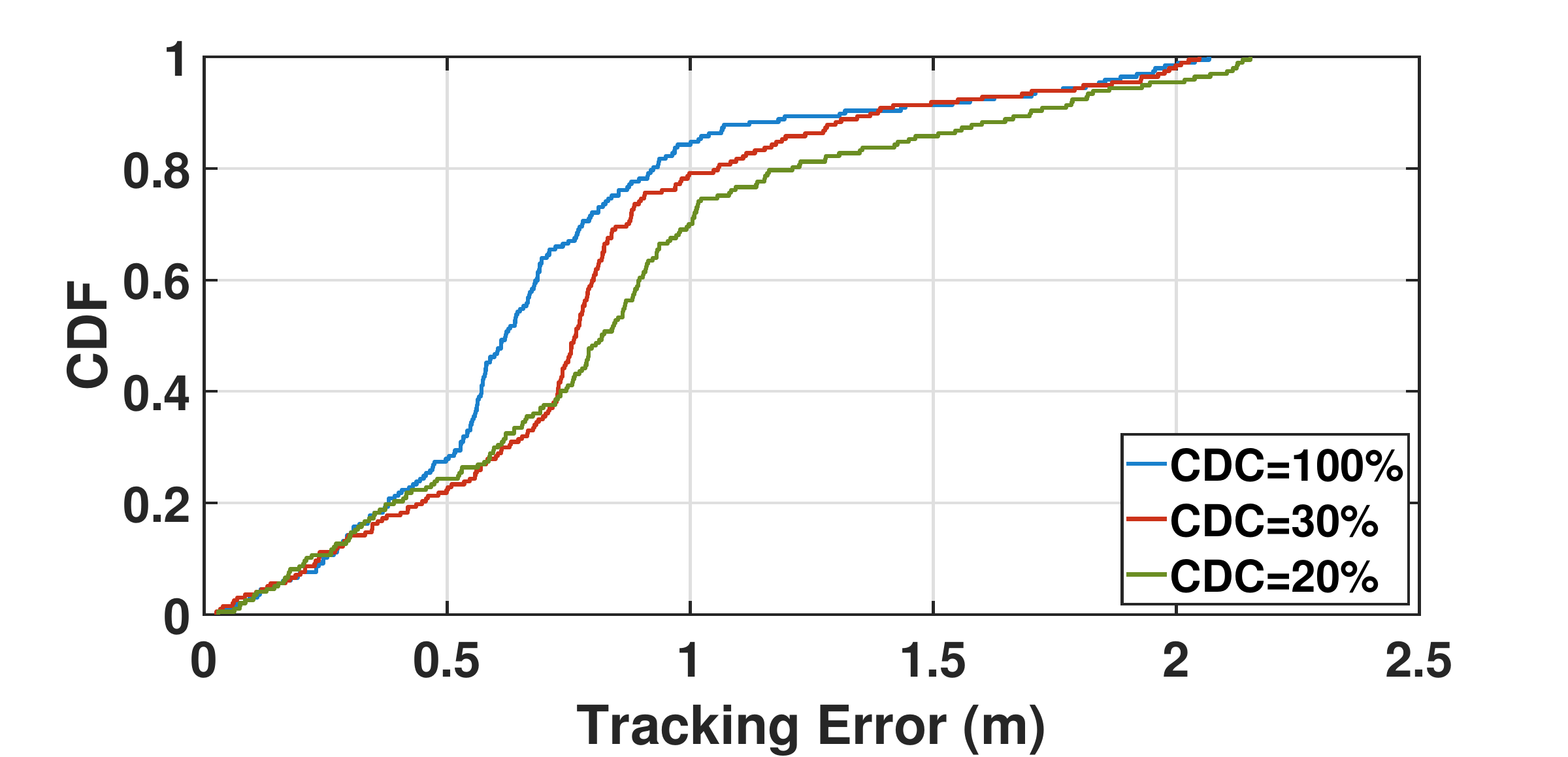}}\label{nlos-cdf}}
\caption{\mynote{Robustness tests.}}
\label{robustness_test}
\vspace{-3mm}
\end{figure*}

\subsection{Robustness Tests}\label{sec:robustness_test}
\textbf{Performance in scenarios with obstacles.}
In real scenarios, obstacles such as furniture can exist in indoor environments.
These obstacles can harm tracking systems' performance since signals reflected by different subjects can interfere with each other when they reach the receivers.
The multi-path effect happens when several delayed versions of the same signal arrive at different times and with different amplitudes.
To test the system's robustness under such conditions, we put a desk (with items on it), a chair and a pile of boxes in the tracking area, as shown in Fig.~\ref{obstacle_scenario}.
% Using 240318-Supplementary: horizon123 rightlean23 square3
\textit{Baton}'s performance is tested with a CDC of $30\%$, and both simple trace shapes (\textit{e.g.} straight) and more complex ones (\textit{e.g.} square) are used.
The median tracking error of \textit{Baton} under this condition is $0.64m$, as shown in Fig.~\ref{obstacle_cdf}, demonstrating a degradation in tracking accuracy because of the multi-path effect caused by obstacles presented.

\textbf{Tracking for non-uniform motion.}
Admittedly, walking is a relatively constant motion.
However, sometimes a person's walking speed might exhibit sudden changes.
To test the system's stability and accuracy, subjects are asked to suddenly accelerate or decelerate when walking.
Under this condition, we observe that the system's performance is unstable when the CDC is set as $20\%$, with some tracking results quite accurate but others deviating from ground truths.
However, when the CDC is increased to $30\%$, the tracking accuracy is rather stable, with a median tracking error of $0.56m$, as depicted by Fig.~\ref{nonuni_cdf}. 
While the \textit{Baton} system's tracking accuracy drops compared with ideal condition, it demonstrates a relatively stable performance when CDC is as low as $30\%$ and outperforms existing solutions.

\textbf{Extending the \textit{Baton} system to NLoS scenarios.}
Xu et al. recently proposed a hyperbolic model for device-free NLoS tracking\mynote{~\cite{xu2024hypertracking}}.
This method, named \textit{HyperTracking}, utilizes a novel differential path length change rate (DPLCR) derived from PLCRs to track by theoretical modeling.
A neural network trained under the supervision of the hyperbolic model can be used to map between DPLCRs and trace predictions.
As the STAP algorithm can be used to compensate for missing PLCR values, the corresponding missing DPLCRs can be predicted meantime.
Therefore, our system can be adjusted to address tracking under severe Wi-Fi feature deficiencies in NLoS scenarios.
Implementation details of the \textit{Baton} system are adjusted according to the hyperbolic model.
As depicted by Fig. \ref{nlos_scenario}, four receivers are set up outside the room, while the transceiver is deployed inside the room (behind the wall).
Trace shapes including straight, turn and square are tested in the tracking area outside the room.
% Use data (4 RX), line 24, line 45, line 48.
As shown in Fig. \ref{nlos-cdf}, the median tracking errors are $0.62m$, $0.76m$ and $0.82m$ when the CDC is $100\%$, $30\%$ and $20\%$, respectively.
This shows the \textit{Baton} system not only successfully compensates for missing Wi-Fi features in LoS scenarios, but can also be extended to NLoS situations where there is no direct line of sight between the transmitter and receivers.
In fact, one advantage of the STAP algorithm is its plug-and-play capability, allowing it to be applied on top of existing data-driven tracking methods when they are faced with Wi-Fi feature deficiencies in real-world scenarios.

% Part 7: Conclusion
\section{Conclusion}\label{sec:conc}
This paper presents \textit{Baton}, the first system capable of accurately tracking targets under severe Wi-Fi feature deficiencies.
We implement the STAP algorithm, a feature compensation mechanism, by leveraging the relations between the motion state and Wi-Fi signal features.
The key idea of the algorithm is to track and predict missing signal features simultaneously, thus incrementally obtaining the complete tracking result.
We conduct extensive experiments on the \textit{Baton} system with COTS Wi-Fi devices.
Results show that \textit{Baton} achieves a median tracking error of $0.46m$ with a low communication duty cycle of merely $20.00\%$, outperforming state-of-the-art approaches.

\section{Acknowledgement}
The work in this paper is supported by National Natural Science Foundation of China (No. 62472307, 62372324, 62402336, 62102282).

\bibliographystyle{IEEEtran}
\bibliography{ref}
\flushend

\newpage
\begin{IEEEbiography}[{\includegraphics
[width=0.9in,height=1.2in,clip,
keepaspectratio]{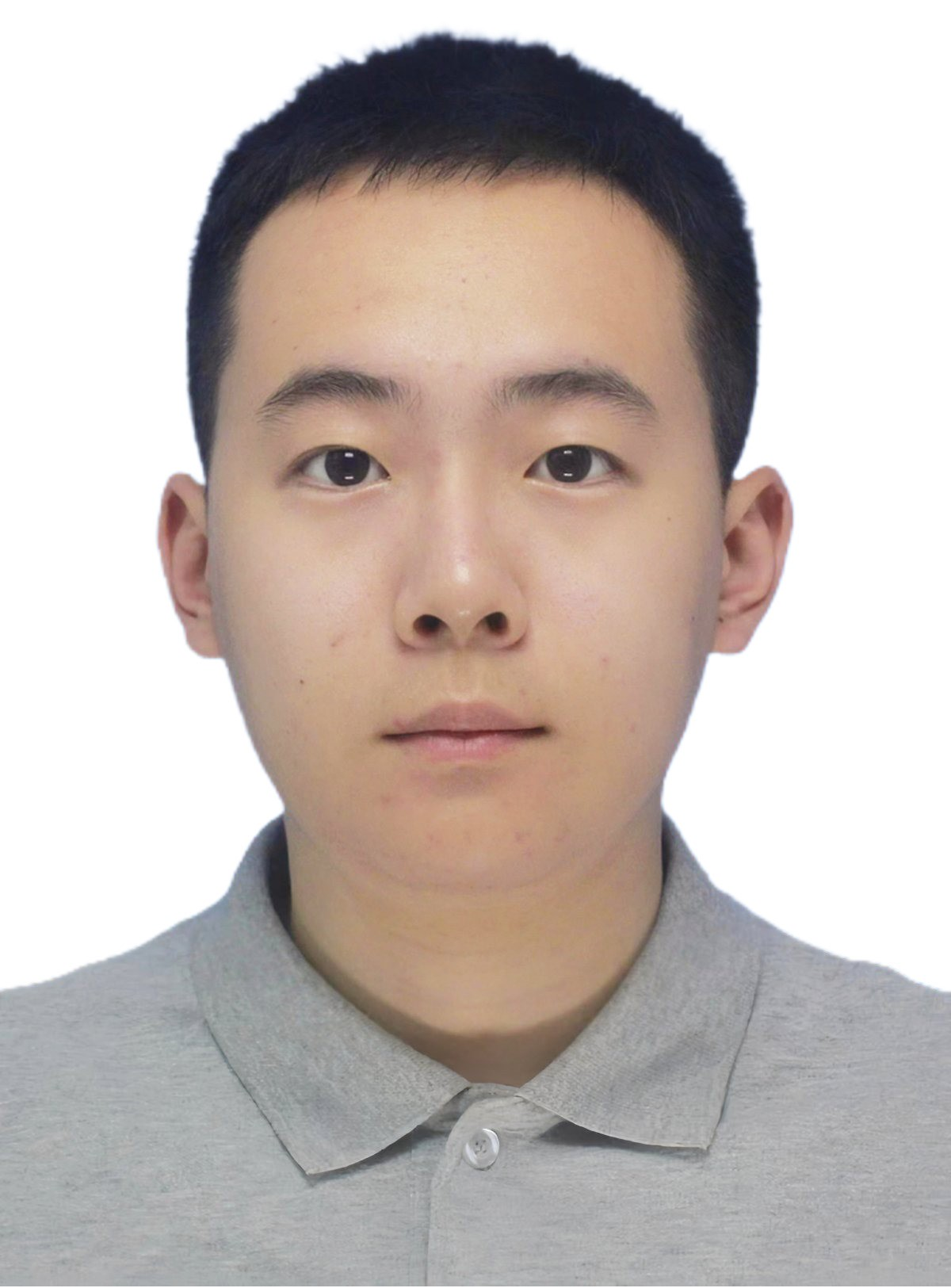}}]
{Yiming Zhao} is a senior undergraduate student at the College of Intelligence and Computing, Tianjin University, China. He is currently pursuing his B.E. degree. His research interests lie in the field of indoor localization.
\end{IEEEbiography}
\vspace{-3mm}
\begin{IEEEbiography}[{\includegraphics
[width=0.9in,height=1.2in,clip,
keepaspectratio]{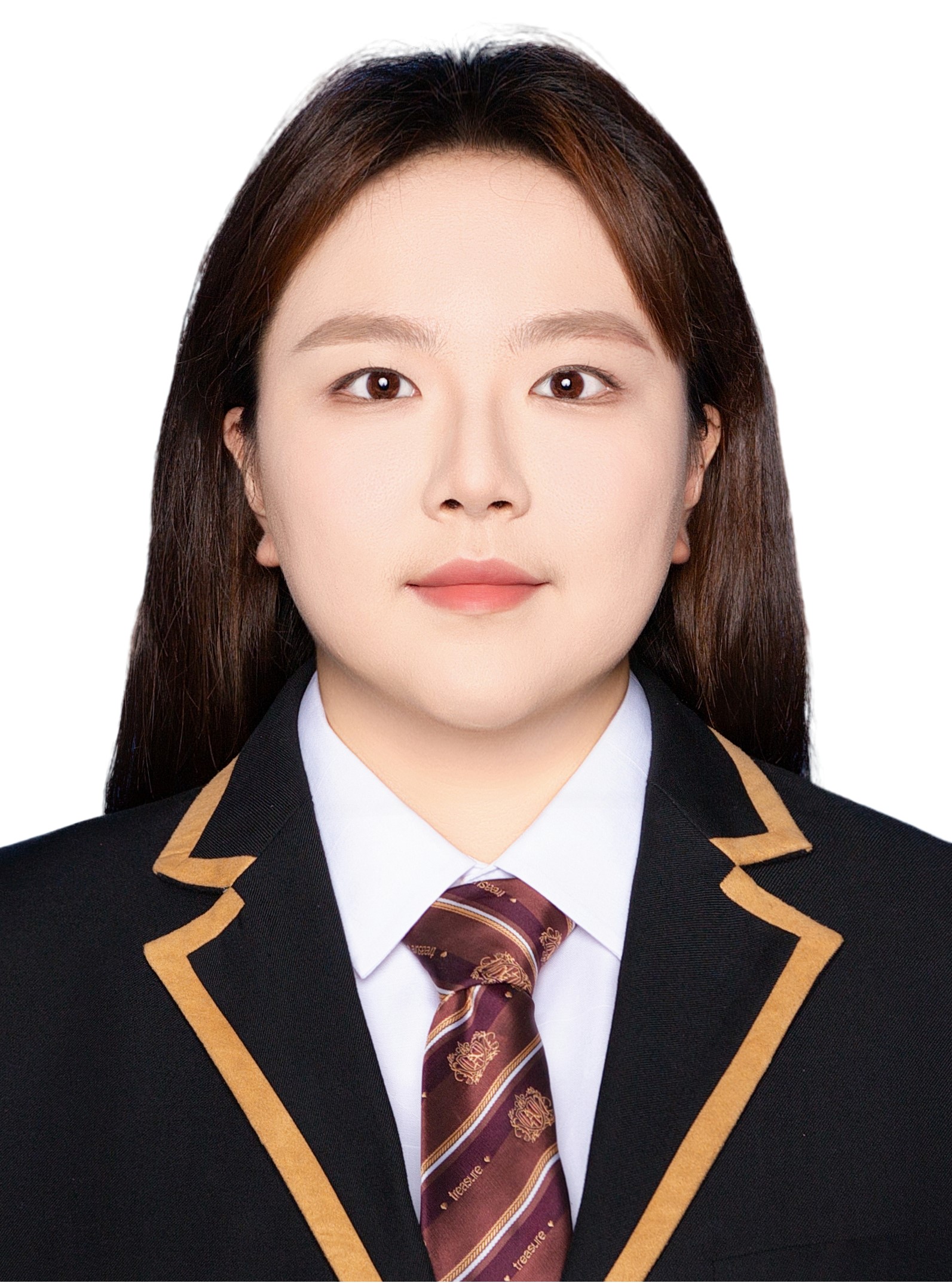}}]{Xuanqi Meng}
	received the B.E degree in computer science from the Dalian University of Technology, Dalian, China, in 2021, and the MS degree from the College of Intelligence  and Computing, Tianjin University, China, in 2024. Currently, she is pursuing her Ph.D. degree in Tianjin University. Her research interests include wireless sensing and localization.
\end{IEEEbiography}
\vspace{-3mm}
\begin{IEEEbiography}[{\includegraphics
[width=0.9in,height=1.2in,clip,
keepaspectratio]{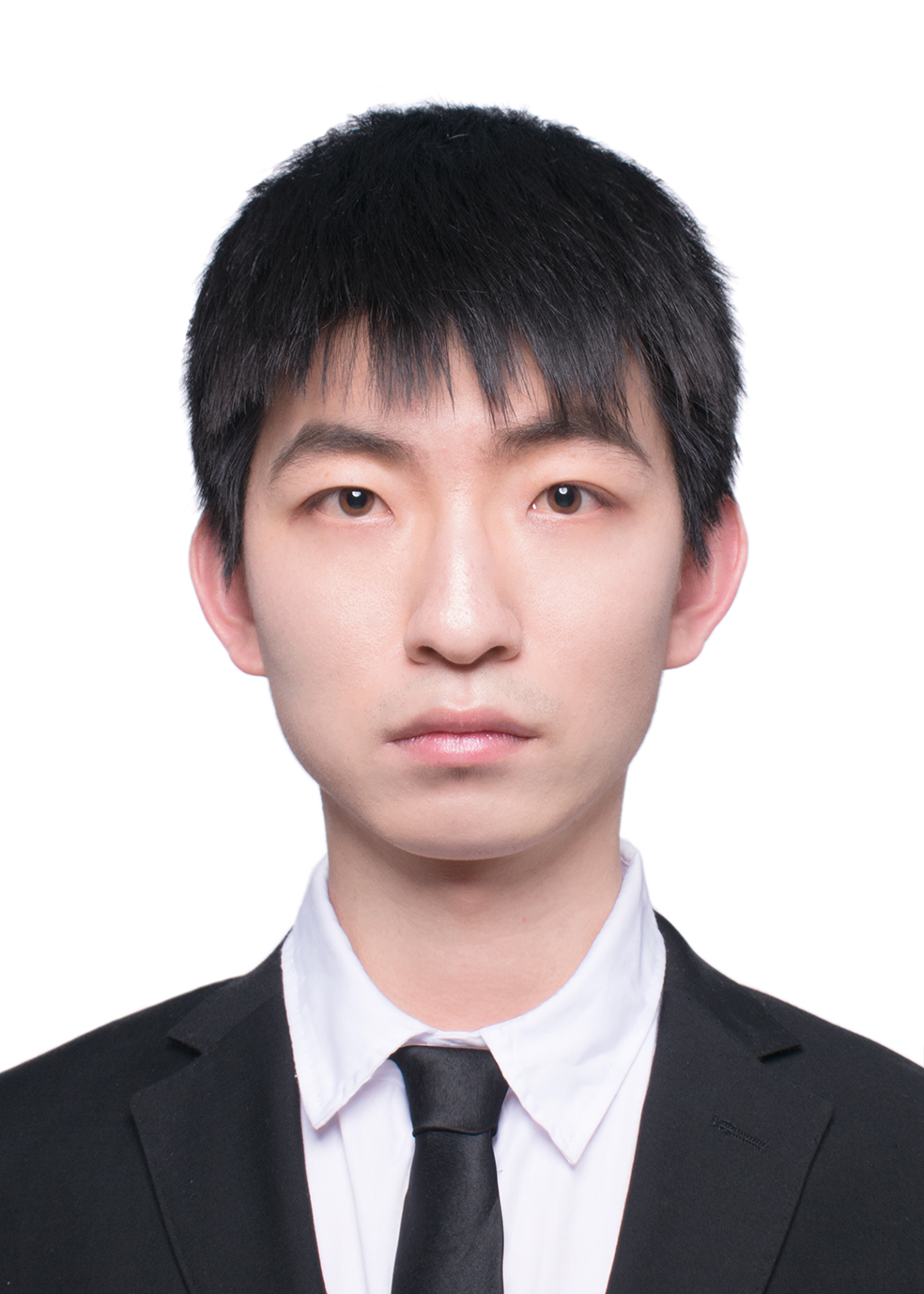}}]{Xinyu Tong}
	received the B.E and Ph.D. degrees in the Department of Electronic Information and Electrical Engineering from Shanghai Jiao Tong University, Shanghai, China in 2015 and 2020. He is currently an associate research fellow in the College of Intelligence and Computing of Tianjin University. His research interests include wireless sensor networks and wireless localization. His research papers were published in many prestigious journals and conferences including the IEEE Transactions on Mobile Computing, IEEE/ACM Transactions on Networking, MobiCom, UbiComp, INFOCOM and MobiSys, \textit{etc}.
\end{IEEEbiography}
\vspace{-3mm}
\begin{IEEEbiography}[{\includegraphics
[width=0.9in,height=1.2in,clip,
keepaspectratio]{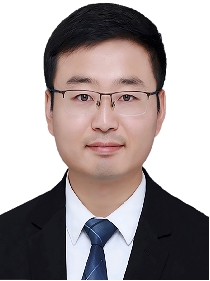}}]{Xiulong Liu}
	received the BE and PhD degrees	from the Dalian University of Technology, Dalian, China, in 2010 and 2016, respectively. He is currently a professor with the College of Intelligence	and Computing, Tianjin University, China. He also worked as a visiting researcher with Aizu University, Japan, a postdoctoral fellow with The Hong Kong Polytechnic University, Hong Kong, and a postdoctoral fellow with the School of Computing Science, Simon Fraser University, Canada. His research interests include wireless sensing and	communication, indoor localization, networking, \textit{etc}. His research papers were published in many prestigious journals and conferences including the IEEE Transactions on Mobile Computing, IEEE/ACM Transactions on Networking, IEEE Transactions on Computers, IEEE Transactions on Parallel and Distributed Systems, IEEE Transactions on Communications, INFOCOM, and ICNP, \textit{etc}.
\end{IEEEbiography}
\vspace{-3mm}
\begin{IEEEbiography}[{\includegraphics
[width=0.9in,height=1.2in,clip,
keepaspectratio]{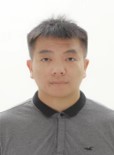}}]{Xin Xie}
	received the B.E. and Ph.D. degrees from the School of Computer Science and Technology, Dalian University of Technology, China, in 2013 and 2019, respectively. He is currently an associate professor with the College of Intelligence and Computing, Tianjin University, China. He also worked as a visiting scholar with the Department of Computer Science, Purdue University, USA, and a postdoctoral fellow with the Department of Computing, The Hong Kong Polytechnic University, Hong Kong SAR. His research interests include IoT, edge intelligence and network systems.
\end{IEEEbiography}
\vspace{-3mm}
\begin{IEEEbiography}[{\includegraphics
[width=0.9in,height=1.2in,clip,
keepaspectratio]{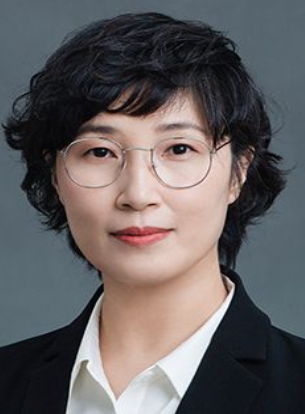}}]{Wenyu Qu}
	received the bachelor’s and master’s degrees from the Dalian University of Technology, China, in 1994 and 1997, respectively, and the Ph.D. degree from the Japan Advanced Institute of Science and Technology, Japan, in 2006. She was a professor with Dalian Maritime University, China, from 2007 to 2015. She was an assistant professor with the Dalian University of Technology, China, from 1997 to 2003. She is currently a professor with the College of Intelligence and Computing, Tianjin University. She has authored over 80 technical articles in international journals and conferences. Her research interests include cloud computing, computer networks, and information retrieval. She is on the committee boards for a couple of international conferences.
\end{IEEEbiography}

\newpage
\end{document}